\documentclass[%
 aip,
 amsmath,amssymb,
 reprint,%
]{revtex4-1}

\usepackage{graphicx}
\usepackage{dcolumn}
\usepackage{bm}
\usepackage{color}
\usepackage[utf8]{inputenc}
\usepackage[T1]{fontenc}
\usepackage{mathptmx}

\usepackage{subcaption}

\DeclareMathOperator{\csch}{csch}
\DeclareMathOperator{\sech}{sech}

\begin{document}

\preprint{AIP/123-QED}

\title[The Singular HI Between Two Spheres In Stokes Flow]{The Singular Hydrodynamic Interactions Between Two Spheres In Stokes Flow}

\author{B. D. Goddard}
\affiliation{School of Mathematics and Maxwell Institute for Mathematical Sciences, University of
Edinburgh, Edinburgh EH9 3FD, United Kingdom}
\author{R. D. Mills-Williams}%
 \email{r.mills@ed.ac.uk}
\affiliation{School of Mathematics and Maxwell Institute for Mathematical Sciences, University of
Edinburgh, Edinburgh EH9 3FD, United Kingdom}%
\author{J. Sun}
 \affiliation{School of Engineering, University of Edinburgh, Edinburgh EH9 3JL, United Kingdom}

\date{\today}

\begin{abstract}
We study exact solutions for the slow viscous flow of an infinite liquid caused by two rigid spheres approaching each either along or parallel to their line of centres, valid at all separations. This goes beyond the applicable range of existing solutions for singular hydrodynamic interactions (HIs) which, for practical applications, are limited to the near-contact or far field region of the flow. For the normal component of the HI, by use of a bipolar coordinate system, we derive the stream function for the flow as $Re\to 0$ and a formula for the singular (squeeze) force between the spheres as an infinite series. We also obtain the asymptotic behaviour of the forces as the nondimensional separation between the spheres goes to zero and infinity, rigorously confirming and improving upon known results relevant to a widely accepted lubrication theory. Additionally, we recover the force on a sphere moving perpendicularly to a plane as a special case. For the tangential component, again by using a bipolar coordinate system, we obtain the corresponding infinite series expression of the (shear) singular force between the spheres. All results hold for retreating spheres, consistent with the reversibility of Stokes flow. We demonstrate substantial differences in numerical simulations of colloidal fluids when using the present theory compared with existing multipole methods. Furthermore, we show that the present theory preserves positive definiteness of the resistance matrix $\bm{R}$ in a number of situations in which positivity is destroyed for multipole/perturbative methods.
\end{abstract}

\maketitle

\section{Introduction}\label{sec:intro}
Hydrodynamic interactions (HI) between bodies immersed in viscous fluid have been shown to be important in modelling many complex fluid phenomena in physics, biology and engineering.  For example, in suspensions of cornstarch and other solid particles of micron sizes at high solid volume fractions, the interplay between HI and particle contacts gives rise to a sudden increase in viscosity with increasing shear stress  \cite{fall2008shear, Lin:2015gf}. HI also affects complex fluid behaviour at many length scales.  At the small scale, the flow properties of suspended particles in emulsions and gels have historically determined their physical and chemical classification. In hemodynamics, blood is a suspension of platelets, white cells and high fractions of red cells in plasma, where fluidity and stability may be significantly altered during disease processes \cite{errill1969rheology}. On larger scales, the formation of topographical features under sea water is due to turbidity currents, where both inertial effects and slow motion of the suspensions are important \cite{bagnold1962auto}. In terms of numerical modelling, such as dynamical density functional theory formalisms for two dimensional colloidal flow, the inclusion of HI is enough to alter the dynamics of the density even when solving for dilute particle collections \cite{goddard2012general}.

The relevance and applicability of HI are therefore well established in many fluid flow problems in science and engineering. 
Many physical models for the flow of particles accounting for such phenomena have largely varying spatial scales which makes their computation challenging. Generally speaking, a numerical model that accurately predicts complex fluid phenomena requires the full knowledge of the HI between the suspended particles.  
In Stokesian dynamics (SD), the quasi-static motion of a suspension of $N$ rigid spherical particles at low Reynolds number is given by \cite{ball1997simulation}
\begin{align}\label{eq:particle_dynamics}
\bm{F}^{\text{diss}}(\vec{r}^N, \vec{v}^N) + \bm{F} = \bm{M}\frac{\mathrm{d}\vec{v}^N}{\mathrm{d}{t}},
\end{align}
where $\bm{M}$ is a mass matrix, $\bm{F}^{\text{diss}}$ is the dissipative force due to the HI of the particles mediated by the solvent fluid, $\vec{r}^N = [\vec{r}_1,\cdots \vec{r}_N]^\top$ is a vector of $6N$ particle position coordinates and $v^N = [\vec{v}_1,\cdots \vec{v}_N]^\top= \mathrm{d}\vec{r}^N/\mathrm{d}t$. The vector $\bm{F}$ accounts for conservative and non-conservative applied forces for example: the force due to gravity and the frictional force applied to the particle surfaces at contact, respectively. By nondimensionalising \eqref{eq:particle_dynamics} with an appropriately defined Reynolds number $Re$, the dissipative forces are taken as linear in the velocity of the particles, and after setting $Re = 0$, equation \eqref{eq:particle_dynamics} reads
\begin{align}\label{eq:slow_flow_particle_conformation_eqn}
-\bm{R}(\vec{r}^N)v^{N}+\bm{F} = 0,
\end{align}
where $\bm{R}$ is the resistance matrix for the conformation of particles with position vector $\vec{r}^N$. As is standard in the theory, $\bm{R}$ is independent of the properties of the solvent fluid, as well as the magnitudes and directions of the particle velocities. Rather, $\bm{R}$ depends only on the particle separations and sizes. Note also that by adding a noise term to \eqref{eq:slow_flow_particle_conformation_eqn}, correlated to the thermal fluctuations of the solvent fluid according to the generalised fluctuation-dissipation theorem \cite{plischke1994equilibrium}, one may obtain the dynamics of Brownian motion. 

In theory $\bm{R}$ has a large bandwidth, owing to $N$- body interactions. As in SD, in order to solve for the particle velocities, one must invert a dense matrix in $O(N^3)$ operations which will be computationally expensive. Approximations to $\bm{R}$ may be made in order to reduce the computational cost for SD simulations. For example, Ball and Melrose \cite{ball1997simulation} showed that $\bm{R}$ is made sparse by approximating the full $N$-body interactions to a two-body formalism of long range forces, with elements decaying as $1/r_{ij}$, where $r_{ij} = |\vec{r}_i-\vec{r}_j|$ is the distance between the centres of sphere $i$ and $j$ (c.f. Rotne-Prager \cite{rotne1969variational}). Such an approximation of $\bm{R}$ is valid for non-dense systems, and in this dilute regime, the hydrodynamic force due to lubrication is dominated by the long range mobility force. 

Conversely, in the highly concentrated regime the lubrication forces will dominate the elements of $\bm{R}$. This may be justified by expanding $\bm{R}$ in moments including the one, two, three, $\cdots$, $n$- body interactions. One finds that the pairwise lubrication forces dominate the expansion and higher order effects found using far-field expansions such as the method of reflections will fade in comparison due to the divergent scalar functions of the lubrication forces \cite{bossis1984dynamic}. 

\subsection*{The Model for the Resistance Matrix $\bm{R}$}
For the present analysis, we are interested in highly concentrated systems. We specify the three approximations we make in our construction of $\bm{R}$. 

\begin{description}
\item[A1] The HI are {\bf{lubrication dominated}}, that is, the divergent interactions between close surfaces dominate the elements of $\bm{R}$ in the highly concentrated regime. 
\item[A2] The HI are {\bf{strongly coupled}} and we neglect $n$-body HI for $n>2$. 
\item[A3] The HI are {\bf{frame-invariant}}; the justification being that the solvent fluid (over large enough distances) comoves with the particles. 
\end{description}

Assumption {\bf{A3}} says that for a steady solvent velocity $\vec{u}$ of a Stokes fluid in a domain $\Omega$ one has
\begin{align}
\frac{1}{|\Omega |}\int_{\Omega}\mathrm{d}\vec{r}\, \vec{u}(\vec{r}) = \frac{1}{N}\sum_{i = 1}^N \vec{v}_i.
\end{align}
Such an assumption is not valid for sedimentation problems, where the solvent velocity $\vec{u} = 0$ in $\Omega$ and the sphere velocities are collinear and non-zero. Non-frame-invariant simulations of Brownian motion in shear flow show shear induced ordering at low volume fractions, deviating from experimental observations \cite{evans1986shear}. We may however relax {\bf{A3}} by rewriting the resistance matrix, as we will in Section \ref{sec:numerics}.

With Assumptions {\bf{A1}}, {\bf{A2}}, {\bf{A3}} we now present our model for the resistance matrix $\bm{R}$. For a finite Reynolds number, and in components, the force balance in \eqref{eq:particle_dynamics}, in the absence of external and contact forces, is given by an equation for the velocity $\vec{v}_i$ of the $i^\text{th}$ particle
\begin{multline}\label{eq:sim_particle_hydro}
Re\,\dot{\vec{v}}_i =-\sum_{j=1}^N a(\vec{n}_{ij})(\vec{v}_i-\vec{v}_j) \cdot \hat{n}_{ij}\otimes \hat{n}_{ij}\\
+b(\vec{n}_{ij})(\vec{v}_i-\vec{v}_j) \cdot (\mathbb{I}-\hat{n}_{ij}\otimes \hat{n}_{ij})
\end{multline}
for $1\leq i\leq N$, where 
$\vec{n}_{ij}$ ($\hat{n}_{ij}$) is the (normalised) vector pointing between the centre of sphere $j$ to $i$, and $\mathbb{I}$ is the identity tensor. 

Here $a(\vec{n}_{ij})$ and $b(\vec{n}_{ij})$ are the normal and tangential components of the hydrodynamic interaction respectively as functions of $\vec{n}_{ij}$. A crucial observation is that in the diffuse system limit, both $a(\cdot)$ and $b(\cdot)$ should decay to unity so that Stokes law is recovered: the total force on particle $i$ is proportional to its velocity with proportionality constant Stokes unit $-\gamma$. In terms of the spectral properties of $\bm{R}$, this means the eigenvalues must be degenerate in the dilute sphere limit, and the general solution to \eqref{eq:sim_particle_hydro} (after setting $Re = 1$) becomes $\vec{v}_i(t) = e^{-\gamma t}\sum_{j = 1}^Nc_j\vec{e}_j$ for $\{e_j\}_{j=1}^N$ a basis of $R^{3N}$ and $c_j$ constants dependent on the initial velocity data.

Additionally, particular if $\vec{v}_i=\vec{v}_j = \vec{c}$ for all $i, j$ then the total HI force on each particle is zero, in the reference frame co-moving at velocity $\vec{c}$ . This is equivalent to saying that $\bm{R}$ has a zero eigenvalue associated with the translation of the entire system of particles at some uniform velocity, or that the interaction model is Galilean invariant. 

We may expand the summation in \eqref{eq:sim_particle_hydro} and collect together terms multiplying $\vec{v}_i$ to define the resistance matrix $\bm{R}$ in block form, here determined by diagonal and off-diagonal submatrices and $\bm{Z}_1$, $\bm{Z}_2$ respectively. We have
\begin{align}
\bm{R} = \begin{pmatrix}
\sum_{i\neq l} \bm{Z}_1(\vec{r}_1,\vec{r}_l) & \bm{Z}_2(\vec{r}_1,\vec{r}_2)& \cdots & \bm{Z}_2(\vec{r}_1,\vec{r}_N)\\
\bm{Z}_2(\vec{r}_2,\vec{r}_1) & \sum_{i\neq l} \bm{Z}_1(\vec{r}_2,\vec{r}_l) & \cdots & \vdots \\
\vdots & \vdots & \ddots & \vdots \\
\bm{Z}_2(\vec{r}_N,\vec{r}_1) & \cdots & \cdots  &  \sum_{i\neq l} \bm{Z}_1(\vec{r}_N,\vec{r}_l)
\end{pmatrix},\label{eq:R_def_blockwise}
\end{align}
where the block matrices $\bm{Z}_1$ and $\bm{Z}_2$ are defined as
\begin{align}
\bm{Z}_1(\vec{r}_i,\vec{r}_l) &= -a(r_{il})\frac{\vec{r}_i\otimes\vec{r}_l}{r_{il}^2} - b(r_{il})\left(\bm{1}-\frac{\vec{r}_i\otimes\vec{r}_l}{r_{il}^2}\right)\\
\bm{Z}_2(\vec{r}_i,\vec{r}_j) &= a(r_{ij})\frac{\vec{r}_i\otimes\vec{r}_j}{r_{ij}^2} + b(r_{il})\left(\bm{1}-\frac{\vec{r}_i\otimes\vec{r}_j}{r_{ij}^2}\right)
\end{align}
and where $r_{il} = |\vec{r}_i-\vec{r}_j|$ and $a(\cdot)$, $b(\cdot)$ are the scalar resistance functions corresponding to the divergent squeezing and shearing lubrication interactions of the close surfaces at high concentrations respectively. We note that the block-wise notation of \eqref{eq:R_def_blockwise} with summations on the diagonal is standard notation in statistical mechanical models of suspensions such as dynamic density functional theories (DDFTs), see \cite{goddard2012general}, \cite{goddard2012unification}. Note that the rows of $\bm{R}$ sum to zero, which implies that whenever $\vec{v}^N =c_0 \vec{e}_{i}$ for some constant vector $c_0\in \mathbb{R}$, and $e_{i}$ a basis vector of $\mathbb{R}^{3N}$, then $\vec{v}^N\in \ker \bm{R}$ and the interaction is Galilean invariant.

With the model for the resistance matrix $\bm{R}$ defined we now discuss the model for the scalar resistance functions which make up the elements of $\bm{R}$. 

\subsection*{The Model for the Scalar Resistance Functions $a(\cdot) $ and $b(\cdot)$}
For short range HI current models use asymptotic formulae for $a(\cdot)$ and $b(\cdot)$, for example the expressions found in Kim \& Karrila \cite{kim2013microhydrodynamics}, valid in a `close' region of particle separation, combined with an arbitrary outer cut-off. It would be preferable to have a formula for both $a(\cdot)$ and $b(\cdot)$ valid at all particle distances so that arbitrary cut-offs are avoided. This property is particularly desirable in continuum formalisms, where the HI appear as convolution integrals with a separate additive Stokes term. The convergence of such integrals requires knowledge of the behaviour and decay of the scalar resistance functions over the entire support of the hard sphere number density for accurate numerical solutions. As such, this paper provides a derivation and analysis of both resistance functions $a(\cdot)$ and $b(\cdot)$ valid at all particle separations. The analytical $b(\cdot)$ for two spheres of unequal radii is not considered in the main text, because we found that in this case, the boundary equations which need to be solved for the final set of series coefficients are an intractable system of coupled recurrence equations requiring dedicated computer algebra. 

We determine $a(\cdot)$ and the corresponding stream function at all particle separations, which, to our knowledge, has not been previously obtained. We restrict the calculations to two non-rotating spheres with opposite velocities. By the linearity of Stokes equations however, the angular component of the stream function for two approaching spheres rotating asymmetrically may be linearly superimposed. 

To compare to existing results, we provide in the following section a history of slow viscous flow problems for two spheres.

\subsection{History of Slow Viscous Flow Problems for Two Spheres}
The singular HIs for each of the scalar resistance functions $a(\cdot)$ and $b(\cdot)$ which are computed in this paper take the general form of infinite series. These are not the same solutions to problems for two spheres in bipolar coordinates previously considered, e.g., \citet{stimson1926motion}, \citet{goldman1966slow}. It is the boundary condition choice, entire regime of validity, and singular nature of the HI that distinguishes this from previous works, described as follows. 

The classical work concerning exact solutions for two spheres with equal velocities in viscous flow was presented by \citet{stimson1926motion} for two \emph{drafting} spheres. Similarly, \citet{goldman1966slow} consider two spheres \emph{settling side by side} for a single mode of tangential interaction. Our derivations use the same formalisms but with opposite velocities, leading to the $\epsilon^{-1}$ and $\log \epsilon^{-1}$ singularities respectively. In that paper~\cite{stimson1926motion}, there are two errata: firstly, for the first equation of their section 4, the factor inside the square bracket $-(1-\mu^2)$ should be $(1-\mu^2)$ (where $\mu = \cos\xi$ in their notation, we use $\mathfrak{x} = \cos \xi$. See List of Notation \ref{sec:list_of_notation}), secondly their equation (37) for $\lambda$, a nondimensional force, is defined as half the correct value as noted in \citet{happel2012low}. While on the subject of errata, we refer the reader to \citet{townsend2018generating} for a discussion and derivation of corrections to the scalar resistance functions computed in \citet{jeffrey1984calculation}.

Not long after the result of \citet{stimson1926motion}, \citet{faxen1927} gave a value of the hydrodynamic force on the two drafting spheres at contact.  Both results have since been validated by \citet{bart1959interaction}, who experimentally measured the force on two equal spheres settling under gravity in viscous fluid and showed good agreement with the theoretical value. Later work by \citet{maude1961end}, adapting \citet{stimson1926motion}, calculated the finite-size effects of a falling-sphere viscometer. Hence the chosen bipolar formalism for exact solutions has good experimental validation as a method to compute flow around two spheres. 

The subsequent history of the mathematical treatment of viscous flow around two spheres can be divided into two classes: exact and approximate. In the exact class, notable results are obtained by employing bipolar coordinates to solve for the fluid velocity and hydrodynamic force. Boundary condition cases include those due to \citet{o1964slow}, considering the parallel motion of a sphere to a plane wall; \citet{o1970asymmetrical} treating the rolling and translating motion parallel to a stationary sphere in viscous fluid; \citet{goldman1966slow} studying the motion of two spheres settling under gravity; and finally \citet{cox1967slow} treating the motion of a sphere normal to a plane wall and considering the asymptotic limits at small separations.  The asymptotic methods presented in this paper are analogous to those in \citet{cox1967slow}, also similar to a treatment by \citet{hansford1970converging}, but therein the work is based on the constants determined by \citet{brenner1961slow}. The asymptotics in the present work go beyond the statement that the $O(1)$ term cannot be obtained by asymptotic analysis (see \citet{kim2013microhydrodynamics}, chapter 7).

There have also been more recent studies and applications of the solutions arising from the bipolar coordinate system, e.g., by \citet{papavassiliou2017exact} which concerns the motion of a sphere in viscous flow near a convex shell. For completeness, the study of droplets should be mentioned: \citet{wacholder1972slow} considered the exact solution to Stokes equations both inside and outside spherical droplets with equal settling velocities, and \citet{haber1973low} generalised the former to two spherical droplets of different viscosities. Both of these studies concern a non-singular hydrodynamic interaction between droplets, which is different to the present boundary condition choice. 

In the approximate class lie techniques such as the method of reflections (a series solution best suited for widely separated spheres \cite{kim2013microhydrodynamics}) and lubrication theory (solving Stokes equations directly by a perturbation expansion). Notable publications are, e.g., by \citet{jeffrey1982low} on which a popular reference for the singular hydrodynamic force between two collinear spheres in viscous fluid \citet{kim2013microhydrodynamics} is based. The derivation by perturbation methods in the latter, apart from algebraic errors not affecting the final result, is not valid as the sphere separation increases. This means arbitrary truncations must be used for numerical implementation \cite{townsend2018anomalous}. The choice of location of the cut-off and convergence of the truncated expressions remains mysterious. A fundamental assumption shared by these formalisms is the choice of scaling ratio between the cylindrical coordinates $z/r\sim\epsilon^{1/2}$ defining a singular perturbation problem, which has not been justified until the analysis in the present work. In particular we show this scaling is correct by expanding the bipolar coordinate system and infinite series around the singular contact point.

An alternative approach is the multipole method. To do this for our chosen sphere configuration, one would compute the velocity and pressure fields using the method of reflections around the two sphere centres, separated by a distance $R$.  Using the addition theorems for spherical harmonics, the pressure and velocity are written as linear combinations of Lamb's solutions to Stokes equations. However, this results in an infinite set of series coefficients for the velocity and pressure, which are obtained only in the form of another series in $R^{-1}$, the coefficients of which satisfy known but non-analytical recurrence relations \cite{kim2013microhydrodynamics}. The method is by no means explicit, only obtaining Taylor series representations of the velocity and pressure fields and requires unavoidable computer algebra. What is more, to compute the hydrodynamic force on two spheres to a given accuracy will require ever more expansion terms as $R$ decreases, making the method computationally unfavourable in the near-contact limit.

In this paper we give the first quantitative comparison, for this particular two-sphere interaction, between the present solution obtained by spherical bipolar methods and the one obtained by the multipole methods\cite{jeffrey1984calculation}. As a result, we are able to highlight the analytical and practical strengths of the present work by implementing both the novel and existing results in a numerical example for colloidal flow.

\subsection{Organisation of the Paper}
This paper presents the rigorous derivation and asymptotic analysis of the singular scalar resistance functions $a(\cdot)$ and $b(\cdot)$, valid for all non-contacting particle separations. In Section \ref{sec:bipolar} we provide the definition of the bipolar coordinate system. Following this, in Section \ref{sec:stokes} we present the steady flow equations. Section \ref{sec:normal_interaction} concerns the steady flow equations for the normal interaction and in Section \ref{sec:force} we calculate the scalar resistance function, $a(\cdot)$, as an infinite series. In Appendix \ref{sec:asymptotics} we derive rigorous small and large argument limits of our expression for $a(\cdot)$, as well as showing agreement with the  perpendicular motion of a sphere and plane. In Section \ref{sec:comparetomultipole} we compare our results for $a(\cdot)$ to the widely used expressions determined by the method of multipole expansions. In Section \ref{sec:tang_interaction} we consider the steady equations for the tangential interaction and in Section \ref{sec:tang_force_compute} we compute the scalar resistance function, $b(\cdot)$, as an infinite series. In Section \ref{sec:positivity} we examine the positivity of $\bm{R}$ built by our scalar resistance functions and existing expressions. Section \ref{sec:numerics} sees the implementation the results of this work in a numerical computation to show substantial differences in flows of colloidal suspensions compared with using existing expressions for scalar resistance functions. In Section \ref{sec:discussion} we make our conclusions and discuss open problems. Finally, in Appendices \ref{appA}, \ref{app:derivation_of_tang_fields} and \ref{sec:list_of_notation} we provide useful formulae, a derivation of the tangential scalar fields and a list of notation. 

\section{Spherical Bipolar Coordinates}\label{sec:bipolar}
The spherical bipolar coordinate system is a convenient setting in which to apply the boundary conditions on both spheres. The coordinate transformation from cylindrical coordinates $\mathfrak{\vec{r}}=(r,\, z,\, \theta)$ to spherical bipolar coordinates $\mathfrak{\vec{q}}=(\eta,\,\xi, \, \theta)$ is
\begin{align}\label{eq:bipolar_trans}
z+ir = ic\cot \tfrac{1}{2}(\xi+i\eta)
\end{align}
where $\theta$ remains unchanged, $i=\sqrt{-1}$ and $c>0$ is a geometrical constant. Every point in $(r,\, z)$ space is represented uniquely in $(\eta, \xi)$ space, so long as $\xi\in[0,\pi]$, $-\infty <\eta <\infty$, $\theta\in[0,2\pi)$. Expanding the cotangent and equating real and imaginary parts one obtains
\begin{align}\label{eq:eqn_for_z_and_r}
z(\eta,\xi) = \tfrac{c\sinh\eta}{\cosh\eta-\cos\xi}, \qquad  r(\eta,\xi) = \tfrac{c\sin\xi}{\cosh\eta-\cos\xi}.
\end{align}

\begin{figure}
\centering
\includegraphics[width=3.3in]{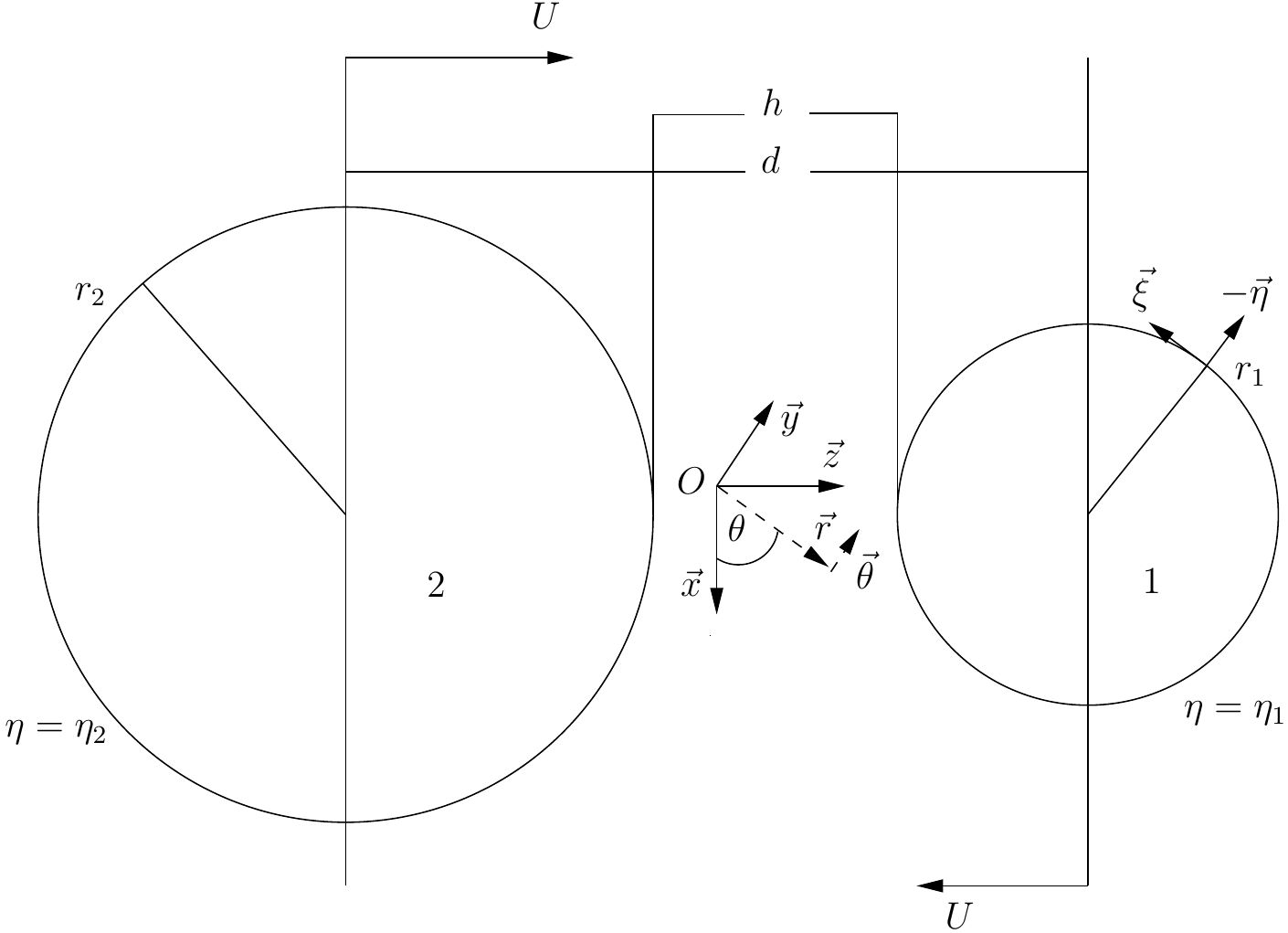}
\caption{Schematic of two unequal spheres of radii $r_1$, $r_2$ converging along their line of centres in viscous fluid. Included in the diagram are the cylindrical and bipolar unit vectors, the dimensional gap distance $h$, and centre to centre distance $d$. Note that $\eta_1$ and $\eta_2$ are implicit functions of $r_1$, $r_2$ and $d$.}\label{fig:two_particle_side_squeeze_bipolar}
\end{figure}

There is a one to one correspondence between $\mathfrak{\vec{r}}$ and $\mathfrak{\vec{q}}$ except at the limiting points $\eta = \pm \infty$ where $\xi$ is multivalued. Geometrically this occurs when the spheres are vanishingly small, or remotely separated. As such, these points indicate the limit direction in which to obtain classical Stokes drag. The surfaces $\eta =$ constant are non-intersecting coaxial spheres with centres at the Cartesian coordinates $(r,z)=(0,\, c \coth \eta)$ and radii $c|\csch\eta|$. Denoting the centre distance from sphere $i$ to the origin $O$ by $d_i$ and its radius by $r_i$,
we identify the bipolar ordinates defining sphere 1 and 2 as
\begin{align*}
\cosh\eta_1:=\tfrac{d_1}{r_1}, \qquad \cosh\eta_2:= \tfrac{d_2}{r_2}.
\end{align*}
Note that $\eta_1>0$ and $\eta_2<0$. The geometry is summarised in Figure \ref{fig:two_particle_side_squeeze_bipolar}.

\section{Steady Flow Equations}\label{sec:stokes}
Consider the steady incompressible Navier-Stokes equations governing the evolution of the fluid velocity $\vec{u}$ and pressure $p$ in an unbounded domain $\Omega$ outside of the spheres:
\begin{equation}
Re\left(\vec{u}\cdot \nabla\right)\vec{u}=-\nabla p +\nabla^2 \vec{u},\qquad 
\nabla\cdot \vec{u}=0 \label{eq:NS_eqn}
\end{equation}
where $Re = \rho\, U L /\mu$ for $U$ a characteristic velocity, $L$ a characteristic length, $\rho$ the fluid density and $\mu$ the dynamic viscosity. Here $\rho$ and $\mu$ are assumed to be constant. 

In the following section we consider the analytical solution of \eqref{eq:NS_eqn} for the case of two approaching collinear spheres.

\section{Normal Interaction $a(\cdot)$}\label{sec:normal_interaction}

For axisymmetric flow the assumed existence of a stream function $\psi$ permits \eqref{eq:NS_eqn} to be recast into
\begin{align}\label{eq:steady_stream_fun_eqn}
\tfrac{1}{Re\, r}L_{-1}^2\psi =\partial_z\psi\, \partial_r (\tfrac{1}{r^2}L_{-1}\psi)-\partial_r\psi\, \partial_z (\tfrac{1}{r^2}L_{-1}\psi).
\end{align}

The differential operator $L_{-1}$ is a member of a class of axisymmetric potential operators $L_k:=\partial_z^2+\partial_r^2+kr^{-1}\partial_r$ for $k\in(-\infty,\infty)$, for which compact formulae hold. In particular,
by use of the chain rule and the Cauchy-Riemann equations for $z$ and $r$,
it is straightforward to obtain $L_k = r^{-k}\mathfrak{h}^{-2}[\partial_\xi(r^k\partial_\xi)+ \partial_\eta(r^k\partial_\eta)]$.
Here $\mathfrak{h}$ is the metrical coefficient arising from the transformation between coordinate systems, defined by $\mathfrak{h}^{2} =(\partial_\xi z)^2 +(\partial_\xi r)^2=(\partial_\eta r)^2 +(\partial_\eta z)^2= c^2/(\cosh\eta-\cos\xi)^2$.
After setting $k=-1$ the first approximation to the flow around $Re=0$ yields the biharmonic equation subject to two no slip and two no flux conditions
\begin{align}
\left[\tfrac{r}{\mathfrak{h}^{2}}\left(\partial_\xi\left(r^{-1}\partial_\xi\right)+ \partial_\eta\left(r^{-1}\partial_\eta\right) \right )\right]^2\psi&=0, \qquad \text{ in } \Omega\label{eq:stokes_eqn_B_op}\\
\psi\pm\tfrac{Ur^2}{2}=0, \quad \partial_n\left(\psi\pm\tfrac{Ur^2}{2}\right) &=0, \label{bc:boundary_cond_stream_fb}
\end{align}
where the positive sign is taken on sphere $1$ and the negative sign on sphere $2$.

\subsection{Solution in Spherical Bipolar Coordinates}\label{sec:sol_in_sph_bip_coor}
To solve the PDE \eqref{eq:stokes_eqn_B_op}--\eqref{bc:boundary_cond_stream_fb} it is sufficient to write $\psi = \psi_1+z\,\psi_2$ where  $L_{-1}\psi_1=L_{-1}\psi_2=0$. This ansatz may be heuristically justified by reference to \citet{payne1960stokes}.
A solution for $\psi_1$ is found by
$\psi_1 = r^{1/2} f(\xi)g(\eta)$ with
\begin{align}\label{eq:sep_of_vars_of_psi}
f''+(\lambda^2-\tfrac{3}{4\sin^2\xi})f=0, \qquad g''-\lambda^2g=0,
\end{align}
where $\lambda^2$ is a separation constant. 
The transformation $f=(\bar{\xi}^2-1)^{1/4}\bar{f}$ yields the Legendre equation 
$$(1-\bar{\xi}^2 )\bar{f}''-2\bar{\xi} \bar{f}'+(\lambda^2-1/4-(1-\bar{\xi^2})^{-1})\bar{f}=0$$
with order $1$ and degree $n=\lambda-1/2$ for $n$ a non-negative integer. Thus
by various recurrence relations of the Legendre functions and the principle of linear superposition, one has
%
\begin{align}\label{eq:complete_summation_for_psi_1}
\psi_1(\xi,\eta)& =\sum_{n=1}^{\infty}\left[a_n\cosh(n+\tfrac{1}{2})\eta+b_n\sinh(n+\tfrac{1}{2})\eta\right] \nonumber\\
& \qquad \times \tfrac{Q_n(\cos\xi)}{\sqrt{\cosh\eta-\cos\xi}}
\end{align}
where $Q_n:=P_{n+1}-P_{n-1}$. It is elementary to see that the $Q_n$ satisfy the ODE 
\begin{align}\label{eq:ode_for_Qn}
(1-\mathfrak{x} ^2) Q_n''(\mathfrak{x} )+n(n+1)Q_n(\mathfrak{x} )&=0
\end{align}
and the recursion relation
\begin{align}\label{eq:Qn_recurrence_relation}
\mathfrak{x} \,Q_n(\mathfrak{x} ) = \tfrac{n+2}{2n+3}Q_{n+1}(\mathfrak{x} )+\tfrac{n-1}{2n-1}Q_{n-1}(\mathfrak{x} ).
\end{align} 
Using equation \eqref{eq:complete_summation_for_psi_1}
the stream function may be constructed as
%
\begin{equation}
\begin{split}\label{eq:stream_fn_good_form}
\psi(\xi,\eta)  &= \left(\cosh\eta-\cos\xi\right)^{-3/2} \chi(\xi,\eta),\\
\chi(\xi,\eta)&:=\sum_{n=1}^{\infty}Q_n(\cos\xi)R_n(\eta)
\end{split}
\end{equation}
where
\begin{multline}\label{eq:expansion_eta_part}
R_n(\eta) := a_n\cosh (n+\tfrac{3}{2})\eta+b_n\sinh (n+\tfrac{3}{2})\eta\\
+c_n\cosh (n-\tfrac{1}{2})\eta +d_n\sinh (n-\tfrac{1}{2})\eta
\end{multline}
and $a_n$--$d_n$ are to be determined by the boundary conditions \eqref{bc:boundary_cond_stream_fb}. 

For later calculations, we provide the following useful relations 
\begin{align}
(2n+1)(1-\mathfrak{x} ^2)P_n(\mathfrak{x} )&=\tfrac{n(n-1)}{2n-1}Q_{n-1}(\mathfrak{x} )-\tfrac{(n+1)(n+2)}{2n+3}Q_{n+1}(\mathfrak{x} ),\label{eq:writing_one_minus_x_squared}\\
\tfrac{2n+1}{n+1}(1-\mathfrak{x} )^2 P_n'(\mathfrak{x} )&=-n\, Q_n(\mathfrak{x} )\label{eq:writing_one_minus_x_squared_der}
\end{align}
and the orthogonality conditions for the polynomials
\begin{align}
\begin{split}  \label{eq:legendre_ortho}
\int_{-1}^{1}P_m(\mathfrak{x} )P_n(\mathfrak{x} )\,\mathrm{d}\mathfrak{x}  &= \tfrac{2}{2m+1}\delta_{m,n},\\
\int_{-1}^{1}P_m(\mathfrak{x} )Q_n(\mathfrak{x} )\,\mathrm{d}\mathfrak{x}  &= \tfrac{2}{2m+1}\delta_{m, n+1}-\tfrac{2}{2m+1}\delta_{m, n-1}
\end{split}
\end{align}
where $\delta_{i,j}$ is the Kronecker delta.

\subsection{Boundary Conditions}
Now that the stream function is in the form \eqref{eq:stream_fn_good_form}, we combine \eqref{bc:boundary_cond_stream_fb} with the expressions for $r$ and $z$ in \eqref{eq:eqn_for_z_and_r} and rescale the stream function $\psi\sim U c^2/2 \,\psi'$ (immediately dropping primes), where $U$ is the instantaneous sphere speed, to obtain the transformed boundary conditions on sphere $j$
\begin{align}\label{bc:chi1_bcs}
\chi(\xi,\eta_j)=\tfrac{(-1)^j \sin^2\xi}{(\cosh\eta_j-\cos\xi)^{1/2}}, \quad
\partial_\eta\chi(\xi,\eta_j)= \tfrac{(-1)^j\sin^2\xi\sinh\eta_j}{2(\cosh\eta_j-\cos\xi)^{3/2}}.
\end{align}
We proceed to find $a_n$, $b_n$, $c_n$ and $d_n$ by using orthogonality of the $P_n$. 

In the case of no slip, using the formula for $\chi$ in \eqref{eq:stream_fn_good_form} and integrating over the interval $\xi\in[0, \pi]$, the sum and integral signs may be commuted using the dominated convergence theorem: Note that the truncated quantity $\int\mathrm{d}\xi\,|\sum_{n=1}^{N}R_n(\eta)\, Q_n(\xi)\, P_m(\xi)\, \sin\xi | \leq C_m |R_{m+1}(\eta)|$ where the constant $C_m$ is independent of $N$.
Writing $\mathfrak{x} =\cos\xi$ one obtains the integral
\begin{multline}
I_{m,j}:=(-1)^j\int_{-1}^{1}\tfrac{(1-\mathfrak{x} ^2) P_m(\mathfrak{x} )}{(\cosh\eta_j-\mathfrak{x} )^{1/2}}\,\mathrm{d}\mathfrak{x} \\
= \tfrac{2}{2m-1}R_{m-1}(\eta_j)-\tfrac{2}{2m+3}R_{m+1}(\eta_j)\label{bc:eta_j_sum_int}
\end{multline}
where we have used \eqref{eq:legendre_ortho}. 

The integrals may be evaluated by considering the Newtonian potential $(\cosh\eta_j-\mathfrak{x} )^{-1/2}=(\zeta^2+\zeta'^2-2\zeta\zeta'\mathfrak{x} )^{-1/2} = \zeta^{-1}\sum_{k=0}^{\infty}(\zeta'/\zeta)^kP_k(\mathfrak{x} )$ where $\zeta'=\zeta^{-1}/2$, $\zeta=e^{(-1)^{j-1}\eta_j/2}/\sqrt{2}$.
Using equation \eqref{eq:writing_one_minus_x_squared} with \eqref{bc:eta_j_sum_int} we find on sphere $j$ 
\begin{align*}
I_{m,j} &=(-1)^j\sqrt{2}\int_{-1}^{1}\mathrm{d}\mathfrak{x} \,(1-\mathfrak{x} )^2P_m(\mathfrak{x} )\tfrac{1}{\zeta}\sum_{k=0}^{\infty}\left(\tfrac{\zeta'}{\zeta}\right)^kP_k(\mathfrak{x} )\\
&=(-1)^j\sqrt{2}\int_{-1}^{1}\mathrm{d}\mathfrak{x} \,P_m(\mathfrak{x} ) \\
&\times\sum_{k=0}^{\infty}\tfrac{k(k+1)e^{(-1)^j(k+1/2)\eta_j}}{2k+1}\left[ \tfrac{(k+2)Q_{k+1}(\mathfrak{x} )}{k(2k+3)} -\tfrac{(k-1)Q_{k-1}(\mathfrak{x} )}{(k+1)(2k-1)}\right]\\
&=(-1)^j\sqrt{2}\sum_{k=0}^{\infty}\tfrac{k(k+1)e^{(-1)^j(k+1/2)\eta_j}}{2k+1}\\
& \times\left\lbrace \tfrac{(k+2)}{k(2k+3)}\left[\tfrac{2}{2k+5}\delta_{m, k+2}-\tfrac{2}{2k+1}\delta_{m, k}\right] 
\right.\\
&\left.  -\tfrac{(k-1)}{(k+1)(2k-1)}\left[\tfrac{2}{2k+1}\delta_{m, k} -\tfrac{2}{2k-3}\delta_{m, k-2}\right] \right\rbrace
\end{align*}
where we have used \eqref{eq:legendre_ortho}. Distributing the sum we find
\begin{multline*}
I_{m,j}=(-1)^j\sqrt{2}\sum_{k'=2}^{\infty}\tfrac{k'e^{(-1)^j(k'-1/2)\eta_j}}{2k'-1}
\tfrac{(k'+1)}{(2k'+1)}\\
\times \left[\tfrac{2}{2k'+3}\delta_{m, k'+1}-\tfrac{2}{2k'-1}\delta_{m, k'-1}\right] \\
-(-1)^{j}\sqrt{2}\sum_{k'=-1}^{\infty}\tfrac{(k'+1)e^{(-1)^j(k'+3/2)\eta_j}}{2k'+3}\tfrac{k'}{(2k'+1)}\\
\times \left[\tfrac{2}{2k'+3}\delta_{m, k'+1} -\tfrac{2}{2k'-1}\delta_{m, k'-1}\right] 
\end{multline*}
where we have made the substitutions $k=k'-1$ and $k=k'+1$ for the former and latter sums respectively. After applying the Kronecker deltas, we therefore find the equality
\begin{multline}\label{eq:eta_1_slip_manip}
\tfrac{(-1)^j2\sqrt{2}m(m+1)}{2m+1} \left[\tfrac{e^{(-1)^j(m-1/2)\eta_j}}{2m-1}-\tfrac{e^{(-1)^j(m+3/2)\eta_j}}{2m+3}\right]\\
 =  \tfrac{2}{2m-1}R_{m-1}(\eta_j)-\tfrac{2}{2m+3}R_{m+1}(\eta_j).
\end{multline}

For the no flux condition a similar dominating argument to that above again permits the interchange of the sum and integral signs. Expediently the no flux condition may be obtained by differentiating through \eqref{eq:eta_1_slip_manip} with respect to $\eta_j$ to find
\begin{multline}\label{eq:eta_1_flux_manip}
\tfrac{\sqrt{2}m(m+1)}{2m+1} \left[e^{(-1)^j(m-1/2)\eta_j}-e^{(-1)^j(m+3/2)\eta_j}\right]
\\ =  \tfrac{2}{2m-1}R_{m-1}'(\eta_j)-\tfrac{2}{2m+3}R_{m+1}'(\eta_j).
\end{multline}

\subsection{Linear System}\label{subsec:linsys}
With the boundary conditions in hand we define the right hand side vector
\begin{multline}
\vec{f}:=[-\tfrac{e^{-(n-1/2)}\eta_1}{2n-1}+\tfrac{e^{-(n+3/2)}\eta_1}{2n+3},\,
\tfrac{e^{(n-1/2)}\eta_2}{2n-1}-\tfrac{e^{(n+3/2)}\eta_2}{2n+3},\\
\tfrac{1}{2}(e^{-(n-1/2)\eta_1}-e^{-(n+3/2)\eta_1}),\,
\tfrac{1}{2}(e^{(n-1/2)\eta_2}-e^{(n+3/2)\eta_2})]^\top.
\end{multline}
The unknowns $\vec{a} = [a_n, b_n, c_n, d_n]^\top$ are determined by inverting the system of equations
$M\vec{a} =\vec{f}$, in particular
\begin{align}
&\Delta(n)\, a_n = (2n+3)\left[(2n+1)(n-\tfrac{1}{2})(\cosh 2\eta_1 - \cosh 2\eta_2 )\right. \nonumber\\
&-\left. 2\left( (2n-1)\sinh (n+\tfrac{1}{2})(\eta_1-\eta_2)\sinh (n+\tfrac{1}{2})(\eta_1+\eta_2)\right.\right. \nonumber\\
& -\left.\left.  (2n+1)\sinh (n+\tfrac{3}{2})(\eta_1-\eta_2)\sinh(n-\tfrac{1}{2})(\eta_1+\eta_2)\right)\right],\\
&\Delta(n)\, b_n  =  -(2n+3)\left[(2n+1)(n-\tfrac{1}{2})(\sinh 2\eta_2-\sinh 2\eta_1)\right.\nonumber\\
& -2\left((2n-1)\sinh (n+\tfrac{1}{2})(\eta_1-\eta_2)\cosh(n+\tfrac{1}{2})(\eta_1+\eta_2)\right.\nonumber\\
& \left.\left.-(2n+1)\sinh (n+\tfrac{3}{2})(\eta_1-\eta_2)\cosh(n-\tfrac{1}{2})(\eta_1+\eta_2) \right)\right. \nonumber\\
& \left.+ 4\cdot \exp\left\lbrace -(\eta_1-\eta_2)(n+\tfrac{1}{2})\right\rbrace\sinh(n+\tfrac{1}{2})(\eta_1-\eta_2)\right.\nonumber\\
& \left. + (2n+1)^2\exp\left\lbrace \eta_1-\eta_2\right\rbrace\sinh(\eta_1-\eta_2)\right], \\
&\Delta(n)\, c_n  = -(2n-1)\left[(2n+1)(n+\tfrac{3}{2})(\cosh 2\eta_1 -\cosh 2\eta_2)\right.\nonumber\\
& \left.+2\left((2n+3)\sinh(n+\tfrac{1}{2})(\eta_1-\eta_2)\sinh (n+\tfrac{1}{2})(\eta_1+\eta_2)\right.\right.\nonumber\\
& \left.\left.+(2n+1)\sinh (n+\tfrac{3}{2})(\eta_1+\eta_2)\sinh (n-\tfrac{1}{2})(\eta_2-\eta_1)\right)\right],\\
&\Delta(n)\, d_n = (2n-1)\left[(2n+1)(n+\tfrac{3}{2})(\sinh 2\eta_1-\sinh 2\eta_2)\right.\nonumber\\ 
&  +2\left((2n+3)\sinh(n+\tfrac{1}{2})(\eta_1-\eta_2)\cosh(n+\tfrac{1}{2})(\eta_1+\eta_2)\right.\nonumber\\
&  \left.\left. +(2n+1)\cosh(n+\tfrac{3}{2})(\eta_1+\eta_2)\sinh(n-\tfrac{1}{2})(\eta_2-\eta_1)\right)\right.\nonumber\\
& +4\cdot \exp\left\lbrace -(\eta_1-\eta_2)(n+\tfrac{1}{2})\right\rbrace\sinh(n+\tfrac{1}{2})(\eta_1-\eta_2)\nonumber\\
& \left.- (2n+1)^2\exp\left\lbrace -(\eta_1-\eta_2) \right\rbrace\sinh(\eta_1-\eta_2) \right].
\end{align}
Note that $M$ is defined explicitly in Appendeix~\ref{appA}, and we have defined 
\begin{multline}
\Delta(n) := \tfrac{(2n+1)(2n-1)(2n+3)}{\sqrt{2}n(n+1)}\\
\times[4\sinh^2(n+\tfrac{1}{2})(\eta_1-\eta_2)-(2n+1)^2\sinh^2(\eta_1-\eta_2)].\label{eq:def_of_Dellta_of_n}
\end{multline}

These coefficients are distinct from those found by \citet{stimson1926motion} because of the present choice in boundary conditions. Note that the method here is generalisable in the boundary conditions, demonstrating the utility of the coordinate system. A corollary of the result is that the calculations are valid for retreating spheres, since the change in boundary conditions is equivalent to the permutation of two rows of $M$, which amounts to a change in the sign of $\text{det}\,M$ and thus a global sign change on $a_n,b_n,c_n,d_n$. This can also be seen as a consequence of the reversibility of Stokes flow.

\section{The Force Experienced by the Spheres}\label{sec:force}
\citet{happel2012low} give exact expression for the force on a sphere in terms of the stream function in cylindrical coordinates, namely
\begin{align}\label{eq:stress_surface_integral}
\mathcal{F}_z=\mu\pi\int_{S}r^3\,\partial_n\left(\tfrac{L_{-1} \psi}{r^2}\right)\,\mathrm{d}s
\end{align}
where $\mu$ is dynamic viscosity, $S$ is a meridian line of the sphere and $\mathrm{d}s$ is an infinitesimal arc length measured in radians.
Assuming the summand decays sufficiently quickly to permit the interchange of the summation sign and two derivatives in $\xi$ and $\eta$, and $\mathfrak{x} =\cos\xi$, and performing the $\xi$ derivatives explicitly, the $n$th term of the integrand in \eqref{eq:stress_surface_integral} (before applying the normal derivative) takes the form
\begin{align}\label{eq:L_minus_one_integrand_res}
\left [\tfrac{L_{-1} \psi}{r^2}\right]_n&= \tfrac{(\cosh\eta-\mathfrak{x} )^{5/2}}{c^4(1-\mathfrak{x} ^2)}\left[Q_n(\mathfrak{x} )\left(R_n''(\eta)\right. \right. \\ 
& \quad \left.\left.  -\tfrac{2\sinh\eta}{\cosh\eta-\mathfrak{x} } R_n'(\eta) +\tfrac{3}{4}\tfrac{3x+\cosh\eta}{\cosh\eta-\mathfrak{x} }  R_n(\eta)\right) \right. \nonumber
\\
& \quad  \left. +(1-\mathfrak{x} ^2) R_n(\eta)\left(Q_n''(\mathfrak{x} )+\tfrac{2}{\cosh\eta-\mathfrak{x} }Q_n'(\mathfrak{x} )\right)\right].\nonumber
\end{align}
%

The infinitesimal line element of the integral \eqref{eq:stress_surface_integral} has a simple explicit form due to the fact that the only contribution to the line element is along  $\mathrm{d}\xi$, in particular
\[
	r^3\mathrm{d}s = -c^4\tfrac{(1-\mathfrak{x} ^2)}{(\cosh\eta-\mathfrak{x} )^4} \mathrm{d}\mathfrak{x} .
\]
Finally the normal derivative in bipolar coordinates is given
by 
\begin{align}
\partial_n= -\mathfrak{h}^{-1}\partial_\eta.
\end{align}
%
%

We are now in a position to calculate the force given by \eqref{eq:stress_surface_integral}. 
%
%
For ease of notation, we reformat \eqref{eq:L_minus_one_integrand_res}:
\begin{multline}\label{eq:reformed_lminus_1_psi}
\left [\tfrac{L_{-1} \psi}{r^2}\right]_n= \tfrac{(\cosh\eta-\mathfrak{x} )^{5/2}}{c^4(1-\mathfrak{x} ^2)} \Big[
R_n''(\eta)Q_n(\mathfrak{x} )
\\
-\tfrac{2\sinh\eta}{\cosh\eta-\mathfrak{x} } R_n'(\eta)Q_n(\mathfrak{x} )  +\tfrac{3}{4}\tfrac{3x+\cosh\eta}{\cosh\eta-\mathfrak{x} }  R_n(\eta)Q_n(\mathfrak{x} ) \\ +(1-\mathfrak{x} ^2) R_n(\eta) Q_n''(\mathfrak{x} )+ \tfrac{2}{\cosh\eta-\mathfrak{x} }(1-\mathfrak{x} ^2)R_n(\eta) Q_n'(\mathfrak{x} ) \Big]. 
\end{multline}
Computing $\mathfrak{h}^{-1}\partial_\eta (\cdot)$ for each of these terms is straightforward, but we make the following remarks.
For the third term, it is useful to rewrite $\cosh \eta = - 3 \cosh \eta + 4\eta$ in the numerator.  For the fourth term we use
the ODE \eqref{eq:ode_for_Qn} to write $Q_n''(\mathfrak{x} )$ in terms of $Q_n(\mathfrak{x} )$.  For the fifth term, we retain $Q_n'(\mathfrak{x} )$ and use 
integration by parts. Finally multiplying the resulting terms by $r^3\mathrm{d}s$ and manipulating one obtains
\begin{align*}
\mathcal{F}_z = \sum_{n=1}^\infty s^{(1)}(n)+s^{(2)}(n)+s^{(3)}(n)+s^{(4)}(n)
\end{align*}
where
\begin{align*}
s^{(1)} &:=\int_{-1}^{1}\mathrm{d}\mathfrak{x} \,\tfrac{Q_n(\mathfrak{x} )}{(\cosh\eta-\mathfrak{x} )^{1/2}}\left[-R_n^{(3)}(\eta)+\tfrac{9}{4}R_n'(\eta)+n(n+1)R_n'(\eta)\right], 
\end{align*}
\begin{align*}
s^{(2)}&:=\int_{-1}^{1}\mathrm{d}\mathfrak{x} \,\tfrac{Q_n(\mathfrak{x} )}{(\cosh\eta-\mathfrak{x} )^{3/2}}\left[-\tfrac{1}{2}R_n''(\eta)\,\sinh\eta-\cosh\eta\,R_n'(\eta)\right.\\
&\qquad \qquad \qquad\left.  +\tfrac{21}{8}\sinh\eta\, R_n(\eta)+\tfrac{5}{2}n(n+1)\sinh\eta\,R_n(\eta)\right]\\
&\quad  -\int_{-1}^{1}\mathrm{d}\mathfrak{x} \, \left[\tfrac{n+2}{2n+3}Q_{n+1}(\mathfrak{x} )+\tfrac{n-1}{2n-1}Q_{n-1}(\mathfrak{x} )\right] \tfrac{4\,R_n'(\eta)}{(\cosh\eta-\mathfrak{x} )^{3/2}}, 
\end{align*}
\begin{align*}
s^{(3)}&:=\int_{-1}^{1}\mathrm{d}\mathfrak{x} \tfrac{Q_n(\mathfrak{x} )}{(\cosh\eta-\mathfrak{x} )^{5/2}}\left[3\sinh^2\eta\, R_n'(\eta)\right .\\
&\qquad \qquad  \qquad \left. -\tfrac{9}{2}\sinh\eta\, \cosh\eta\, R_n(\eta)\right]\\
&\quad   -\int_{-1}^{1}\mathrm{d}\mathfrak{x} \, \left[\tfrac{n+2}{2n+3}Q_{n+1}(\mathfrak{x} )+\tfrac{n-1}{2n-1}Q_{n-1}(\mathfrak{x} )\right]\tfrac{6\, \sinh\eta\,R_n(\eta)}{(\cosh\eta-\mathfrak{x} )^{5/2}}
\\
&\quad +\int_{-1}^{1}\mathrm{d}\mathfrak{x} \,\left[-\tfrac{(n+2)(n+3)}{(2n+3)(2n+5)}Q_{n+2}(\mathfrak{x} ) +\tfrac{2n\,(n+1)}{(2n-1)(2n+3)}  Q_n(\mathfrak{x} ) \right. \\
&\left. \qquad \qquad \qquad \qquad -\tfrac{(n-1)(n-2)}{(2n-1)(2n-3)}Q_{n-2}(\mathfrak{x} )\right] \times\tfrac{3\,R_n'(\eta)}{(\cosh\eta-\mathfrak{x} )^{5/2}}, 
\end{align*}
\begin{align*}
s^{(4)}&:=\int_{-1}^{1}\mathrm{d}\mathfrak{x} \,\left[-\tfrac{(n+2)(n+3)}{(2n+3)(2n+5)}Q_{n+2}(\mathfrak{x} )
+\tfrac{2n\,(n+1)}{(2n-1)(2n+3)}  Q_n(\mathfrak{x} )\right. \\
& \qquad  \qquad  \qquad \left.-\tfrac{(n-1)(n-2)}{(2n-1)(2n-3)}Q_{n-2}(\mathfrak{x} )\right]\tfrac{15\sinh\eta\,R_n(\eta)}{2(\cosh\eta-\mathfrak{x} )^{7/2}}.
\end{align*}

The exact evaluation of each of the above integrals is detailed in Appendix \ref{appA}, in particular by use of the functions $\mathcal{I}_{p/2}$. Now by redimensionalising the stream function, substituting the explicit formulae for corresponding $\mathcal{I}_{p/2}$ and the expression for $R_n(\eta)$ in \eqref{eq:expansion_eta_part} and simplifying one obtains the dimensional force experienced by either sphere 
\begin{align}
F_{z}^{1}&=-\sqrt{2}c\pi \mu U\sum_{n=1}^{\infty}(2n+1)(a_n+b_n+c_n+d_n), \,\, \text{on sphere 1} \label{eq:inf_series_force_expressions_1}\\
F_{z}^{2}&=\sqrt{2}c\pi \mu U\sum_{n=1}^{\infty}(2n+1)(-a_n+b_n-c_n+d_n),\,\, \text{on sphere 2} \label{eq:inf_series_force_expressions_2}.
\end{align}
Note that nowhere in such a calculation is any information on the $a_n$, $b_n$, $c_n$, $d_n$ required. In particular, alternative boundary condition choices amount to a different linear system to be solved and a redefinition of these series coefficients.

\subsection{Reduction to a Sphere and Plane}
The limit of the second sphere radius tending to infinity $\beta=r_2/r_1\to \infty$ corresponds to a plane wall. It is of interest how the present theory compares to existing formulae for the slow motion of a sphere perpendicular to a plane wall. Consider the formula \eqref{eq:alpha_asymptotics_Fz}. Assuming the limit exists one obtains
\begin{equation}\label{eq:sphere_plane_limit}
\lim_{\beta\to +\infty}F_{z}^\ast(\alpha,\beta)
= 4\alpha^{-2}-\tfrac{4}{5}\log \alpha +K_3 + o(1)
\end{equation}
where $K_3:=\tfrac{4}{5}(\gamma+\log 2)+\tfrac{16}{15}+\tfrac{2}{3}\lim_{\beta\to +\infty}(C_1+C_2)$. The first five terms in the expansion \eqref{eq:sphere_plane_limit} differ from [2.45] of  \citet{cox1967slow} by a total factor of two, originating from the motion of the plane in our analysis. All that remains is to study $C_1+C_2$ under the limit $\beta\to +\infty$. Observe that
\begin{align}
\lim_{\beta\to +\infty}l(x,\beta)=\tfrac{4 e^{2 x} x^2+4 e^{2 x} x+2 e^{2 x}-2}{-2 e^{2 x} \left(2 x^2+1\right)+e^{4 x}+1}=\tfrac{\sinh 2x +2x}{\cosh 2x -1-2 x^2}-1
\end{align}
where $x=n\alpha$ is an intermediate variable for $\alpha$ vanishingly small and $n$ ever increasing. A more in depth discussion of the asymptotic variable $x$ can be found in Section \ref{sec:asymptotics}. This expression for $l(x,\beta)$ is precisely the integrand for the numerical constants [2.43] of \citet{cox1967slow}. Therefore, up to errors of order $O(\beta^{-1})$, the sphere-plane limit is recovered exactly as $\beta\to +\infty$.

\section{Comparison With Existing Methods}\label{sec:comparetomultipole}
In this section we compare our expression for $a(\cdot)$ to the results obtained using multipole and perturbative methods. We make use of computer code which computes expansion coefficients for the multipole method available online \cite{jeffrey_online_code}.  We show that our results are significantly more accurate and efficient to compute, and cannot be reproduced by the multipole expansion programme. It is widely accepted that for two sphere problems, when tractable, spherical bipolar coordinates will yield the most accurate method to calculate the force. We refer the reader to previous publications making reference to this, which instead use multipole and lubrication methods to carry out the calculations \cite{jeffrey1984calculation, jeffrey1992calculation, kim2013microhydrodynamics}. Whilst such spherical bipolar methods have been used in previous studies of hydrodynamic interactions, we can find no reference to their use in the singular problem studied here.  

We do this because we have identified the absence of any analytical calculations reducing corresponding expressions available in spherical bipolar coordinates to asymptotic expansions for the force in the separation distance.  Previous such `asymptotic' results, such as those in Kim and Karilla \cite{kim2013microhydrodynamics}, are, in fact, not asymptotic and contain divergent terms both as the spheres approach (which is physically reasonable), and as the spheres become widely-separated (which is completely unphysical).  This introduces a need for artificial cutoffs, or matching procedures.

Up until now there has been no ratification of the expressions for the widely used resistance functions $X^A_{11}$, $X^{A}_{22}$, for the force on sphere 1 and 2 respectively, as defined in \cite{jeffrey1984calculation} against spherical bipolar coordinates.  There is simply (unquantified) wisdom concerning the inefficiency of the computation of the $X^A_{11}$, $X^{A}_{22}$ as the separation distance tends to zero \cite{jeffrey1992calculation}. Pertaining to this, we provide the numerical comparison and identify the short comings in using the series representations of $X^A_{11}$, $X^{A}_{22}$ for practical applications.  

\subsection{Inner Region Lubrication Theory}

\begin{figure}
    \centering
    \begin{subfigure}[b]{0.48\textwidth}
		\includegraphics[width=\textwidth]{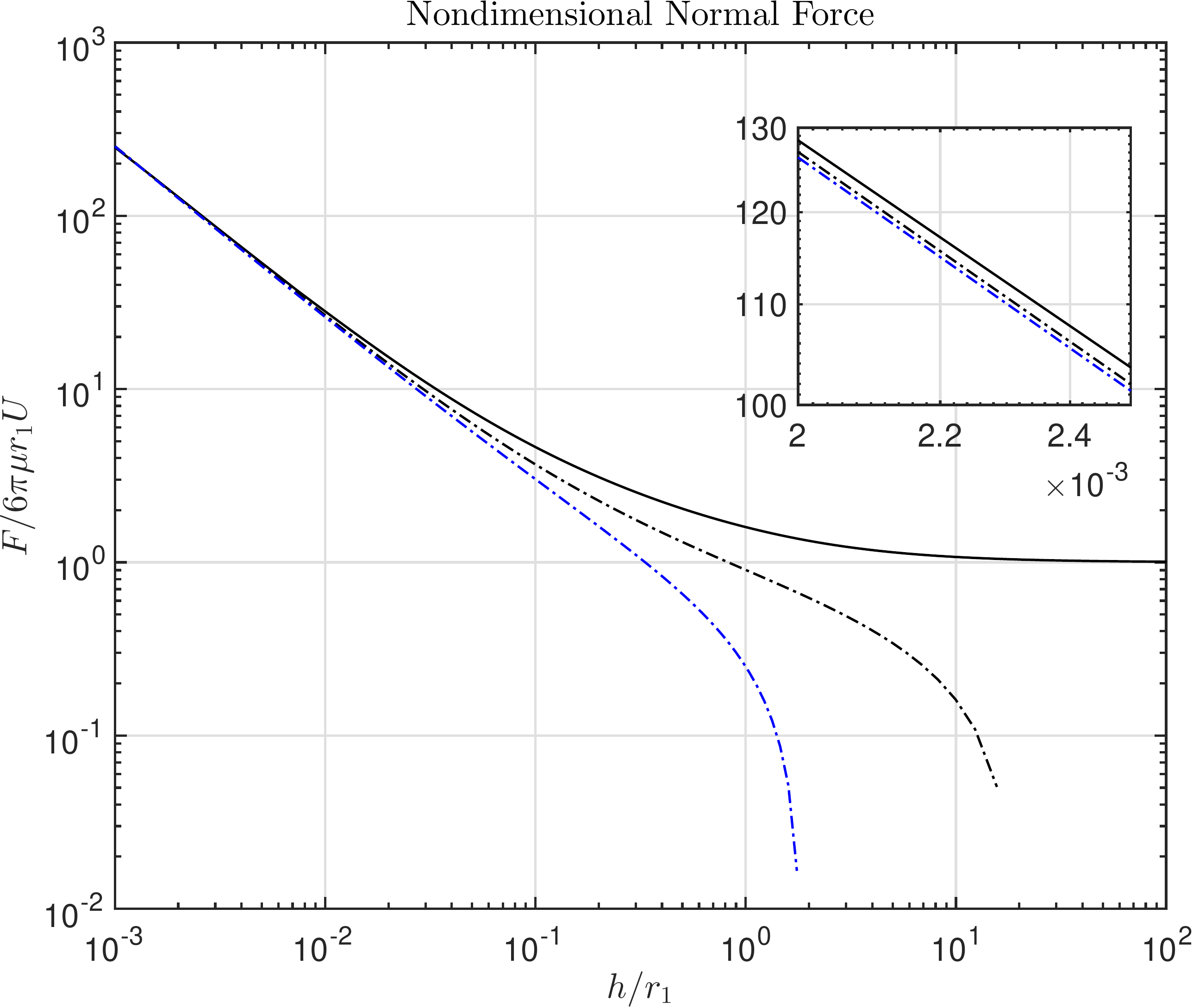}
        \caption{A comparison of the force for equal spheres $r_1=r_2$. $F_{z}^1$ ({\bf{solid-black}}), $F^e_z$ ({\bf black-dot-dashed}), $F_{z,l}$ ({\color{blue}blue-dot-dashed}). Inset: magnified view of each solution as $h/r_1\to 0$.}
        \label{fig:force_biexact_asymp_kk}
    \end{subfigure}
    \,\,
    ~ 
       \begin{subfigure}[b]{0.48\textwidth}
		\includegraphics[width=\textwidth]{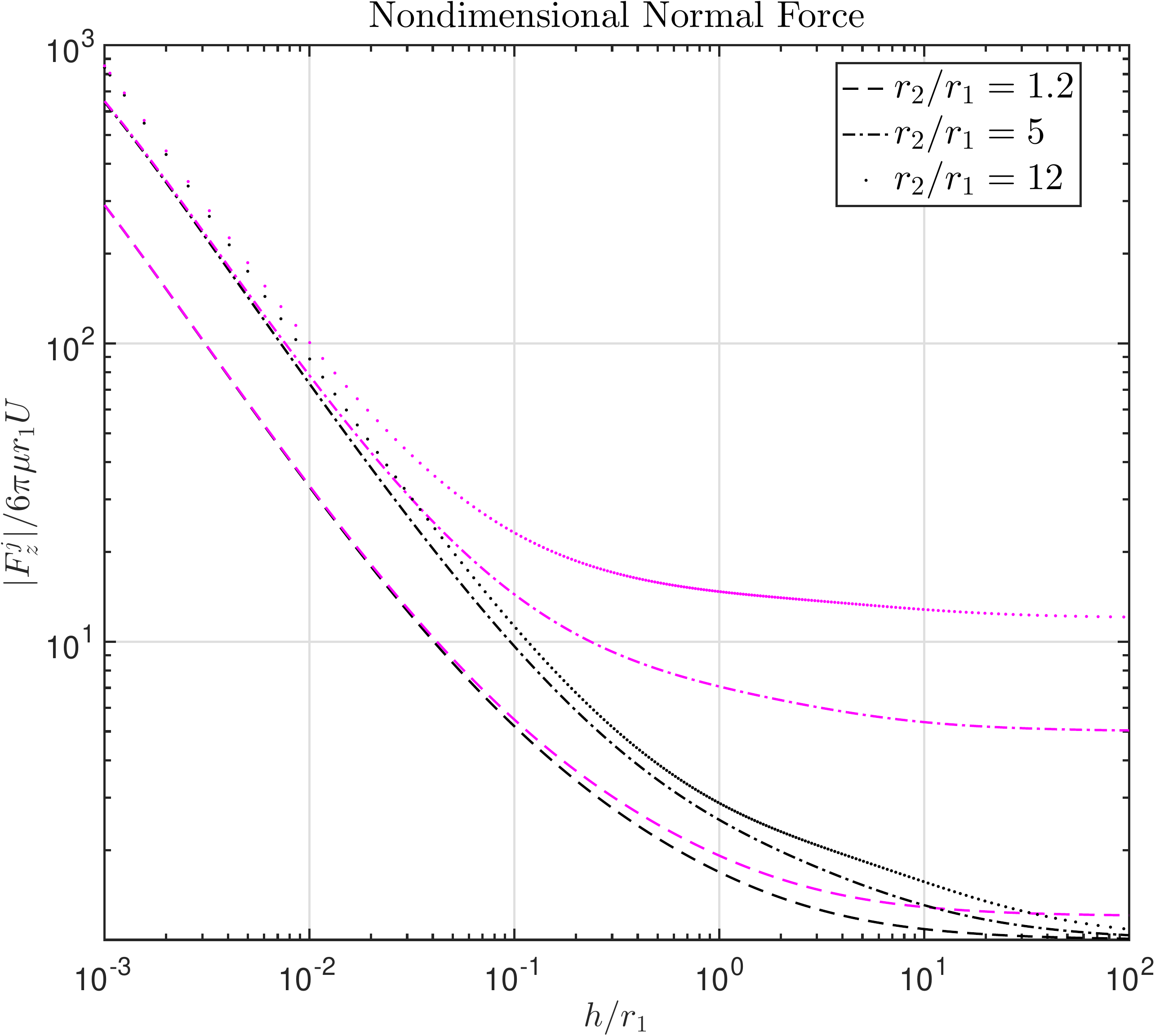}
        \caption{The hydrodynamic force on sphere 1 ({\bf{black}}) and sphere 2 ({\color{magenta}magenta}) for three different radii ratios $r_2/r_1$ using the present theory and formulas \eqref{eq:inf_series_force_expressions_1}, \eqref{eq:inf_series_force_expressions_2}.\linebreak\linebreak}
        \label{fig:force_r1r2}
    \end{subfigure}
          \caption{Plots of the present theory and the lubrication results.}\label{fig:interpolant_kk_gms_different_betas}
      \end{figure}
      
\begin{figure}
    \centering      
    \begin{subfigure}[b]{0.48\textwidth}
		\includegraphics[width=\textwidth]{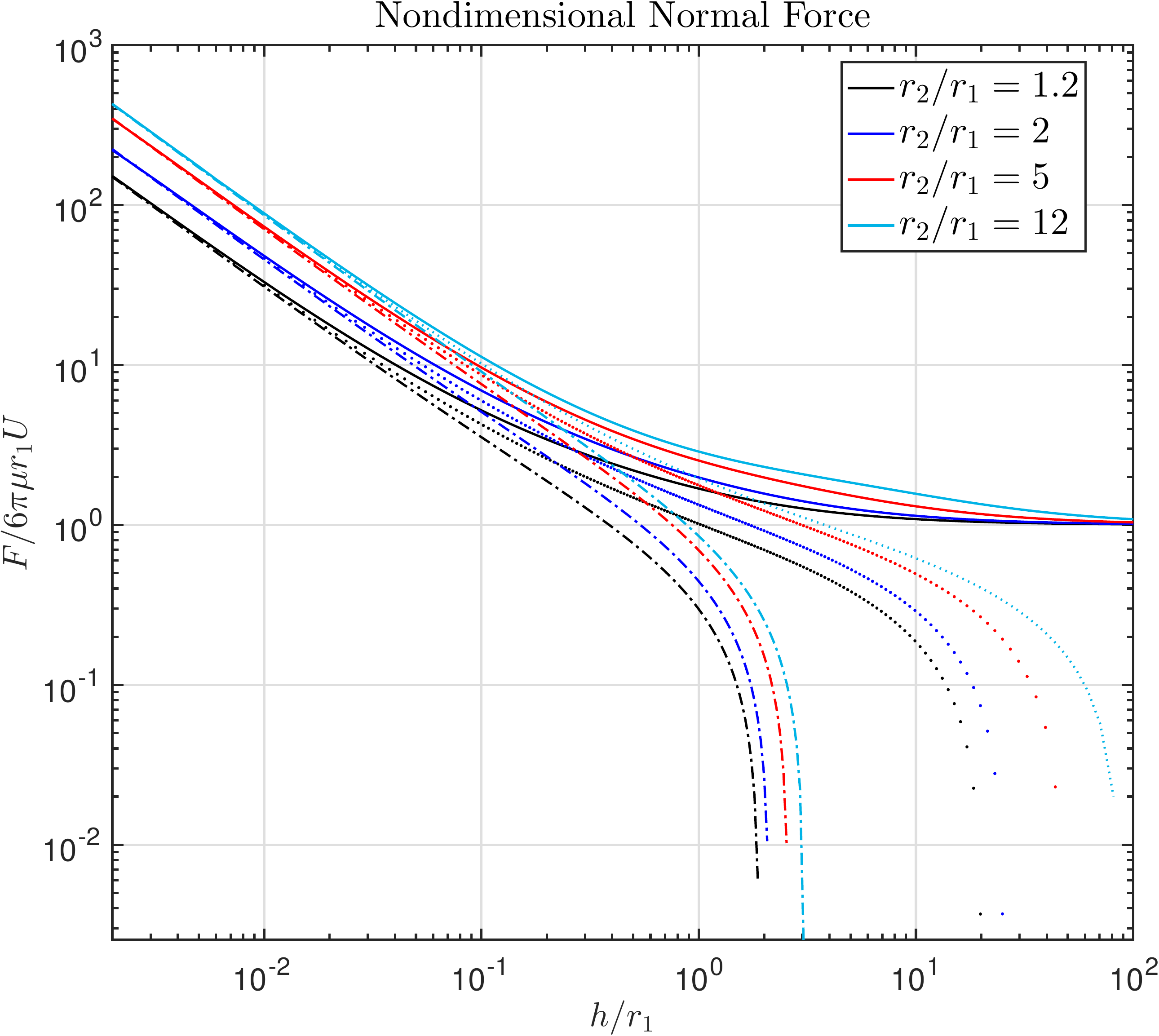}
        \caption{The force on sphere 1 for different radii ratios $r_2/r_1\geq 1$. Exact $F^1_z$ (solid), $F_{z,l}$ (dot-dashed) and $F_{z}^e$ (dot).}
        \label{fig:force_bi_asymptotics_r1r2}
    \end{subfigure}
    \,
    \begin{subfigure}[b]{0.48\textwidth}
		\includegraphics[width=\textwidth]{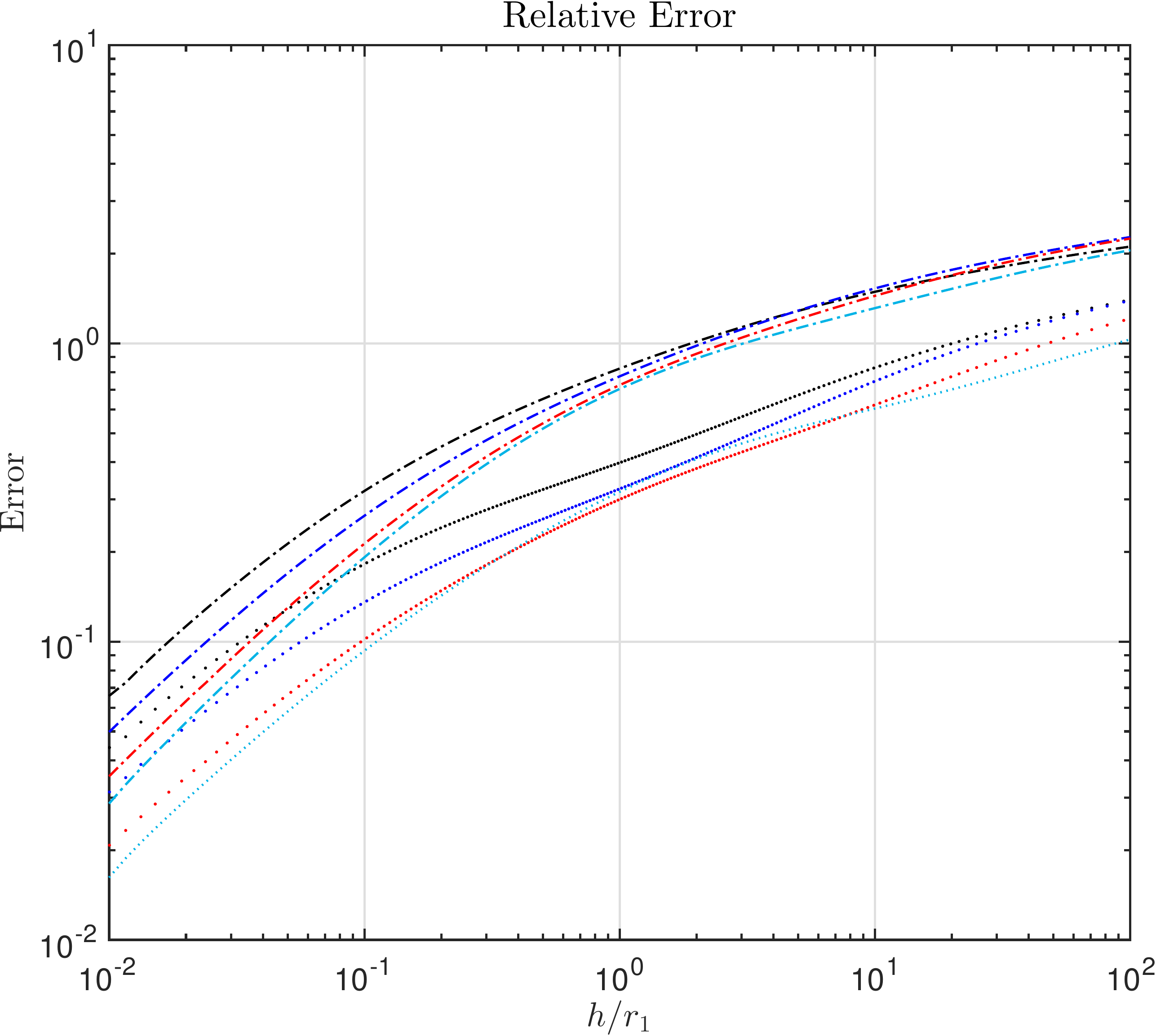}
        \caption{The relative error to the exact force for each the two asymptotic formulae, $F_{z,l}$ (dot-dashed) and $F_{z}^e$ (dot). Colours as in (a).}
        \label{fig:rel_err_asymp}
    \end{subfigure}

    \caption{A comparison of force formulae with varying radii ratios. For exact unequal spheres \eqref{eq:trans_eqns_for_etas} was solved numerically to obtain corresponding $\eta_1$, $\eta_2$ ordinates before summing the functionals \eqref{eq:inf_series_force_expressions_1}, \eqref{eq:inf_series_force_expressions_2} and truncating the infinite series to within machine precision.}\label{fig:animals}
\end{figure}

In this section we present a comparison between the exact \eqref{eq:inf_series_force_expressions_1}--\eqref{eq:inf_series_force_expressions_2} (valid for all separation and sphere sizes), and asymptotic formulae \eqref{eq:alpha_asymptotics_Fz}--\eqref{eq:euclidean_asymptotics_Fz} (valid for all sphere sizes) determined by the present work and the existing lubrication theory \cite{kim2013microhydrodynamics}. In Table \ref{tab:force_comp} and Figure \ref{fig:animals} we compare $F_z^{i}$ (\eqref{eq:inf_series_force_expressions_1}--\eqref{eq:inf_series_force_expressions_2}), $F_z^e$ \eqref{eq:euclidean_asymptotics_Fz}, and $F_{z,l}$ by defining the `lubrication theory' formula
\begin{align}\label{eq:kk_F_z}
 F_{z,l}:=\tfrac{2\beta^2\epsilon^{-1}}{(1+\beta)^2}+\tfrac{2\beta(1+7\beta+\beta^2)}{5(1+\beta)^3}\log\epsilon^{-1}.
\end{align} 
We have truncated this expression to $\log(\cdot)$, omitting terms equal to and higher than $\epsilon\log\epsilon$ because those higher order terms are based on the expansion of a stream function at $r=\infty$ without proper control on the convergence of the force integral used to compute $F_{z,l}$. The exact force, as given by \eqref{eq:inf_series_force_expressions_1}, as well as an interpolant produces a hydrodynamic force varying smoothly between the small and large argument limits, as seen in Figure \ref{fig:force_biexact_asymp_kk} for two equal spheres. In Figure \ref{fig:force_r1r2} we plot the functions \eqref{eq:inf_series_force_expressions_1}, \eqref{eq:inf_series_force_expressions_2} for different radii ratios.

The force calculated from the asymptotic formula \eqref{eq:euclidean_asymptotics_Fz} deviates from the exact solution and becomes unphysical at large separation, as expected. However, from Figure \ref{fig:force_biexact_asymp_kk} (with inset), we observe that our asymptotic formula $F_{z}^e$ agrees more closely with the exact formula $F^1_z$ than $F_{z,l}$. In particular $F_{z}^e$ is barely visible on top of the black curve. This is true even for distances up to one radius, $r_1$, whilst $F_{z,l}$ agrees with $F_{z}^1$ only for distances less than one tenth of $r_1$. 

We also demonstrate the applicability of the exact and asymptotic formulae to unequal spheres of various size ratios in Figure \ref{fig:force_r1r2}, \ref{fig:force_bi_asymptotics_r1r2} and \ref{fig:rel_err_asymp}. In each of these figures $h=d-r_1-r_2$ is dimensional. We remark that in Figure \ref{fig:force_r1r2} we plot the magnitude of the force on either sphere for different radii ratios, and note that as $\alpha\to 0$ the forces are equal and opposite as we would expect by Newton's third law. That is, once the forces $F^j_z$ are scaled by the same Stokes constant, they collapse onto each other for all $r_2/r_1>0$. This may be seen more rigorously by repeating the analysis of Section \ref{sec:small_arg_behaviour} on sphere 2; one finds $F_{z}^{2}/(6\pi\mu U r_2)\sim -4\beta^2/(1+\beta)^3 \alpha^{-2}$. The force magnitude, however, increases as the radii ratio increases; see Figures \ref{fig:force_r1r2}, \ref{fig:force_bi_asymptotics_r1r2}. The relative error for the present asymptotic formula in Figure \ref{fig:rel_err_asymp} using \eqref{eq:alpha_asymptotics_Fz} improves monotonically as $r_2/r_1$ becomes larger. This was observed to hold for even larger ratios (not shown for clarity).

\begin{table}
    \centering
\caption{Comparison of exact and approximate nondimensional forces with $r_2/r_1=5$.}
\label{tab:force_comp}
\begin{tabular}{ccccccc}
\hline\vspace*{7pt}
$h$ &  \multicolumn{2}{c}{$\tfrac{\text{Centre Distance}}{\text{Diameter}}$} & $F_{z}^{1}\cdot 10^4$ & $-F_{z}^{2}\cdot 10^3$ & $F_{z}^{\ast}\cdot 10^4$ & $F_{z,l}\cdot 10^4$  \\
&  sphere 1 & sphere 2 & &  & sphere 1 & sphere 1 \\
\hline
0.0001 & 3.0000 & 0.6000 & 1.3896 & 2.7801 & 1.3896 & 1.3894 \\
0.0212 & 3.0106 & 0.6021 & 0.0069 & 0.0148 & 0.0069 & 0.0068 \\
0.1008 & 3.0504 & 0.6101 & 0.0017 & 0.0043 & 0.0017 & 0.0015 \\
0.3217 & 3.1609 & 0.6322 & 0.0007 & 0.0023 & 0.0007 & 0.0005\\
1.1291 & 3.5646 & 0.7129 & 0.0003 & 0.0016 & 0.0003 & 0.0001\\
9.9660 & 7.9830 & 1.5966 & 0.0002 & 0.0011 & 0.00004& -0.0001\\
$\infty$ & $\infty$ &$\infty$&0.0001 &0.0010 & - & -\\\hline
\end{tabular}
\vspace*{-4pt}
\end{table}


\subsection{The Multipole Expansion Functions}
In this section we examine the behaviour of the multipole scalar resistance functions $X^A_{ij}$ as defined in \citet{jeffrey1984calculation}. Local to this section only we define some notation to be consistent with \citet{jeffrey1984calculation}: $a_1$, $a_2$ are the radii of spheres 1 and 2 respectively, $\lambda$ is the sphere radii ratio, $s$ is a nondimensional separation parameter, $\xi$ is $s$ shifted by two, and $h$ is the dimensional sphere surface separation. The following hold
\begin{align*}
\lambda = \frac{a_2}{a_1}, \quad s-2 = 2\frac{h/a_1}{1+\lambda}, \quad \xi = s-2.
\end{align*}

\begin{figure}[t!]
    \centering
    \begin{subfigure}[b]{0.48\textwidth}
        \includegraphics[width=\textwidth]{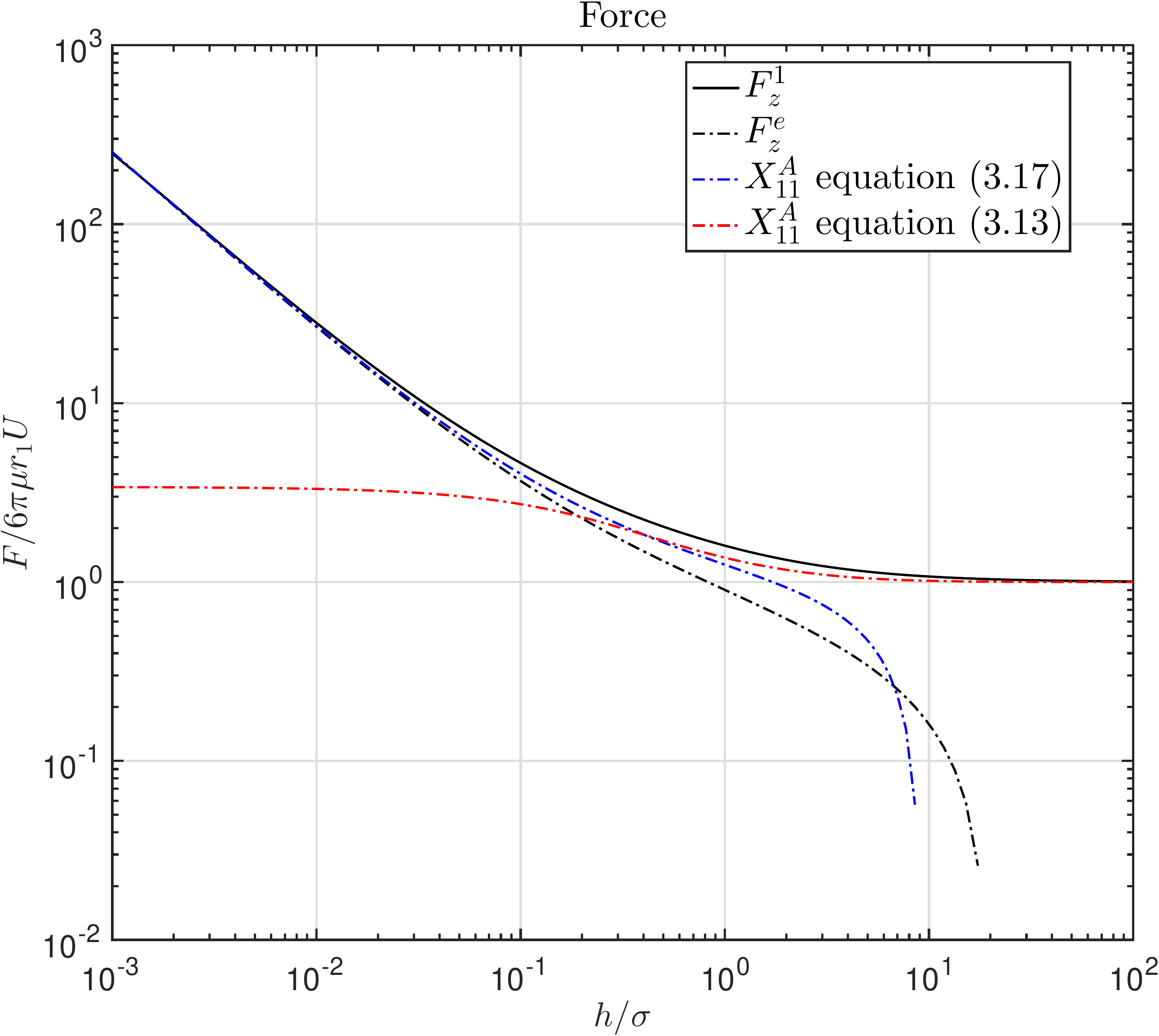}
        \caption{The force on sphere 1 comparing various formalisms with $r_2/r_1 = 1$. Key: spherical bipolar coordinates ({\bf{black-solid}}), asymptotic expansion $F_{z}^e$ obtained by spherical bipolar coordinates ({\bf black-dot-dashed}), $X^A_{11}$ ({\color{red}red-dot-dashed}) as computed using provided 300 terms in \citet{jeffrey_online_code} for the equation (3.13) in  \citet{jeffrey1984calculation}, the expansion of $X^A_{11}$ including the order 1 term $A^X_{11}$ equation (3.17) of \citet{jeffrey1984calculation}  ({\color{blue}blue-dot-dashed}).}
        \label{fig:comparison_with_jeffrey_onishi_300_terms_beta_1}
    \end{subfigure}
    \,
    \centering      
    \begin{subfigure}[b]{0.48\textwidth}
        \includegraphics[width=\textwidth]{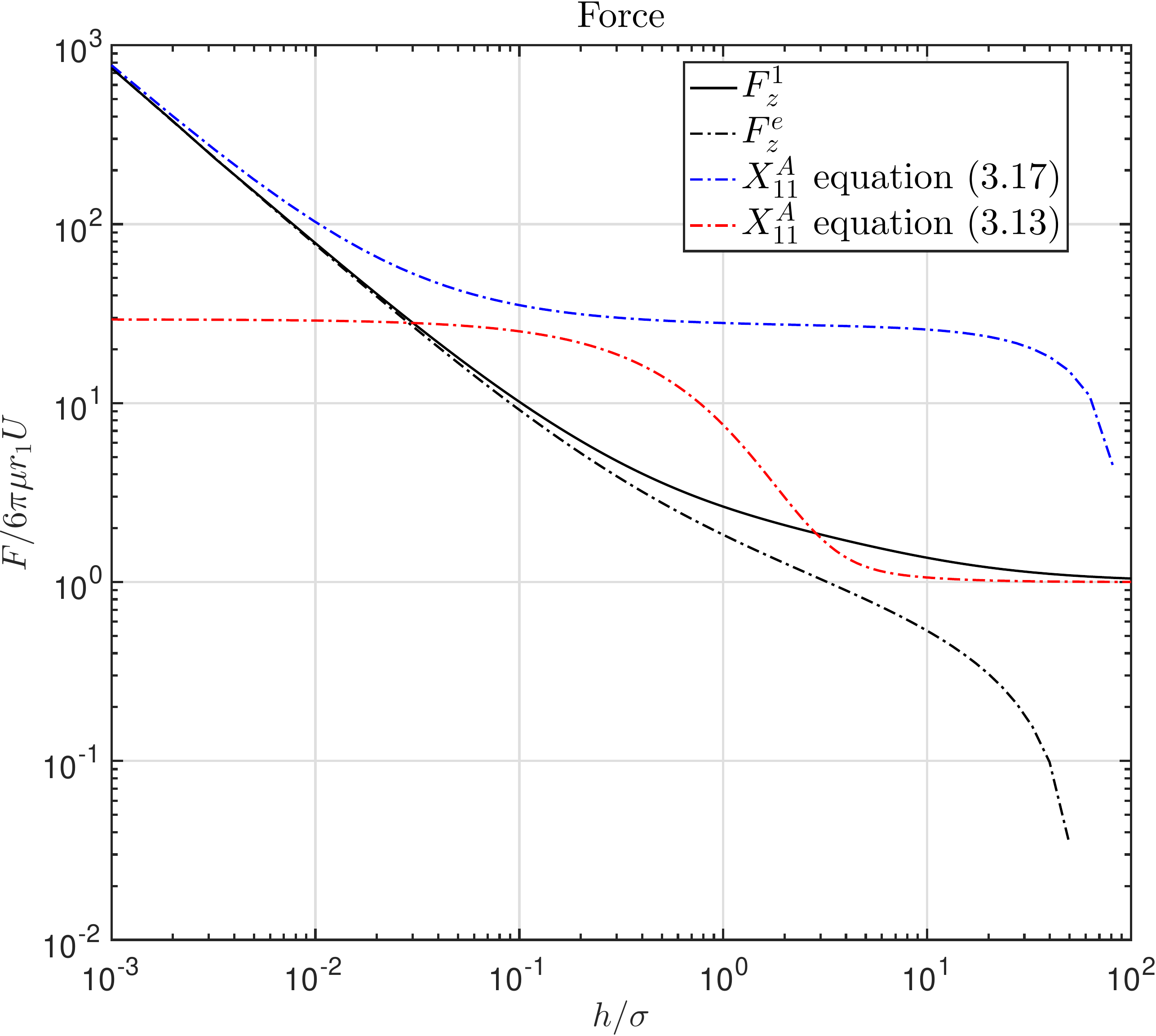}
        \caption{The force on unequal spheres for $r_2/r_1=2\pi$: spherical bipolar coordinates ({\bf{black-solid}}), asymptotic expansion $F_{z}^e$ obtained by spherical bipolar coordinates ({\bf{black-dashed -dot}}), $X^A_{ij}$ as computed using 15 terms in the equation (3.13) of
        \citet{jeffrey1984calculation} ({\color{red}red-dot-dashed}), the expansion of $X^A_{11}$ including the order 1 term $A^X_{11}$ equation (3.17) of
        \citet{jeffrey1984calculation} ({\color{blue} blue-dot-dashed}).} \label{fig:comparison_with_jeffrey_onishi_15_terms_beta_4}
\end{subfigure}
\caption{A comparison of the normal component of the HI as obtained by the present theory and multipole methods.}
\end{figure}

The existing programs consist of Fortran code for the resistance functions $X^A_{11}$ (and $X^{A}_{22}$) as defined in \citet{jeffrey1984calculation}, provided by D. J. Jeffrey~\cite{jeffrey_online_code}. The functions $X^A_{11}$ and $X^{A}_{22}$ are expressions for the force normal to the sphere surfaces due to sphere 1 and sphere 2, respectively. We now demonstrate that our corresponding functions $F_{z}^{1}$ and $F_{z}^2$ are more accurate than $X^A_{11}$ and $X^{A}_{22}$ in computing the force both for arbitrary sphere size ratios and for arbitrary sphere separations.  See Figures \ref{fig:comparison_with_jeffrey_onishi_300_terms_beta_1} and \ref{fig:comparison_with_jeffrey_onishi_15_terms_beta_4}. 

For Figure \ref{fig:comparison_with_jeffrey_onishi_300_terms_beta_1} we computed both the $X^A_{11}$ by using equation (3.13) of \citet{jeffrey1984calculation} and via the asymptotic form (3.17) of \citet{jeffrey1984calculation} using the first 300 terms $f_m$ as provided by the code \cite{jeffrey_online_code_300} and compared to the results obtained by spherical bipolar coordinates. Indeed Figure \ref{fig:comparison_with_jeffrey_onishi_300_terms_beta_1} shows a substantial difference in the singular behaviour between the spherical bipolar and multipole formalisms, particularly in the small argument region where many summand terms are required for an accurate representation of $X^A_{11}$. The largest shortcoming of the multipole method is that the coefficients of summand of $X^A_{11}$, denoted $f_k(\lambda)$, are not all known for all $\lambda$ and require large computing resources \cite{jeffrey_online_code,jeffrey1992calculation}. When computing more $f_m(\lambda)$ the authors found the solution of recurrence relation (3.9) \citet{jeffrey1984calculation} increasingly difficult for both $m$, $\lambda\to\infty$. For the expanded version of $X^A_{11}$ given by eq (3.17) of \citet{jeffrey1984calculation} the behaviour can be understood by closely looking at the formula for the order 1 term $A^X_{11}$ (3.17)
\begin{multline*}
A^X_{11} = 1-\tfrac{1}{4}g_1\\
+\sum_{\substack{m=2\\m\text{ even}}}^{\infty} [2^{-m}(1+\lambda)^{-m}f_m(\lambda) - g_1 -2m^{-1}g_2 + 4m^{-1}m_1^{-1}g_3]
\end{multline*}
where $g_1(\lambda)$, $g_2(\lambda)$, $g_3(\lambda)$,  $m_1(m)$ are all known. We see that this series has a divergent term, namely $-2m^{-1}g_2$. Using this formula for $A^X_{11}$ and the expansion as $\xi\to 0$ one has
\begin{multline*}
X^{A}_{11} = g_1(\lambda)\xi^{-1}+g_2(\lambda)\log \xi^{-1}\\
+ A^X_{11}(\lambda)+ g_3(\lambda)\xi \log\xi^{-1} \quad \text{ as } \xi\to 0,
\end{multline*}
which we compare to the expansion $F_{z}^e$ \eqref{eq:euclidean_asymptotics_Fz} as well as the formula $F_{z}^1$ valid for arbitrary separations.

It is apparent from Figure \ref{fig:comparison_with_jeffrey_onishi_15_terms_beta_4} (using the first 15 terms as provided by \citet{jeffrey_online_code}) that when using the infinite series formula for $X^A_{11}$ to compute the force for a larger aspect ratio $r_2/r_1 = 2\pi$ we see a considerable disagreement with the calculations obtained in spherical bipolar coordinates. The $X^{A}_{11}$ may perform better in the near field when more $f_m$ are known, but computing these coefficients is inefficient for practical applications, and more so for larger aspect ratios $\lambda\to \infty$, as we found when calculating more than 15 $f_m$ for the purposes of this work. 

We are confident in the calculation of $X^A_{11}$ used to produce Figures \ref{fig:comparison_with_jeffrey_onishi_300_terms_beta_1}-\ref{fig:comparison_with_jeffrey_onishi_15_terms_beta_4} because we were able to reproduce the tabulated values of $A^X_{11}(1)$ as listed in section 3.3
in \citet{jeffrey1984calculation}. Meanwhile the spherical bipolar formalism gives an explicit formula for all summand terms and provides the correct decay structure both as the centre distance decreases and increases. We therefore contend that the results obtained using spherical bipolar coordinates are more efficient, accurate and cannot already be produced with existing methods.

\section{Tangential Interaction $b(\cdot)$}\label{sec:tang_interaction}

\begin{figure}
\centering
\includegraphics[width=3.3in]{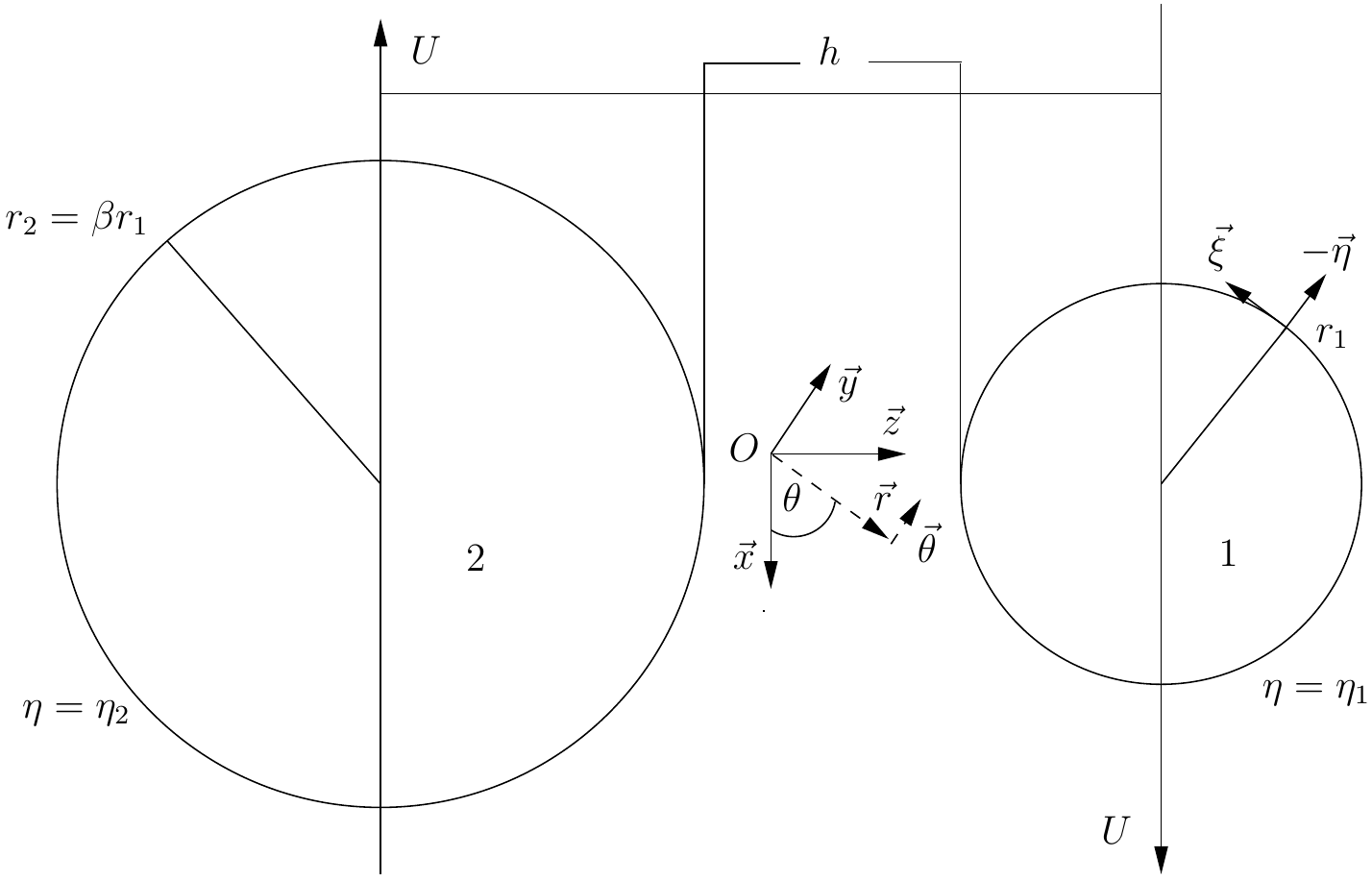}
\caption{Schematic of two unequal spheres of radii $r_1$, $r_2$ converging parallel to their line of centres in viscous fluid. Included in the diagram are the cylindrical and bipolar unit vectors, the dimensional gap distance $h$, and centre to centre distance $d$. Note that $\eta_1$ and $\eta_2$ are implicit functions of $r_1$, $r_2$ and $d$.}\label{fig:two_particle_bipolar_up}
\end{figure}

In this section we write and solve Stokes equations in spherical bipolar coordinates with the view to obtain an expression for $b(\cdot)$, the force on the spheres to the shearing interaction.

\subsection{Stokes Equations}
For this interaction the fluid velocity around the spheres can not be expressed as the curl of a scalar field, since the flow is not axisymmetric. We instead consider the full 3d equations \eqref{eq:NS_eqn} (neglecting inertial effects). We have
\begin{align}\label{eq:NS_eqns_redimensionalised}
\begin{split}
\mu^{-1}\nabla p& =\nabla^2 \vec{u},\\
\nabla\cdot \vec{u}&=0.
\end{split}
\end{align}
The appropriate boundary conditions will be seen to be
\begin{align}
\begin{split}
\vec{u}&=U\hat{\vec{e}}_x \qquad \text{on sphere } 1,\\
\vec{u}&=-U\hat{\vec{e}}_x \,\quad \text{on sphere } 2
\end{split}
\end{align}
along with the far field condition
\begin{align}
\vec{u}\to 0 \qquad \text{ as } \qquad |\vec{x}|\to \infty.
\end{align}
We solve equations \eqref{eq:NS_eqns_redimensionalised} in circular cylindrical coordinates. The equations governing fluid pressure and the three velocity fields read
\begin{align}
\mu^{-1}\partial_r p&=\left(\nabla^2-\frac{1}{r^2}\right)u_r-\frac{2}{r^2}\partial_\theta u_\theta,\label{eq:p_r}\\
\mu^{-1}r^{-1}\partial_\theta p&=\left(\nabla^2-\frac{1}{r^2}\right)u_\theta+\frac{2}{r^2}\partial_\theta u_r,\label{eq:p_theta}\\
\mu^{-1}\partial_zp&=\nabla^2u_z \label{eq:p_z}
\end{align}
where it is emphasised that $\nabla^2$ is the anisotropic Laplacian in circular cylindrical coordinates
\begin{align}
\nabla^2 = \partial_r^2+\partial_z^2 + r^{-1}\partial_r+r^{-2}\partial_\theta^2.
\end{align}
The incompressibility condition becomes
\begin{align}
\partial_r u_r+r^{-1}u_r+r^{-1}\partial_\theta u_\theta+\partial_z u_z=0
\end{align}
and the boundary conditions are
\begin{align}
\begin{split}\label{bc:shearing_spheres}
u_r=U\cos\theta,& \quad u_\theta=-U\sin\theta,  \quad u_z = 0, \,\quad \text{on sphere } 1\\
u_r=-U\cos\theta,& \quad u_\theta=U\sin\theta, \quad u_z=0, \qquad \text{on sphere } 2.
\end{split}
\end{align}
Note that these boundary conditions impose equal and opposite velocities on the spheres, which is the reverse of the case given in \citet{goldman1966slow}. 

\subsection{Derivation of $u_r$, $u_\theta$, $u_z$ and $p$}
For a complete derivation of the pressure and velocity fields we refer the reader to Appendix \ref{app:derivation_of_tang_fields}. 

As already stated, the reduction in symmetry for the tangential interaction means that Stokes equations cannot be solved via a stream function approach. The velocity and pressure fields may, however, be decomposed into four fields which correspond to stream functions for a set of dual axisymmetric flows. For example, $L_1Y = 0$, where $L_1=\partial_z^2+\partial_r^2+r^{-1}\partial_r$ is the \emph{isotropic} Laplacian in cylindrical coordinates.

In brief, $p$, $u_r$, $u_\theta$ and $u_z$ are decomposed into the representations \eqref{eq:p_decomp}, \eqref{eq:u_r_vel_decomp}, \eqref{eq:u_theta_vel_decomp} and \eqref{eq:u_z_vel_decomp} via linear combinations of four scalar fields $W(r,z), X(r,z), Y(r,z), Z(r,z)$. The field $W(r,z)$ is essentially the nondimensional pressure, where $\mu U/c$ defines the viscous pressure scale, recalling that $c$ is the focal length comparable to a sphere diameter. The angular dependence of the expansions are inspired by the boundary conditions \eqref{bc:shearing_spheres} and the compatibility with the Stokes equations \eqref{eq:p_r}--\eqref{eq:p_z}. Theses auxiliary functions are then obtained in spherical bipolar coordinates,  \eqref{eq:eqn_for_W_SB}, \eqref{eq:eqn_for_X_SB}, \eqref{eq:eqn_for_Y_SB}, \eqref{eq:eqn_for_Z_SB} up to a set of arbitrary constants $A_n$--$H_n$ which are determined by the no-flux and no-slip boundary conditions on either sphere. 

We now describe how we obtain the summation coefficients $A_n$--$H_n$. 

\subsection{Boundary Conditions}
The boundary conditions \eqref{bc:shearing_spheres} are transformed into the corresponding conditions on the auxiliary fields  $W(r,z), X(r,z), Y(r,z), Z(r,z)$. By the expressions for  $u_r$, $u_\theta$, $u_z$  \eqref{eq:u_r_vel_decomp}, \eqref{eq:u_theta_vel_decomp}, \eqref{eq:u_z_vel_decomp} respectively we obtain on sphere $1$
\begin{align}
\frac{1}{c}r^{(1)}W_1 + X_1 + Y_1 &= 1, \label{bc:bcs_shearing_transformed_A_r}\\
  X_1-Y_1&=-1, \label{bc:bcs_shearing_transformed_A_theta}\\
  z^{(1)}W_1+ 2cZ_1 &=0, \label{bc:bcs_shearing_transformed_A_z}
\end{align}
and on sphere 2
\begin{align}
\frac{1}{c}r^{(2)}W_2 + X_2 + Y_2 &= -1,\label{bc:bcs_shearing_transformed_B_r}\\
 X_2-Y_2 &=1,\label{bc:bcs_shearing_transformed_B_theta}\\  
 z^{(2)}W_2+ 2cZ_2 &=0,\label{bc:bcs_shearing_transformed_B_z}
\end{align}
where
\begin{align*}
 z^{(1)}&=c\frac{\sinh\eta_1}{\cosh\eta_1-\cos\xi}, \quad \,\,\,\,\,\, r^{(1)}=c\frac{\sin\xi}{\cosh\eta_1-\cos\xi},\\
z^{(2)}&=c\frac{\sinh\eta_2}{\cosh\eta_2-\cos\xi}, \quad  r^{(2)}=c\frac{\sin\xi}{\cosh\eta_2-\cos\xi}.
\end{align*} 
In the singular case, when two spheres are converging perpendicular to there line of centres we do not expect the fluid pressure to remain bounded. Since we expect a divergent pressure field for small separations along the $z$ axis, equivalently $\cos\xi=\pm 1$, the general solution to \eqref{eq:eqn_for_W}--\eqref{eq:incomp_rewrite} is found by setting $H_n = 0$ for every $n$. The six boundary conditions along with the incompressibility condition \eqref{eq:incomp_rewrite} form seven equations for the seven unknowns $A_n$--$G_n$.

\subsection{Equal Spheres}
We now obtain the unknown constants for the case of equal spheres. A set of recurrence relations for the unequal sphere case are presented in Appendix \ref{app:unequal_sphere_derivation} but are not solved due to algebraic complexity in the relations, which may be overcome with computer algebra. As such we set $\eta_1 = -\eta_2 = \alpha>0$. In the equal sphere cases, the cylindrical polar ($r$, $z$) and spherical bipolar ($\eta$, $\xi$) coordinates in the right and left hand planes are related by
\begin{align*}
 z^{(1)}&=c\frac{\sinh\alpha}{\cosh\alpha-\cos\xi}, \quad \,\,\,\,\,\, r^{(1)}=c\frac{\sin\xi}{\cosh\alpha-\cos\xi},\\
z^{(2)}&=-c\frac{\sinh\alpha}{\cosh\alpha-\cos\xi}, \quad  r^{(2)}=c\frac{\sin\xi}{\cosh\alpha-\cos\xi},
\end{align*} 
where $\alpha\in (0,\infty)$ is the spherical bipolar coordinate which draws a sphere of radius $r_1 = c|\csch \alpha|$ in the right and left hand planes. Additionally, $\alpha$ and is a proxy for the sphere centre distance where $\cosh \alpha = d/r_1$ where $d$ is the centre distance of the spheres.  

\subsubsection*{Recurrence Relations for $A_n$--$G_n$}
We now determine $A_n$--$G_n$. By subtracting \eqref{bc:bcs_shearing_transformed_B_z} from \eqref{bc:bcs_shearing_transformed_A_z} we find
\begin{align*}
B_n=0
\end{align*}
for every $n$. Similarly by adding together \eqref{bc:bcs_shearing_transformed_A_r}, \eqref{bc:bcs_shearing_transformed_B_r}, similarly \eqref{bc:bcs_shearing_transformed_A_theta} to \eqref{bc:bcs_shearing_transformed_B_theta} we find
\begin{align*}
D_n=F_n=0
\end{align*}
for every $n$. Note that these zero conditions are the complement of the of the zero conditions found in \cite{goldman1966slow}. Using the Bonnet recursion formula
\begin{align*}
(n+1)P_{n+1}(\mathfrak{x} )=(2n+1)xP_n(\mathfrak{x} )-nP_{n-1}(\mathfrak{x} )
\end{align*}
along with the integration formula
\begin{align*}
(2n+1)P_n(\mathfrak{x} )=P'_{n+1}(\mathfrak{x} )-P'_{n-1}(\mathfrak{x} )
\end{align*}
one can derive
\begin{align}\label{eq:bonnet_third_recursion_formula}
xP'_n(\mathfrak{x} )=\frac{n+1}{2n+1}P'_{n-1}(\mathfrak{x} )+\frac{n}{2n+1}P'_{n+1}(\mathfrak{x} ).
\end{align}
By adding \eqref{bc:bcs_shearing_transformed_A_z} to \eqref{bc:bcs_shearing_transformed_B_z} we find
\begin{align}
&\sinh\alpha\sum_{n=1}^{\infty}C_n\sinh(n+\tfrac{1}{2})\alpha P'_n(\mathfrak{x})\nonumber\\
&+2\cosh\alpha\sum_{n=1}^\infty\cosh(n+\tfrac{1}{2})\alpha P'_n(\mathfrak{x})\nonumber\\
&-2\sum_{n=1}^{\infty}A_n\cosh(n+\tfrac{1}{2})\alpha\nonumber\\
&\quad\times\left[\frac{n+1}{2n+1}P'_{n-1}(\mathfrak{x})+\frac{n}{2n+1}P'_{n-1}(\mathfrak{x})\right]  =0
\end{align}
and we obtain a relation for $C_n$ in terms of $A_n$
\begin{align}
C_n = 2A_{n+1}\frac{n+1}{2n+3}\left[\gamma_n+1\right] -2\gamma_n A_n + 2A_{n-1}\frac{n-1}{2n-1}\left[\gamma_n-1\right] 
\end{align}
where $\gamma_n = \coth\alpha\coth(n+\tfrac{1}{2})\alpha$ and we have used \eqref{eq:bonnet_third_recursion_formula}. Note the definition of $\gamma_n$ is different to that in \cite{goldman1966slow}. By subtracting \eqref{bc:bcs_shearing_transformed_B_r} from \eqref{bc:bcs_shearing_transformed_A_r} and subtracting \eqref{bc:bcs_shearing_transformed_B_theta} from \eqref{bc:bcs_shearing_transformed_A_theta} and finally adding \eqref{bc:bcs_shearing_transformed_B_z} to \eqref{bc:bcs_shearing_transformed_A_z} we obtain
\begin{align}
\frac{\sin\xi}{\cosh\alpha-\mathfrak{x}}\left[W_1-W_2\right]  +X_1-X_2+Y_1-Y_2&=2,\label{eq:aux_bc_1}\\
X_1-X_2-[Y_1-Y_2]&=-2,\label{eq:aux_bc_2}\\
\frac{\sinh\alpha}{\cosh\alpha-\mathfrak{x}}\left[W_1-W_2\right]  +2 [Z_1+Z_2]&=0.\label{eq:aux_bc_3}
\end{align}
Adding together \eqref{eq:aux_bc_1} and \eqref{eq:aux_bc_2} we find 
\begin{align*}
\sum_{n=2}^\infty G_n\sinh(n+\tfrac{1}{2})\alpha P''_n(\mathfrak{x} )=\csch\alpha\sum_{n=1}^{\infty}A_n\cosh(n+\tfrac{1}{2})\alpha P'_n(\mathfrak{x} )
\end{align*}
and using the integration formula
\begin{align*}
(2n+1)P'_n(\mathfrak{x} )=P''_{n+1}(\mathfrak{x} )-P''_{n-1}(\mathfrak{x} )
\end{align*}
we obtain a relation for $G_n$ in terms of $A_n$
\begin{align}
G_n = \frac{A_{n-1}}{2n-1}\left[\gamma_n-1\right]  - \frac{A_{n+1}}{2n+3}\left[\gamma_n+1\right].
\end{align}
Finally by subtracting \eqref{eq:aux_bc_2} from \eqref{eq:aux_bc_1} we obtain
\begin{align}
&-\frac{\sin^2\xi}{\sinh\alpha}(\cosh\alpha-\mathfrak{x})^{1/2}\sum_{n=1}^\infty A_n\cosh(n+\tfrac{1}{2})\alpha P'_n(\mathfrak{x})\nonumber\\
&+(\cosh\alpha-\mathfrak{x})^{1/2}\sum_{n=1}^\infty E_n\sinh(n+\tfrac{1}{2})\alpha P_n(\mathfrak{x})=2
\end{align}
and upon using the generating function
\begin{align*}
(\cosh\alpha-\mathfrak{x})^{-1/2}=\sum_{n=0}^\infty s_nP_n(\mathfrak{x})
\end{align*}
where $s_n =\sqrt{2}e^{-(n+\tfrac{1}{2})\alpha}$ along with the identity
\begin{align*}
(1-\mathfrak{x}^2)  P_n'(\mathfrak{x} ) = \frac{n(n+1)}{(2n+1)}\left[P_{n-1}(\mathfrak{x} )-P_{n+1}(\mathfrak{x} )\right]  
\end{align*}
we obtain a relation for $E_n$ in terms of $A_n$
\begin{align}
E_n &= 2\sqrt{2}e^{-(n+\tfrac{1}{2})\alpha}\csch(n+\tfrac{1}{2})\alpha\nonumber\\
&\quad+A_{n+1}\frac{(n+1)(n+2)}{2n+3}\left[\gamma_n+1\right]-A_{n-1}\frac{n(n-1)}{2n-1}\left[\gamma_n-1\right].
\end{align}

We have now obtained six equations involving the seven unknowns $A_n$--$G_n$. The only condition thus far unused is the incompressibility condition \eqref{eq:incomp_rewrite}, which we will use to identify $A_n$. Since the incompressibility condition \eqref{eq:incomp_rewrite} is invariant in the choice of boundary conditions we may use the relation (3.56) \cite{cox1967slow} with our redefined constants. 

The incompressibility condition \eqref{eq:incomp_rewrite} transformed to spherical bipolar coordinates evaluated on the surface of the sphere $\eta=\alpha$ may be written in the form
\begin{align}\label{eq:relation_for_the_An}
\vec{q}\cdot\left[A_{n-1},A_n,A_{n+1}\right]^\top = p_n(\alpha),
\end{align}
where 
\begin{multline}\label{eq:three_term_An_recurrence}
\vec{q}_n(\alpha):= \left[(n-1)(\gamma_{n-1}-1)-\tfrac{(n-1)(2n-3)}{2n-1}(\gamma_{n}-1),\right.\\  
-\tfrac{n(2n-1)}{2n+1}(\gamma_{n-1}+1)+(2n+1)-5\gamma_n+\tfrac{(n+1)(2n+3)}{2n+1}(\gamma_{n+1}-1),\\
\left.\tfrac{(n+2)(2n+5)}{2n+3}(\gamma_n+1)-(n+2)(\gamma_{n+1}+1)\right].
\end{multline}
and the right hand side vector may be obtained in a similar way, with the exception that $\sech$'s are substituted for $\csch$'s.
\begin{align}
&p_n(\alpha):=-\sqrt{2}e^{-(n+1/2)\alpha}\nonumber\\
&\quad \times \left[e^{\alpha}\csch(n-\tfrac{1}{2})\alpha-2\csch(n+\tfrac{1}{2})\alpha+e^{\alpha}\csch(n+\tfrac{3}{2})\alpha\right]\label{eq:rhs_three_term_An_recurrence}
\end{align}
Both equations \eqref{eq:three_term_An_recurrence} and \eqref{eq:rhs_three_term_An_recurrence} were checked with computer algebra. 
\section{The Force Experienced by the Spheres}\label{sec:tang_force_compute}

There is an exact expression for the force on either sphere for the spherical bipolar coordinate system, first obtained by O'Neil \cite{o1964slow} in general form and applied to the case of a single sphere moving parallel to a plane wall. We may use the expression for a two sphere problem, albeit with different summation coefficients owing to the present choice of boundary conditions. We have for equally sized spheres, in dimensional form, 
\begin{align}
\mathcal{F}_x^{1}&=-\sqrt{2}\pi \mu U c \sum_{n=1}^{\infty}E_n, + n(n+1)C_n\,\quad \text{on sphere 1} \label{eq:inf_series_force_expressions_tangential_1}\\
\mathcal{F}_x^{2}&=\sqrt{2}\pi \mu U c \sum_{n=1}^{\infty}E_n, + n(n+1)C_n\,\quad \text{on sphere 2} \label{eq:inf_series_force_expressions_tangential_2}.
\end{align}
These expressions may be nondimensionalised with the characteristic drag scale $6\pi\mu U r_1$, recalling that $r_1 = c |\csch \alpha|$. 

The three term recurrence relation \eqref{eq:relation_for_the_An} is solved along with the decay condition that $A_N = 0$ for some $N$ sufficiently large. This condition will be seen to be appropriate since if \eqref{eq:inf_series_force_expressions_tangential_1}, \eqref{eq:inf_series_force_expressions_tangential_2} are to converge one must have $A_n\to 0$ as $n\to 0$. This decay assumption on $A_n$ allows \eqref{eq:relation_for_the_An} to be written as a tridiagonal linear system, which may be solved with Gaussian elimination. Where fast solvers are required, for example in direct numerical simulations of hard spheres, one might wish employ a Thomas algorithm \citet{trefethen1997numerical}.

\begin{figure}
    \centering      
    \begin{subfigure}[b]{0.475\textwidth}
		\includegraphics[width=\textwidth]{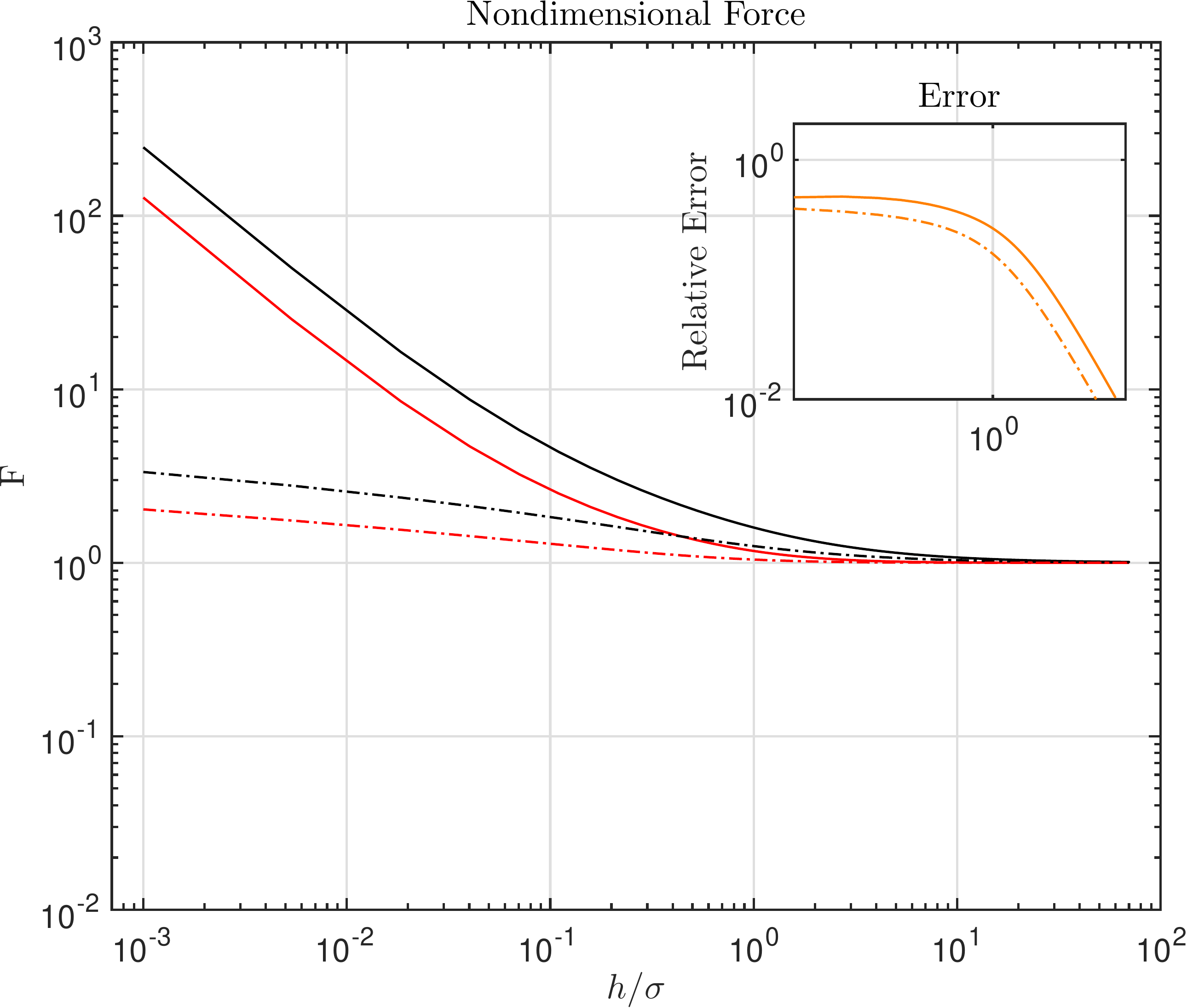}
        \caption{Using the formulae valid for arbitrary separation of Jeffrey \& Onishi the correct far-field behaviour is obtained but the formulae fail in the boundary layer $h/r_1 = \epsilon^{1/2}$.}
        \label{fig:nondimensional_force_arb_inset}
    \end{subfigure}
    \,
    \begin{subfigure}[b]{0.475\textwidth}
		\includegraphics[width=\textwidth]{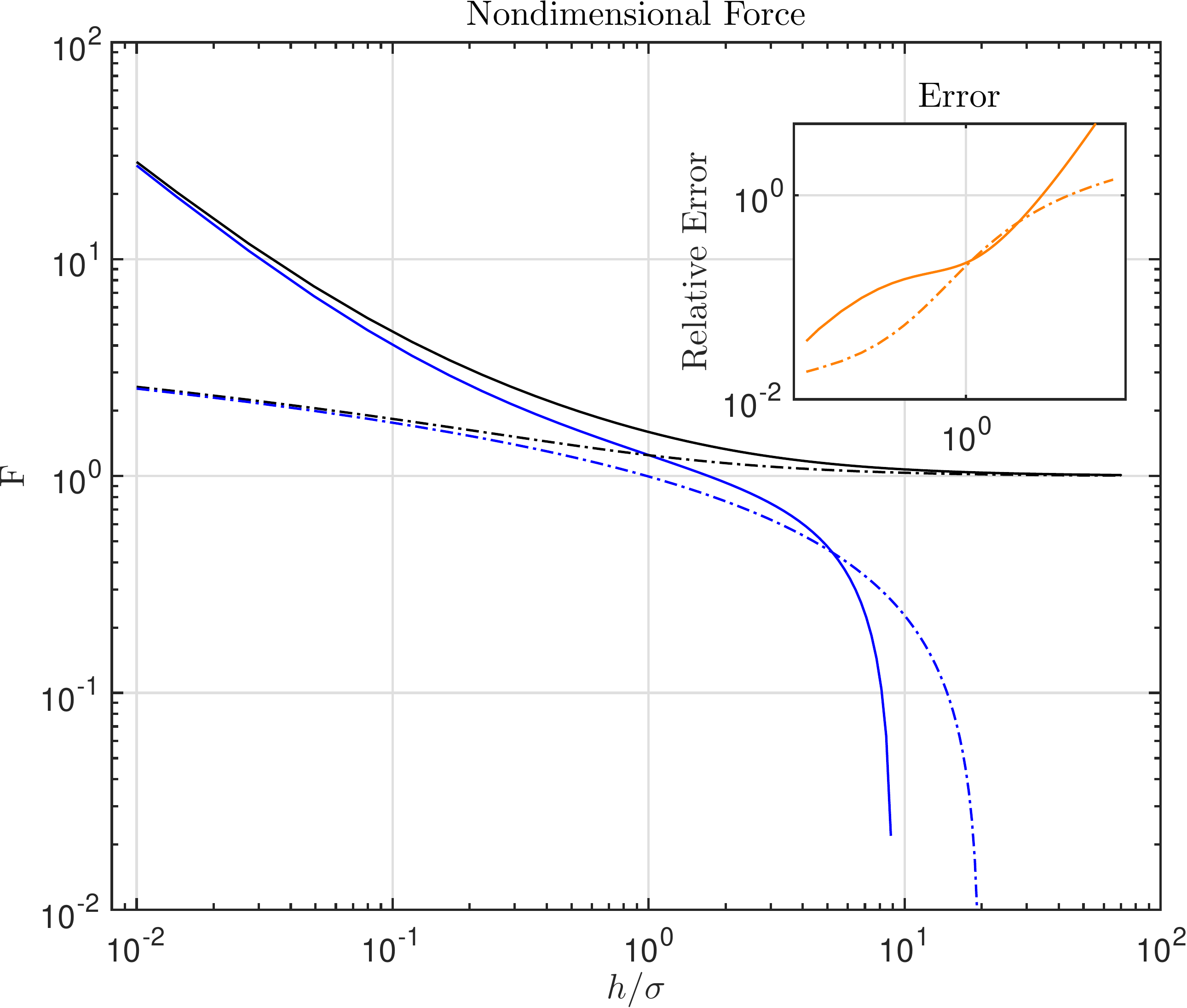}
        \caption{Using the inner region formulae of Jeffrey \& Onishi, the correct singular behaviour is obtained but the far-field is not valid.}
        \label{fig:nondimensional_force_inset}
    \end{subfigure}
    \caption{A comparison of the normal (solid) and tangential (dashed) forces and the inset relative error between the present work and multipole methods. The forces computed using {\bf GMS} and force computed using multipole methods {\color{red} Jeffrey \& Onishi} and perturbative methods {\color{blue}Kim \& Karrila}.}
    \end{figure}

\section{Positivity of $\bm{R}$}\label{sec:positivity}

\begin{figure}
    \centering      
    \begin{subfigure}[b]{0.36\textwidth}
		\includegraphics[width=\textwidth]{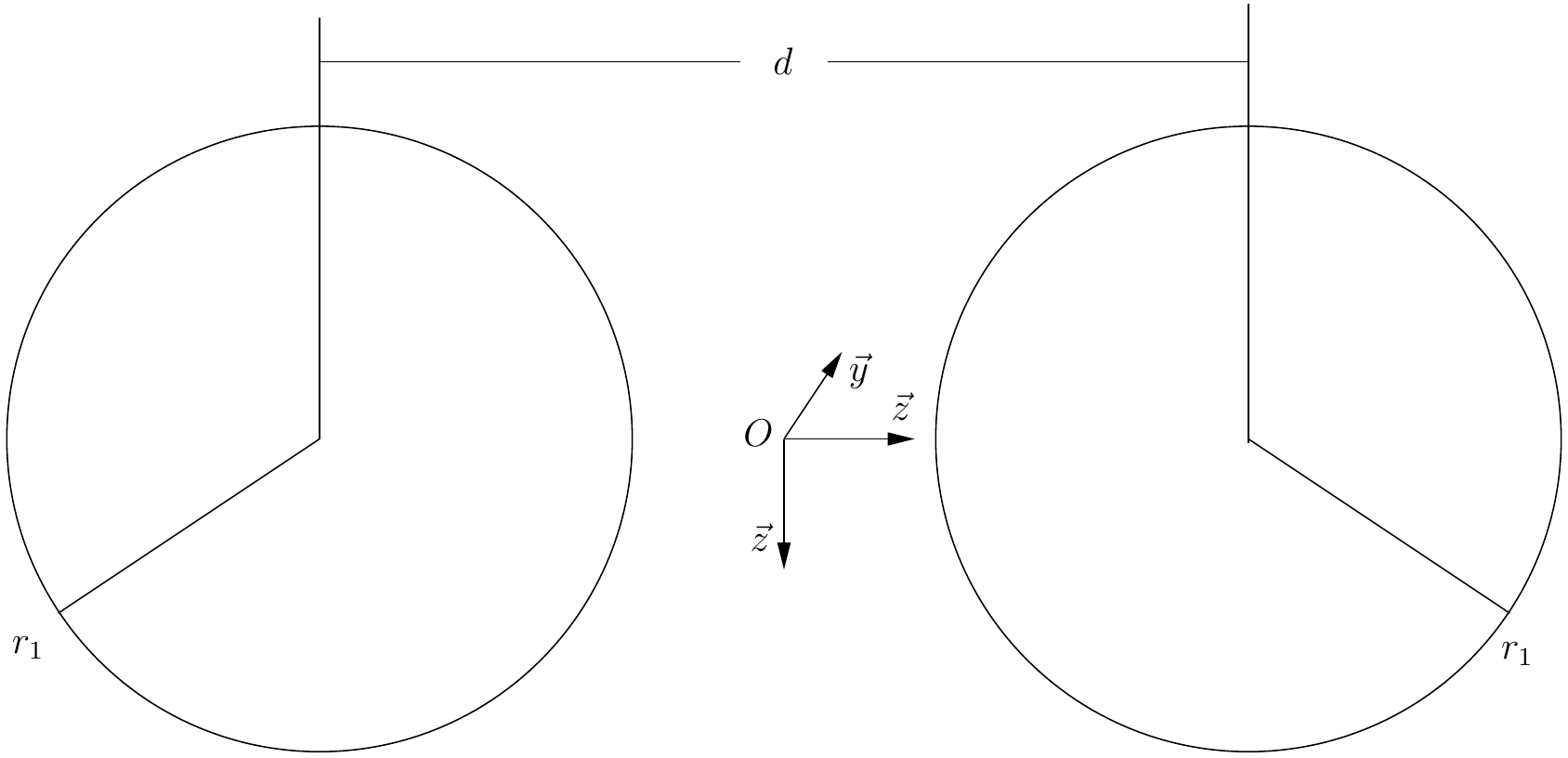}
        \caption{Schematic of the two sphere system where $d$ is varied between $\sigma<d<\infty$.}
        \label{fig:two_sphere}
    \end{subfigure}
    \,
    \begin{subfigure}[b]{0.475\textwidth}
		\includegraphics[width=\textwidth]{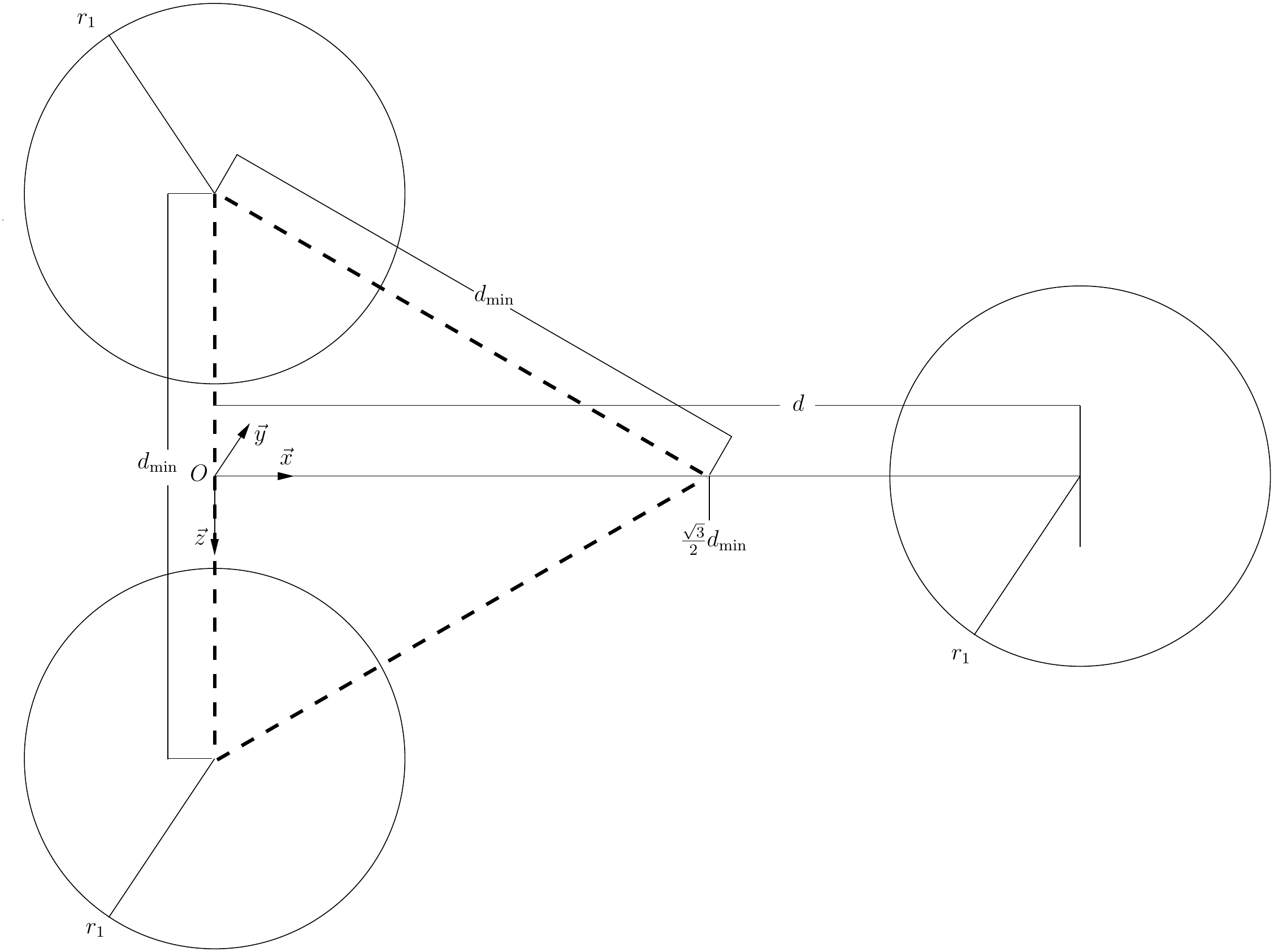}
        \caption{Schematic of the three sphere system where $d$ is varied between $\frac{\sqrt{3}}{2}d_{\min}<d<\infty$ and $d_{\min}$ is held fixed.}
        \label{fig:three_sphere_schematic}
    \end{subfigure}
    \caption{Schematics of a) a two sphere system and b) a three sphere system for the eigenvalues of $\bm{R}$.}
    \end{figure}

\begin{figure}
    \centering      
    \begin{subfigure}[b]{0.475\textwidth}
		\includegraphics[width=\textwidth]{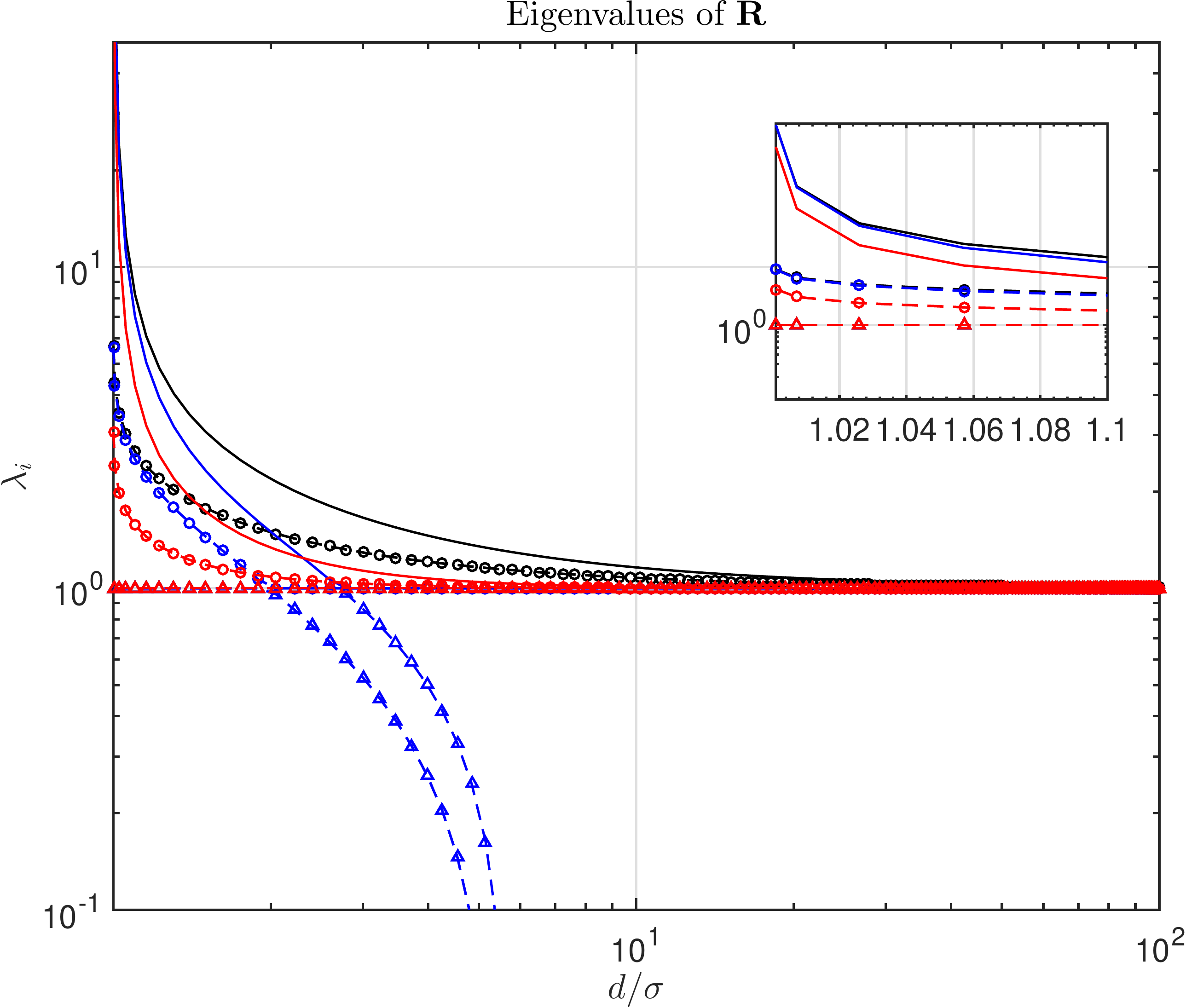}
        \caption{Eigenvalues of $\bm{R}$ for the two sphere system in Figure \ref{fig:two_sphere}.}
        \label{fig:eigs_of_R_2body}
    \end{subfigure}
    \,
    \begin{subfigure}[b]{0.475\textwidth}
		\includegraphics[width=\textwidth]{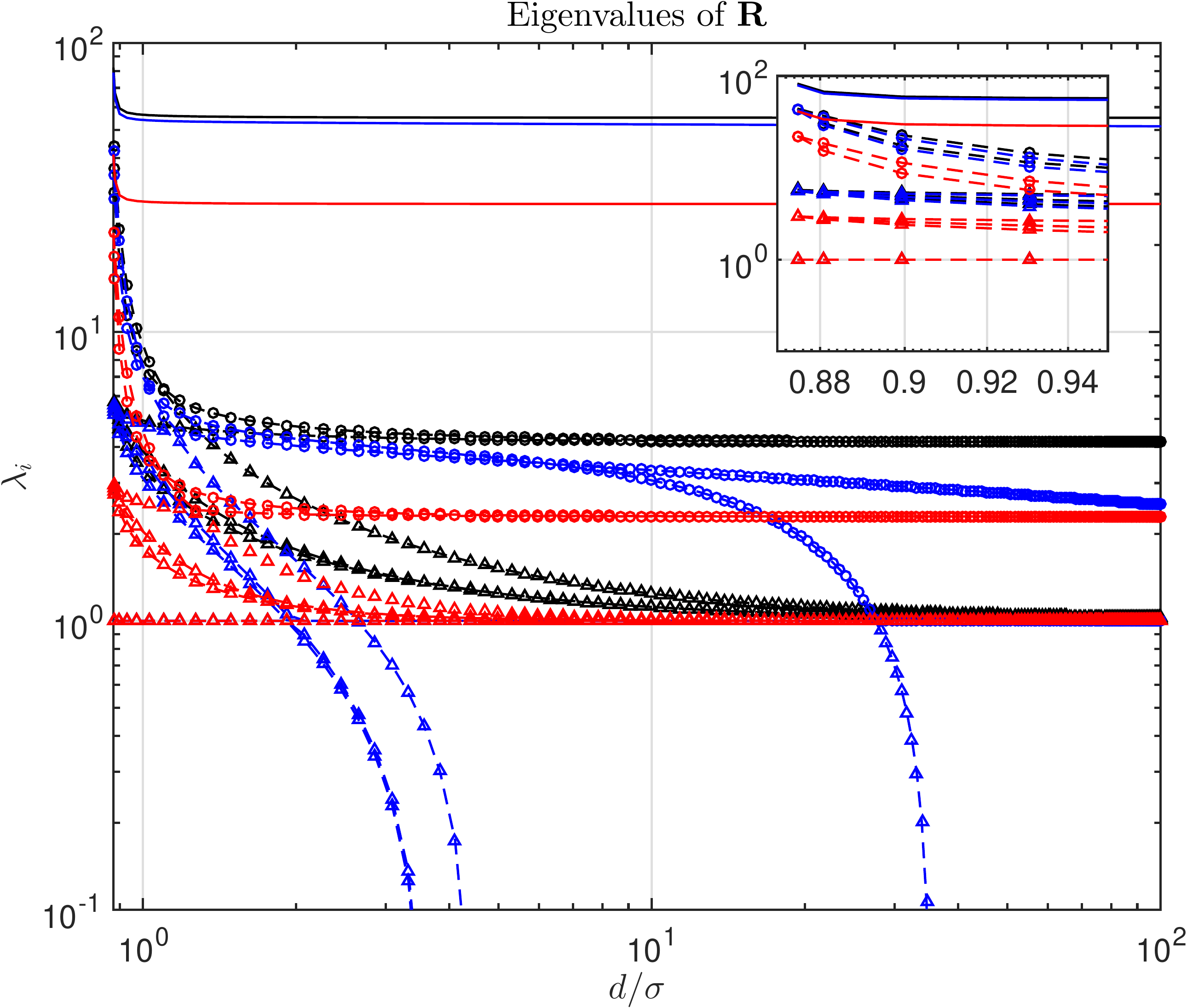}
        \caption{Eigenvalues of $\bm{R}$ for the three sphere system in Figure \ref{fig:three_sphere_schematic}. }
        \label{fig:eigs_of_R_two_body_3spheres}
    \end{subfigure}
    \caption{Eigenvalues of $\bm{R}$ as a function of the centre distance $d$ for a) a two sphere system and b) a three sphere system. Key: {\bf GMS} ($F_{z}$), {\color{blue} Kim \& Karrila} (equiv. $X_{A}^{11}$ (3.17) and $Y_{A}^{11}$ (4.15)) and {\color{red} Jeffrey \& Onishi} (equiv. $X_{A}^{11}$ (3.20) and $Y_{A}^{11}$ (4.19)). Symbols indicate multiplicity of the eigenvalues: solid = 1, circles = 2, triangles = 3. The insets show the failure of {\color{red} Jeffrey \& Onishi} to capture the correct eigenvalues in the singular limit.}
    \end{figure}

The positivity of the resistance matrix is an important property for many computational applications of the HI including Monte Carlo simulations of stochastic particle dynamics. In particular, for Langevin dynamics of colloids, one must compute $\bm{R}^{1/2}$, which is defined by the diagonalisation
\begin{align}
\bm{R}^{1/2} = \bm{S}\Lambda^{1/2}\bm{S}^{-1},
\end{align}
where $\Lambda$ is a diagonal matrix consisting of the eigenvalues of $\bm{R}$ and $\bm{S}$ is a unitary matrix consisting of columns of orthonormal eigenvectors of $\bm{R}$. Such a diagonalisation is ensured to exists when $\bm{R}$ is symmetric and real. Mathematically speaking, the positivity of $\bm{R}$ ensures the existence and uniqueness of $\bm{R}^{1/2}$. Meanwhile, in the sampling of such Langevin trajectories, the positivity is related to the fact that for a particle undergoing friction in a thermostated bath, the rate of mechanical energy dissipation should be positive. A non-positive definite resistance matrix would allow the non-physical situation that a given particle may gain kinetic energy under drag.

In this section we demonstrate that our construction of $\bm{R}$, using scalar resistance functions determined in spherical bipolar coordinates, conserves positivity as a function of sphere separation for the selected sphere set-ups considered, whereas, the alternative constructions given by assembling $\bm{R}$ with entries originating from perturbation (Kim \& Karrila) or multipole methods (Jeffrey \& Onishi), in general do not. For each formalism we obtain numerically the eigenvalues of $\bm{R}$ such that
\begin{align}
\bm{R}\vec{e}_i = \lambda_{i}\vec{e}_i
\end{align}
for $i = 1,\cdots, 3N$ where $\lambda_i$ are smooth functions of the intersphere distance for sequence of particle numbers, $N = 2, 3,\cdots$. We use \textsc{matlab}'s built in function \textsf{eig}, which is a robust eigenvalue solver based on QZ iteration for symmetric matrices. The function \textsf{eig} uses a Cholesky decomposition when $\bm{R}$ is positive definite, however for the present work, the definiteness of the resistance matrices for each of the different scalar function assemblies is not known \emph{a priori}, and in particular, one may suspect $\bm{R}$ may not be positive for some of particle separations (c.f. Oseen tensor\cite{rotne1969variational} as an approximation to the mobility tensor $\bm{R}^{-1}$) depending on the model used to construct it. 

We compute the eigenvalues of $\bm{R}$ for a) a two sphere system and b) a three sphere system, the schematic for both systems are depicted in Figure \ref{fig:two_sphere} and Figure \ref{fig:three_sphere_schematic}. For both cases we fix $\sigma = 1$. 

\subsection*{Two Sphere System}
We refer the reader to Figure \ref{fig:two_sphere} for the following discussion. For the two sphere system, the eigenvalues $\{\lambda_i\}_{i=1}^6$ are computed for $d$ varied between $0<d<\infty$ by using {\bf GMS} (present work, formulae \eqref{eq:inf_series_force_expressions_1},\eqref{eq:inf_series_force_expressions_2} and \eqref{eq:inf_series_force_expressions_tangential_1},\eqref{eq:inf_series_force_expressions_tangential_2}), {\color{blue} Kim \& Karrila}\cite{kim2013microhydrodynamics} and {\color{red} Jeffrey \& Onishi}\cite{jeffrey1984calculation} and are presented in Figure \ref{fig:eigs_of_R_2body}. In this case the eigenvalues correspond to 6 modes: three shearing interactions, two squeezing interactions, and 1 co-translating interaction of multiplicity 3, 2, and 1, respectively, owing to the repeated ways in which shearing and squeezing may occur in each of the three dimensions (recalling that from to the reversibility of Stokes flow, retreating spheres are hydrodynamically equivalent to squeezing ones). Hence, in Figure \ref{fig:eigs_of_R_2body}, repeated eigenvalues are plotted on top of each other. 

We observe that the asymptotic behaviour of the {\color{blue} Kim \& Karrila} eigenvalues agree with the {\bf GMS} eigenvalues as $d/r_1\to 0$ (as we expect since the inner region theories agree) but diverge in the far field (as we expect as the lubrication approximation breaks down). Both the {\bf GMS} and {\color{red} Jeffrey \& Onishi} eigenvalues remain positive for all $d/\sigma>1$, in particular both sets of eigenvalue converge to unity as $d/\sigma\to\infty$, which corresponds to the intrinsic Stokes drag at infinity included in both formalisms. However we know by Figure \ref{fig:nondimensional_force_arb_inset} that {\color{red} Jeffrey \& Onishi} does not provide the correct singular behaviour in the limit $d/\sigma\to 1$, and in particular we observe the eigenvalues are mismatched to both {\bf GMS} and {\color{blue} Kim \& Karrila} in the inner regime. 

\subsection*{Three Sphere System}
We refer the reader to Figure \ref{fig:three_sphere_schematic} which is a schematic for the three sphere configuration. We consider three spheres confined to the plane $y = 0$ with a minimum mutual separation $d_{\min}$ forming the edge of an equilateral triangle where two of the spheres are held fixed. For the eigenvalues $\{\lambda_i\}_{i=1}^9$ we move the location of a third sphere towards the former fixed pair by varying $d$ such that $\tfrac{\sqrt{3}}{2}d_{\min}<d<\infty$ and compute the eigenvalues of $\bm{R}$ as a function of $d$. 

In Figure \ref{fig:eigs_of_R_two_body_3spheres} we plot the eigenvalues and preserve the labelling {\bf GMS}, {\color{blue} Kim \& Karrila} and {\color{red} Jeffrey \& Onishi}. As in the two sphere case, we obtain repeated curves owing to the multiplicity of the eigenvalues. In Figure \ref{fig:eigs_of_R_two_body_3spheres} we report a similar property in the eigenvalue distribution, that the {\bf GMS} are uniformly positive, and, the eigenvalues corresponding to the pairwise interactions between the third free and the two fixed spheres converge to unity for large as $d\to\infty$. {\color{blue} Kim \& Karrila} do not preserve positivity for the three sphere system, in particular, we see that the eigenvalues diverge, and in particular, in a smaller regime of $d$ than in the two sphere system. {\color{red} Jeffrey \& Onishi} preserves positivity however, as in the two sphere system, {\color{red} Jeffrey \& Onishi} does not provide the correct singular behaviour in the limit $d/\sigma\to 1$. 

The emergence of multiple constant eigenvalues as $d\to \infty$ corresponds to convergence to the isolated pair system as the third free sphere is sufficiently separated. The disagreement in the constant eigenvalues of {\bf GMS} and {\color{red} Jeffrey \& Onishi} is a consequence of the inefficient computation of the singular term by {\color{red} Jeffrey \& Onishi}.

\subsection*{Larger Systems}

In assembling the resistance matrix for an arbitrary monodisperse system, the main parameters are the inter-sphere distances and the number of spheres. As the number of spheres increases so does the dimension of the resistance matrix. The inter-sphere distances dictate how the eigenvalues are distributed. Positivity may not necessarily be obtained for an arbitrary system. However, we may obtain some formal results about the spectral properties of $\bm{R}$ by examining a few regular systems. We let $S_N$ denote the set of all possible states of the system of $N$ spheres in a confining box. Additionally we let $X_\phi\in S$ denote the regular sphere packing at some volume fraction $\phi\in (0,\phi_g)$ for $\phi_g = \tfrac{\pi}{3\sqrt{2}}$ Gauss' constant such that for each sphere in $S$, the centre to centre distance of each nearest neighbour is $d_{\min}$. 
$X_\phi$ is a natural configuration to consider because it represents the lowest entropy state of the system at the hydrodynamic diameter $d_{\min}$. Therefore if the spectrum of $\bm{R}(X_\phi)$ may be controlled, i.e., bounded from below, one expects to be able to control $\bm{R}(X_\phi+\varepsilon)$, where $\varepsilon\in S_N$ represents a perturbation from $X_\phi$.

\begin{figure}
    \centering      
    \begin{subfigure}[b]{0.4\textwidth}
		\includegraphics[width=\textwidth]{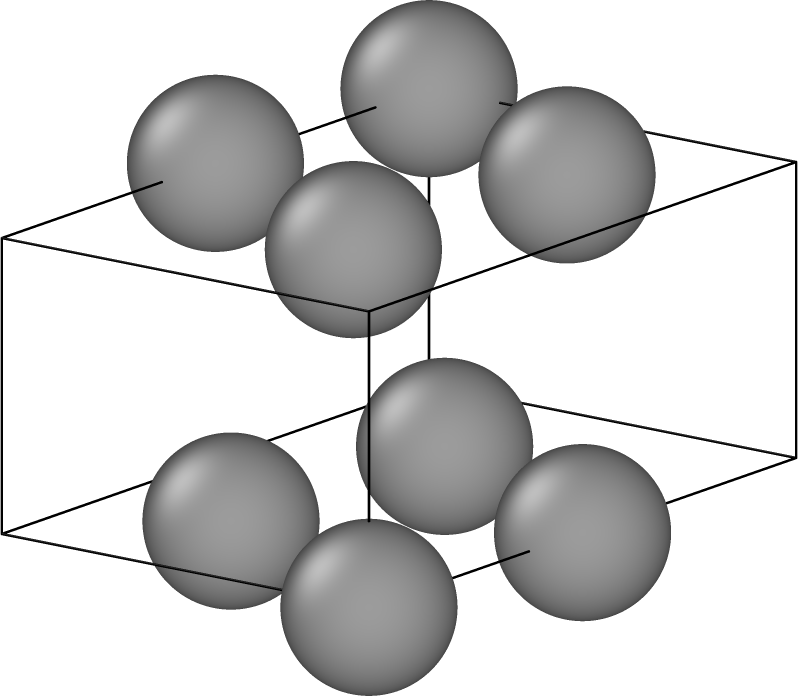}
        \caption{Configuration of $N = 8$ spheres in a regular arrangement with $d_{\min} = 2$ corresponding to a packing fraction of $\phi = 0.370$, or about 50\% of maximum packing.}
        \label{fig:test_sphere_regular_packing_8_2}
    \end{subfigure}
    \,
    \begin{subfigure}[b]{0.475\textwidth}
		\includegraphics[width=\textwidth]{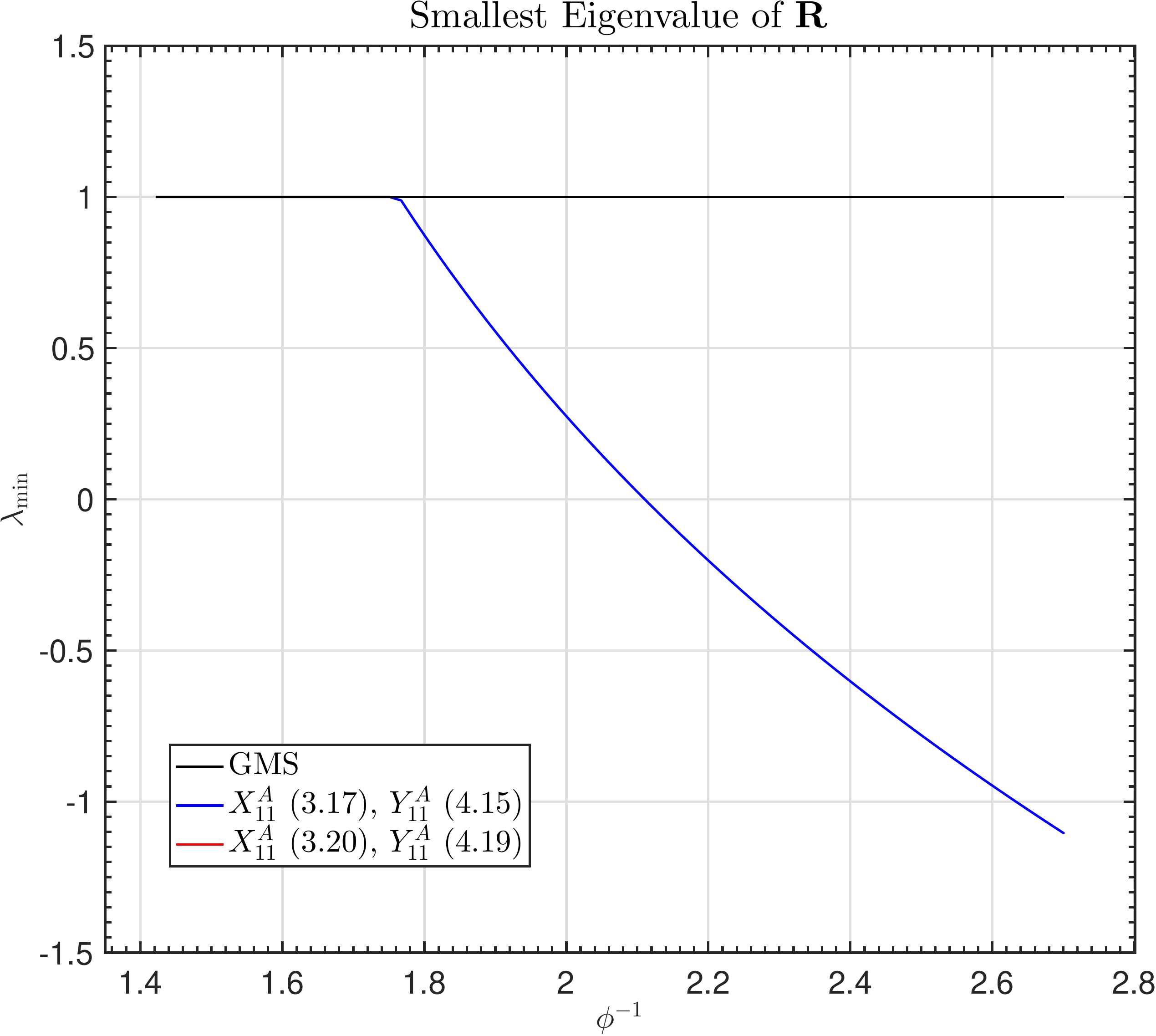}
        \caption{A plot of the smallest eigenvalue of $\bm{R}$ for the $N = 8$ sphere configuration in Figure \ref{fig:test_sphere_regular_packing_8_2}. The {\bf GMS} and {\color{red} Jeffrey \& Onishi} curves are indistinguishable where as the {\color{blue} Kim \& Karrila} curve starts to diverges at a volume fraction around $\phi = 57\%$.}
        \label{fig:eigs_of_R_as_fun_of_volFrac}
    \end{subfigure}
    \caption{a) A regular configuration of $N = 8$ spheres of diameter $\sigma = 1$ with nearest neighbour centre to centre distance $d_{\min}$ and b) The smallest eigenvalue of $\bm{R}$ with configuration as in a) with varying volume fraction.}
    \end{figure}

We may investigate the spectral properties of $\bm{R}$ for larger systems by computing the eigenvalues of $\bm{R}(X_\phi)$ as a function of $\phi\in(0,\phi_g)$ using the different scalar resistance functions. Note that $\phi = \phi_g$ corresponds to contact and is the singular limit of $\bm{R}$, which cannot be evaluated. Since for each $N$, $\bm{R}$ has $3N$ eigenvalues, in order to examine positive definiteness we need only compute the smallest eigenvalue $\lambda_{\min}$ for each formalism. Figure \ref{fig:test_sphere_regular_packing_8_2} shows a unit cell of $S_8$ which may be repeated to produce hexagonal close packing at a hydrodynamic diameter of $d_{\min} = 2$ for spheres of diameter $\sigma = 1$. The hydrodynamic diameter $d_{\min}$ and the volume fraction are related by $\phi = \pi/(3\sqrt{2})(\sigma/d_{\min})$ hence as $d_{\min}$ increases $\phi$ decreases and vice versa. In Figure \ref{fig:eigs_of_R_as_fun_of_volFrac} we plot the smallest eigenvalue of each formalism {\bf GMS}, {\color{blue} Kim \& Karrila} and {\color{red} Jeffrey \& Onishi} verses $\phi^{-1}$ for $S_8$ (so that large $\phi^{-1}$ correspond to dilute $S_8$). We report that both {\bf GMS} and {\color{red} Jeffrey \& Onishi} preserve positivity and that the smallest eigenvalue of {\color{blue} Kim \& Karrila} starts to diverge at volume fractions around $\phi = 57\%$.

Plotting the smallest eigenvalue as a function of $\phi$ gives a rough estimate for the volume fraction at which the Lubrication theory of {\color{blue} Kim \& Karrila} becomes invalid. The theory becomes invalid for volume fractions smaller than $57\%$ because the singular eigenvalues of {\color{blue} Kim \& Karrila} begin to deviate from the exact eigenvalues of {\bf GMS} at much smaller $\phi$ (not shown). We present only the smallest eigenvalues to forgo plotting 24 eigenvalues on a single axes. Additionally, the authors stopped computing the spectra of $\bm{R}$ for each {\bf GMS}, {\color{blue} Kim \& Karrila} and {\color{red} Jeffrey \& Onishi} at $N = 8$, since, beyond this sphere number, the computation time for computing the eigenvalues for regular configurations of $S_N$ outstrips gains in insight of the positivity of $\bm{R}$.

We expect the positivity to be preserved by {\bf GMS} for each $S_N$ since the boundary layer in the inner region of the resistance functions occurs only for nearest neighbours in the configuration, and the squeezing and shearing forces quickly decay to unity for centre distances of order of a sphere diameter. Additionally, the property that the rate of mechanical energy dissipation should be positive is essentially a consequence of the fact that the total solvent fluid velocity may be partitioned into the velocity fields created by the motions of the individual spheres (see Section 8--5 
\emph{Generalized treatment of multiparticle systems} \citet{happel2012low}), which is intrinsic to the spherical bipolar formalism. This cannot be said to hold rigorously for the asymptotic formalisms (Kim \& Karrila, multipole methods) since the velocity fields as found by those methods are valid only in local flow regimes (for example near to or far from sphere surfaces). The advantage of {\bf GMS}, therefore, over the formulae provided by multipole methods of {\color{red} Jeffrey \& Onishi}, is to more efficiently obtain the correct singular behaviour in the close sphere surface flow regime.

\section{A Numerical Application In DDFT}\label{sec:numerics}

\begin{figure}
    \centering      
    \begin{subfigure}[b]{0.48\textwidth}
		\includegraphics[width=\textwidth]{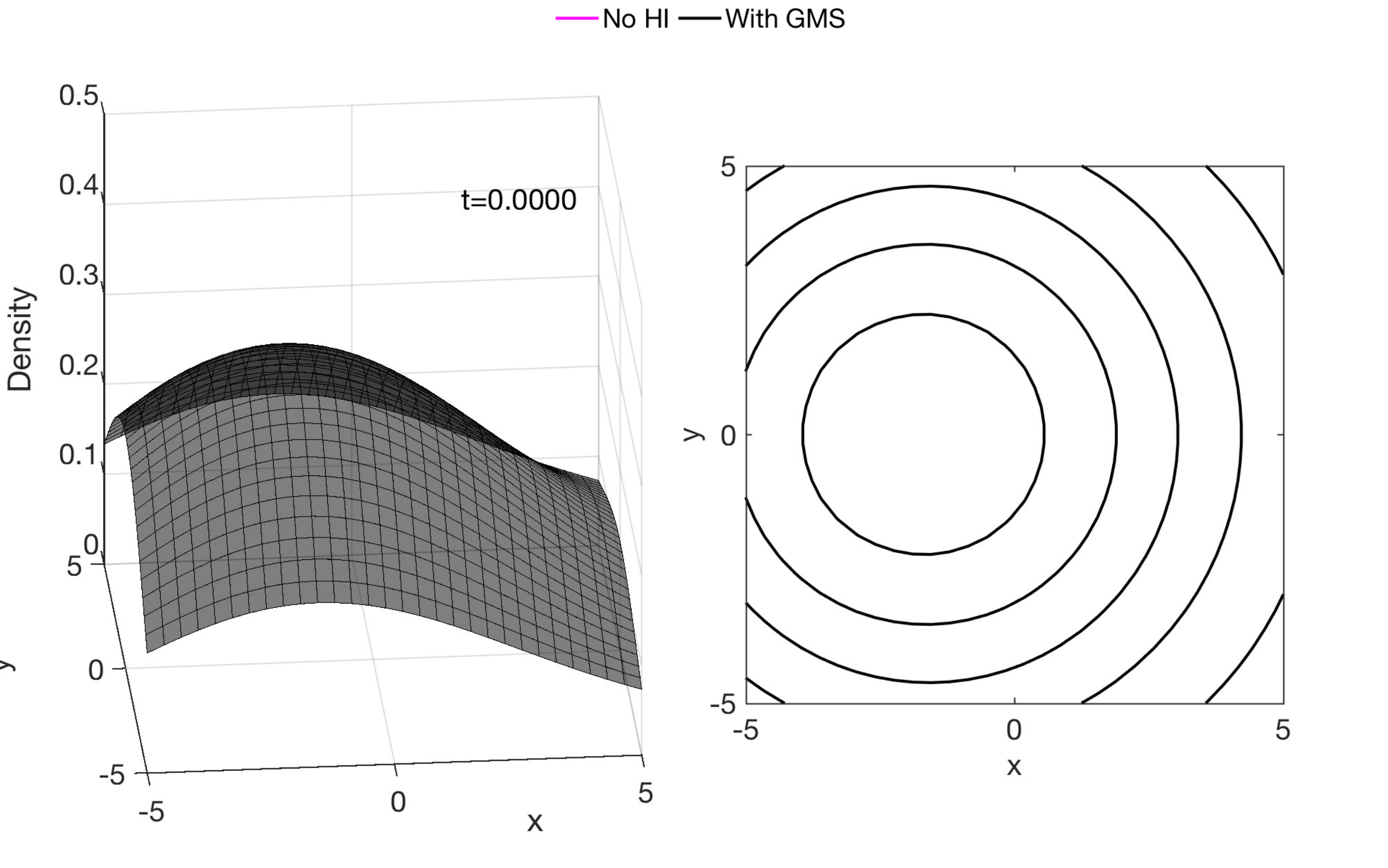}
        \caption{$t = 0.00$}\label{fig:gms_initial}
    \end{subfigure}
    \,
    \begin{subfigure}[b]{0.475\textwidth}
		\includegraphics[width=\textwidth]{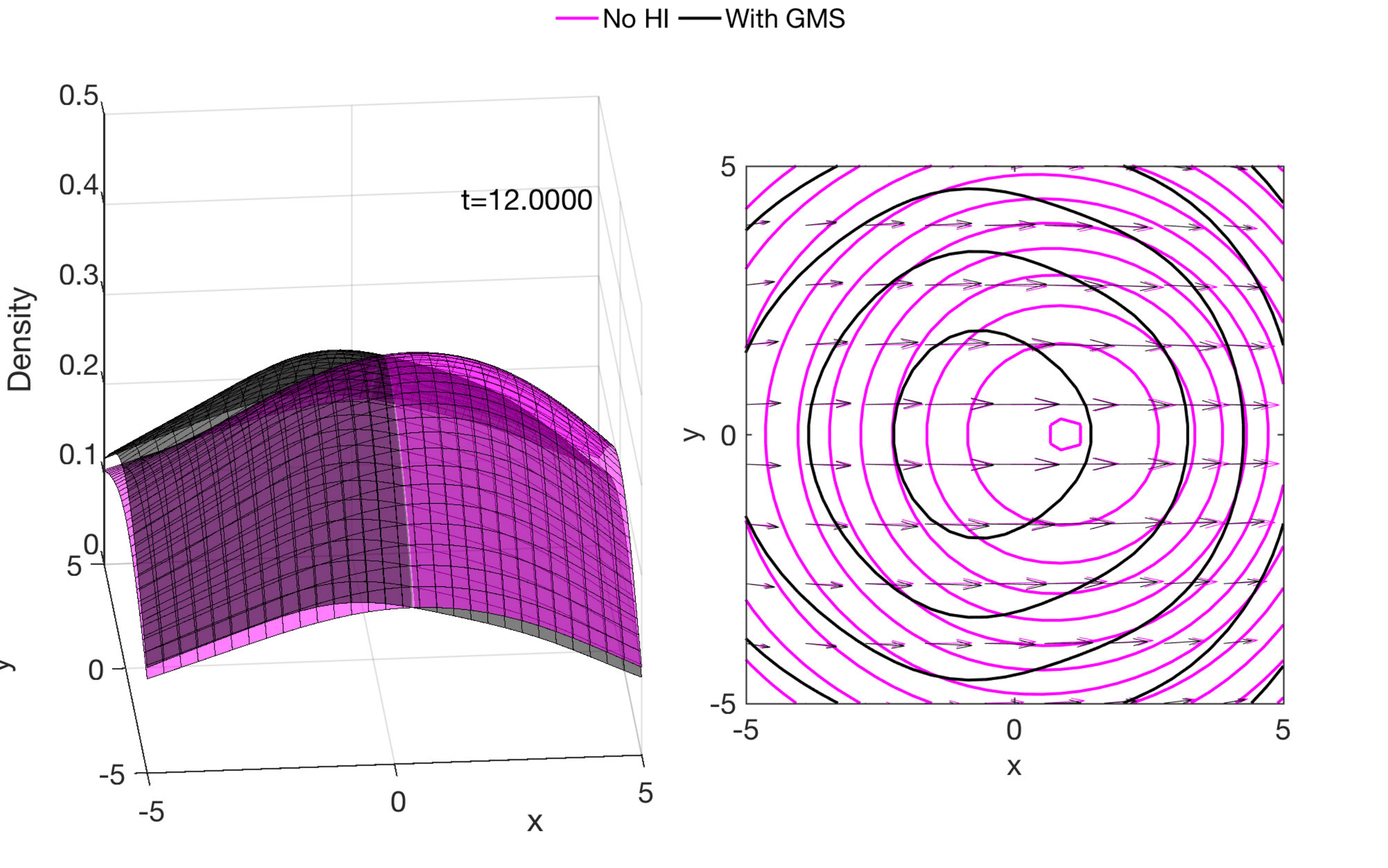}
		\caption{$t= 12.00$}\label{fig:gms_final}
    \end{subfigure}

    \caption{Numerical solution of a DDFT (\citet{goddard2012unification}) including the present theory ({\bf GMS}) compared to a reference solution without HI ({\color{magenta} magenta}). The left hand panels show the density of colloids, $\varrho(\vec{r},t)$, trapped inside a confining poential on the 2D plane. The right hand panels show the level curves of the density and the bulk velocity of the colloidal system (given by the arrows). Figure \ref{fig:gms_initial} shows the initial density and velocity meanwhile Figure \ref{fig:gms_final} shows the evolved density under a translation the confining potential over a period of $t = 12$ time units.}\label{fig:gms_ddft}
\end{figure}

\begin{figure}
    \centering      
    \begin{subfigure}[b]{0.48\textwidth}
		\includegraphics[width=\textwidth]{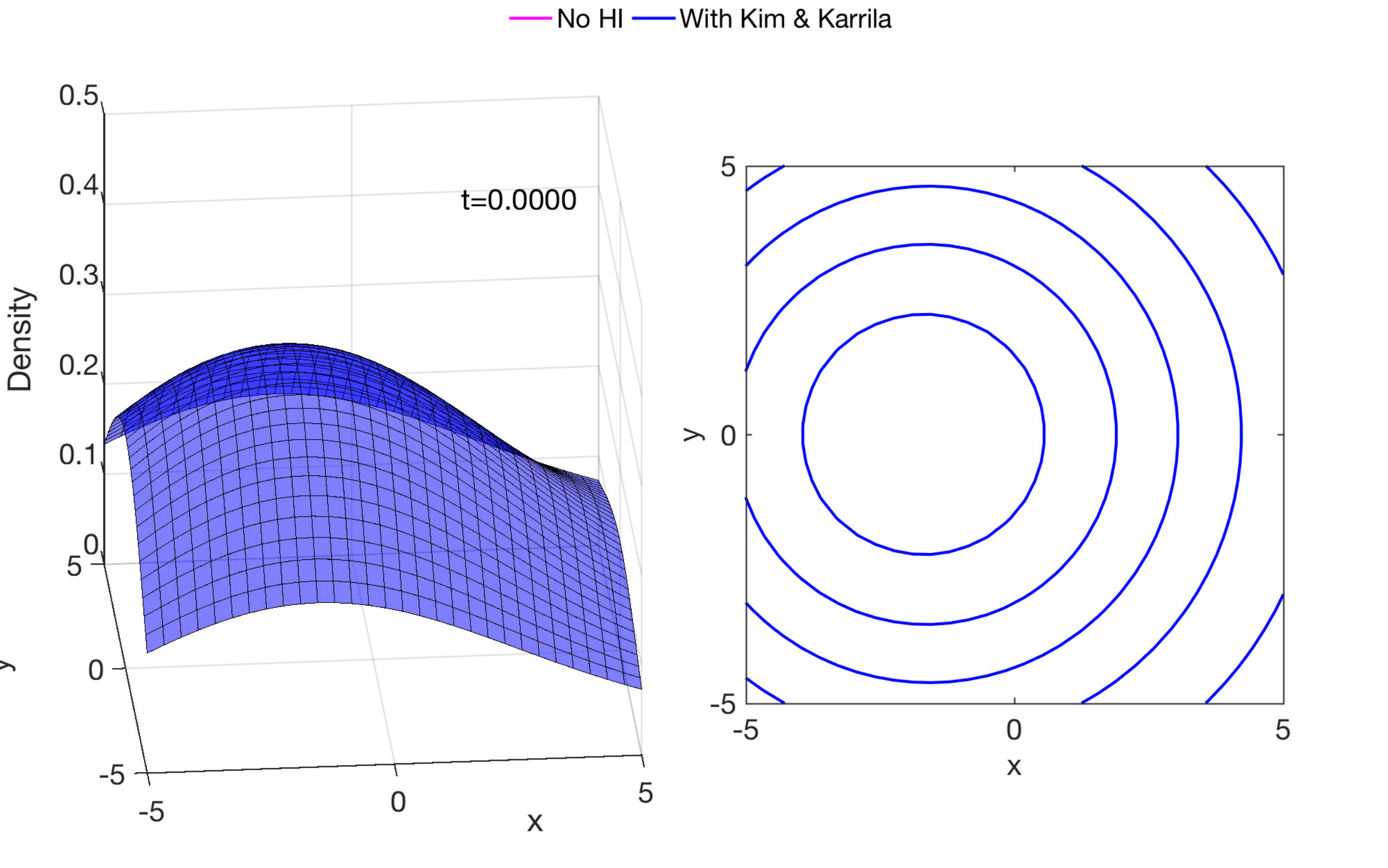}
        \caption{$t = 0.00$}\label{fig:kk_initial}
    \end{subfigure}
    \,
    \begin{subfigure}[b]{0.475\textwidth}
		\includegraphics[width=\textwidth]{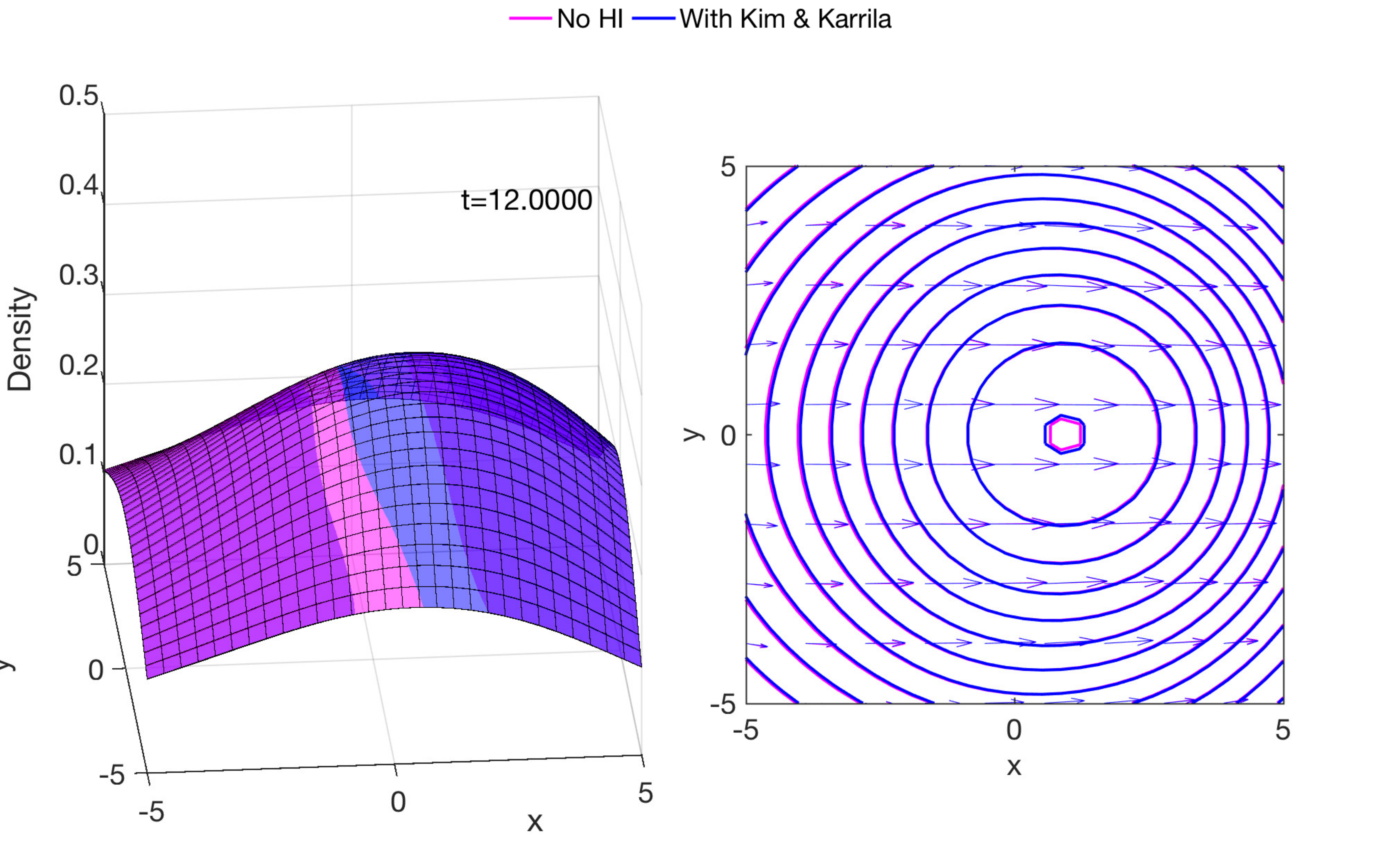}
        \caption{$t = 12.00$}\label{fig:kk_final}
    \end{subfigure}

    \caption{Numerical solution of a DDFT (\citet{goddard2012unification}) including the existing theory ({\color{blue} Kim \& Karrila}) compared to a reference solution without HI ({\color{magenta} magenta}). The left and right hand pannels are as described in Figure \ref{fig:gms_ddft} with same initial density and velocity, that is, Figures \ref{fig:gms_initial} and \ref{fig:kk_initial} would be indistinguishable if plotted on top of each other. After the same time period the density and velocity are substantially different to \ref{fig:gms_final} owing to the underestimate in the lubrication effect by the truncation inherent in the series $F_z$.}\label{fig:kk_ddft}
\end{figure}

\begin{figure}
    \centering      
    \begin{subfigure}[b]{0.48\textwidth}
		\includegraphics[width=\textwidth]{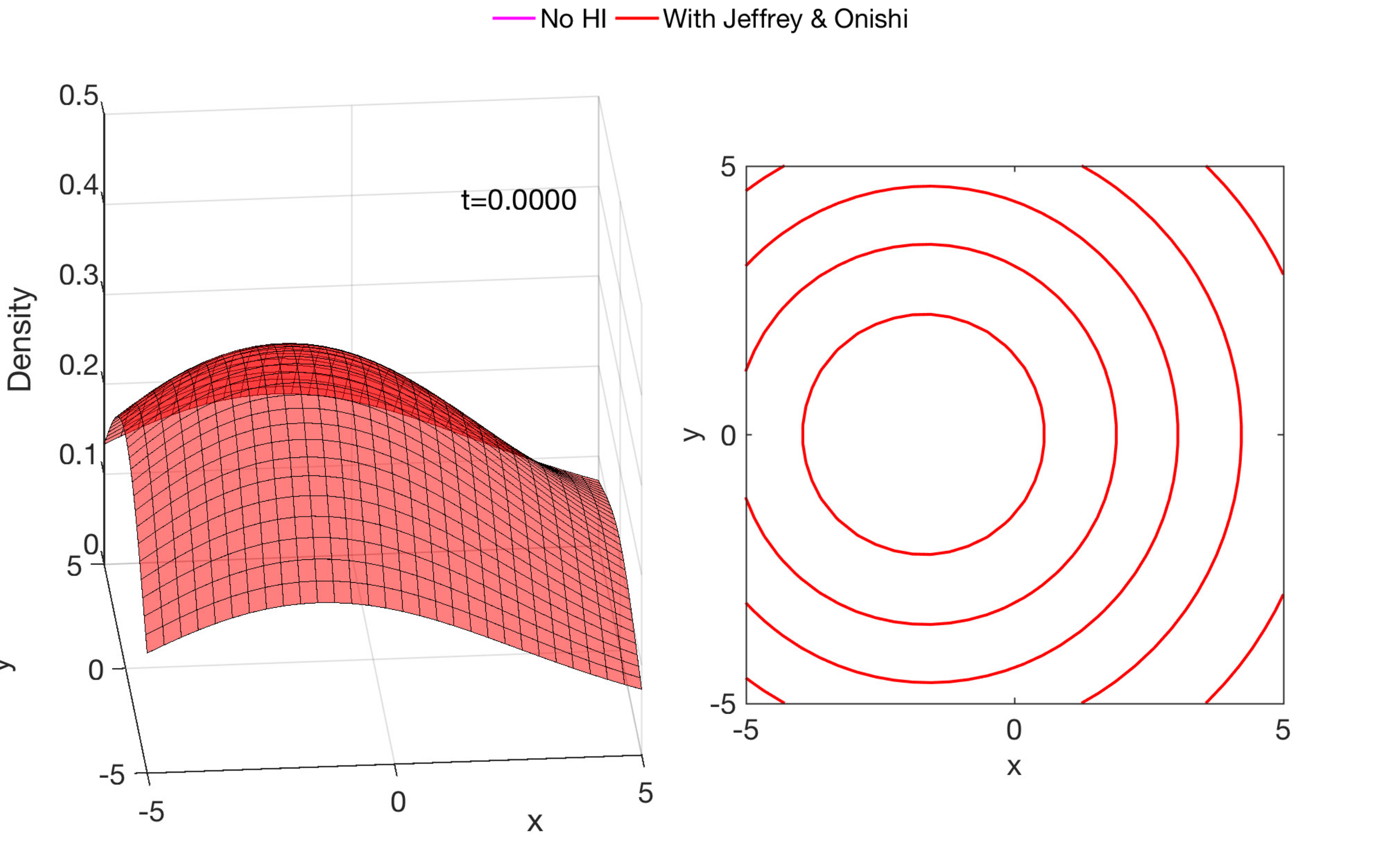}
        \caption{$t = 0.00$}\label{fig:jo_initial}
    \end{subfigure}
    \,
    \begin{subfigure}[b]{0.475\textwidth}
		\includegraphics[width=\textwidth]{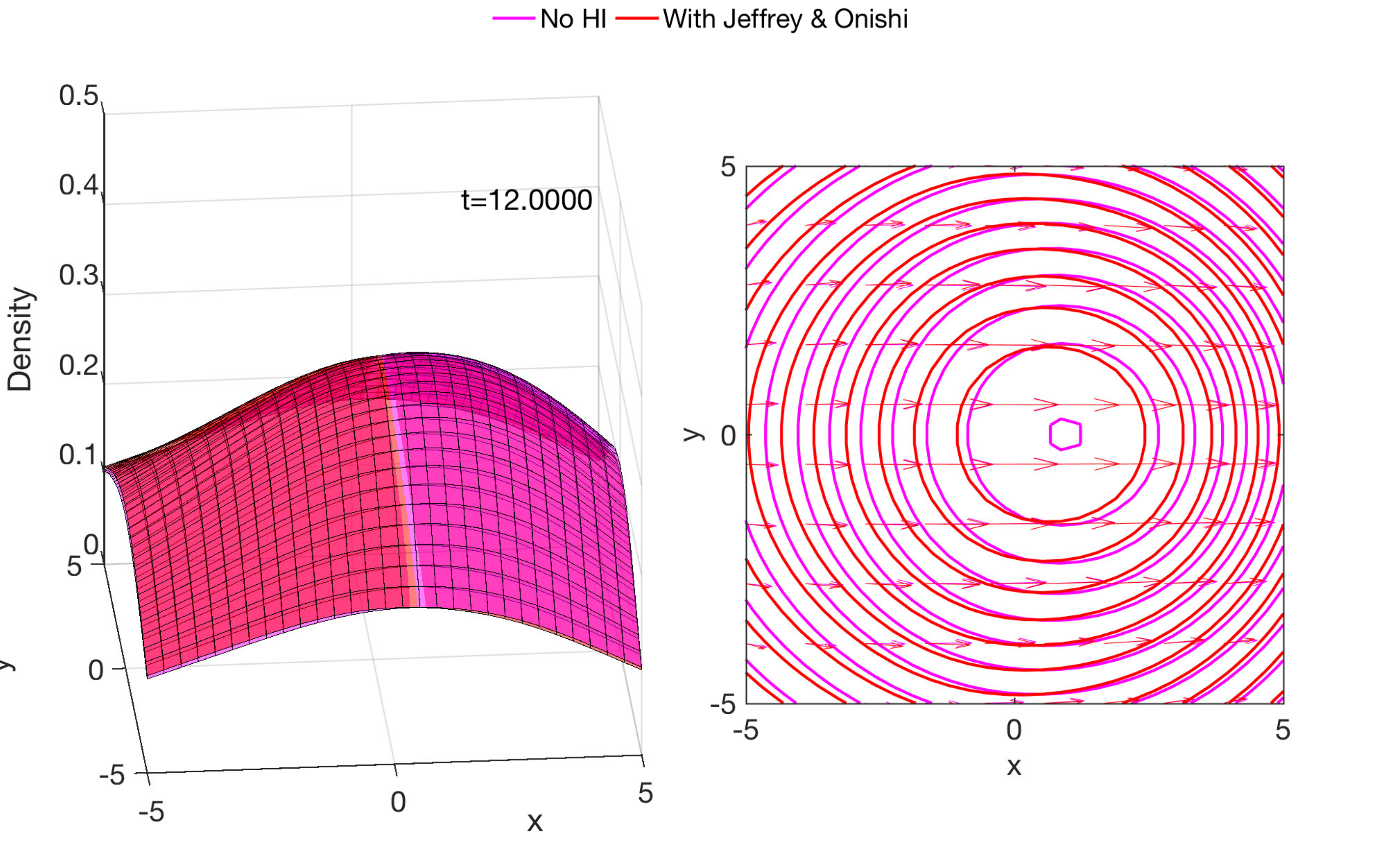}
        \caption{$t = 12.00$}\label{fig:jo_final}
    \end{subfigure}

    \caption{Numerical solution of a DDFT (\citet{goddard2012unification}) including the existing theory ({\color{red} Jeffrey \& Onishi}) compared to a reference solution without HI ({\color{magenta} magenta}). The left and right hand pannels are as described in Figure \ref{fig:gms_ddft} with same initial density and velocity, that is, Figures \ref{fig:gms_initial}, \ref{fig:kk_initial}, and \ref{fig:jo_initial} would be indistinguishable if plotted on top of each other. After the same time period the density and velocity are substantially different to \ref{fig:gms_final} owing to the underestimate in the lubrication effect in the inner region by the multipole functions $X^A_{11}$, $Y^A_{11}$.}\label{fig:jo_ddft}
\end{figure}

In this section we present a practical application of the results of the present work by considering numerical solutions of a dynamic density functional theory (DDFT) to the lubrication forces. A fully formed, in depth numerical study of solutions to DDFTs with these extensions will be considered in a separate publication. The aim of this section is to elucidate to the reader the differences which may be observed between the present and previous theory in a practical computational setting. 

We consider the probability distribution for the positions of a large collection of hard spherical particles immersed in a background bath of many more, much smaller and much lighter bath particles treated essentially as a continuum. The larger particles cause fluid flows in the bath, in turn causing forces on all other particles. These forces are considered to be the short range HI mediated by the bath and are prescribed by the resistance tensor $\bm{R}$. For the following discussion we assume {\bf{A1}}--{\bf{A3}}.

In the DDFT setting it is commonplace to separated out the column space of $\bm{R}$ corresponding to isolated spheres diffusing at infinity. In particular we write  $\bm{R}_{ij} = \bm{\Gamma}_{ij} = \gamma\bm{1} +\gamma\tilde{\bm{\Gamma}}_{ij}$ where $\gamma$ is the friction coefficient (Stokes constant) and $\tilde{\bm{\Gamma}}_{ij}$ are the nondimensional two body HI tensors. The first tensor takes into account Stokes drag on the $i$th particle and the second determines the HI between particle $i$ and particle $j$. 

In DDFT, $\bm{Z}_1$ and $\bm{Z}_2$ are the translational matrix components of $\bm{R}$. We refer the reader to \citet{goddard2012unification} for a longer discussion on the equations of motion that are now examined. In short, inertial DDFTs are nonlinear, nonlocal, integro-partial differential equations in 3D for the one-body density $\varrho(\vec{r},t)$ and one-body velocity $\vec{v}(\vec{r},t)$ describing conservation of mass and momentum of a fluid with non-constant number density. In particular, we consider the numerical solution of
\begin{align}
&\partial_t\varrho(\vec{r}_1,t)+\nabla_{\vec{r}_1}\cdot\left(\varrho(\vec{r}_1,t)\vec{v}(\vec{r}_1,t)\right) =0, \label{eq:DDFT_mass}\\
&\partial_t\vec{v}(\vec{r}_1,t)+(\vec{v}(\vec{r}_1,t)\cdot\nabla_{\vec{r}_1})\vec{v}(\vec{r}_1,t)+\tfrac{1}{m}\nabla_{\vec{r}_1}\frac{\delta \mathcal{F}[\varrho]}{\delta \varrho}(\vec{r}_1,t)\nonumber\\
&\quad+\gamma\vec{v}(\vec{r}_1,t)\nonumber\\
&\quad +\gamma\int\mathrm{d}\vec{r}_2\,\left[ \bm{Z}_1(\vec{r}_1,\vec{r}_2)\vec{v}(\vec{r}_1,t)
)+\bm{Z}_2(\vec{r}_1,\vec{r}_2)\vec{v}(\vec{r}_2,t)
)\right]\nonumber\\
&\quad\quad\times\varrho(\vec{r}_2,t)g(\vec{r}_1,\vec{r}_2,[\varrho])  = 0.\label{eq:DDFT_mom}
\end{align}

For the system of PDEs \eqref{eq:DDFT_mass}--\eqref{eq:DDFT_mom} there are 5 required inputs: 
\begin{enumerate}
\item Initial density and velocity data,
\item Free energy functional $\mathcal{F}$, 
\item Friction coefficient $\gamma$,
\item Pairwise HI tensors  $\bm{Z}_j$,
\item Correlation function $g(\vec{r}_1,\vec{r}_2,[\varrho])$.
\end{enumerate}
 
The initial data are found by solving an equilibrium DFT problem, which amounts to solving the nonlinear functional equation $(\delta \mathcal{F})/(\delta \varrho)[\varrho] = \mu_c$ where $\mu_c$ is the chemical potential of the hard sphere species. The free energy functional $\mathcal{F}[\varrho]$ is modelled with fundamental measure theory (FMT), which provides the functional form of the free energy density taking account of the entropy reduction produced by hard sphere exclusion (see Rosenfled \cite{rosenfeld1989free} or Roth \cite{roth2010fundamental}). Additionally, in $\mathcal{F}$, one may include external potentials such as gravity as well as interparticle potentials, for electrostatic interactions. The friction coefficient $\gamma$ may be varied as a proxy for the solvent viscosity. The pairwise resistance tensors take into account the HI, which we will construct using the resistance functions of the present work, as well as the existing perturbative and multipole counterparts. The correlation function is not known exactly and must ultimately be obtained from the microscopic dynamics, but for a hard sphere fluid may be approximated by $g(|\vec{r}-\vec{r}'|) = 0$ for $|\vec{r}-\vec{r}'|<\sigma$ (denoting exclusion) and unity otherwise. Such an approximation has been shown to give good agreement with comparative stochastic simulations of the underlying Langevin dynamics \citet{goddard2012general},\citet{goddard2012unification}, \citet{goddard2013multi}.

We solve \eqref{eq:DDFT_mass}--\eqref{eq:DDFT_mom} with the pseudospectral collocation scheme 2DChebClass \cite{DDFTCode}. For a more detailed analysis on the numerical method, including the basic quadrature technique of the convolutions of the HI matrices, see \cite{nold2017pseudospectral}. We present two solutions: one labelled {\bf GMS} to denote $\bm{Z}_1$, $\bm{Z}_2$ constructed with the scalar resistance function \eqref{eq:inf_series_force_expressions_1} obtained by present work (Figure \ref{fig:gms_ddft}) and one denoted {\color{blue}Kim \& Karrila} (Figure \ref{fig:kk_ddft}) which uses the well known, widely used expression \eqref{eq:kk_F_z} to construct $\bm{Z}_1$, $\bm{Z}_2$. A reference solution in both cases with $\bm{Z}_1 = \bm{Z}_2 = 0$ is shown in {\color{magenta}magenta}. For the solution using {\color{blue}Kim \& Karrila}, a necessary outer cuttoff was chosen at 2 sphere diameters which is accepted in the community as standard~\cite{townsend2018anomalous}. For {\bf GMS} no outer cut off is required. 

We solve the DDFT equations \eqref{eq:DDFT_mass}, \eqref{eq:DDFT_mom} in a 2D planar geometry confining the colloids in a weak quadratic background potential before driving the colloids from left to right with a potential flow. We take $\gamma = 2$, with $50$ colloids, but many more may be included since the dimensionality of DDFT is independent of the number of colloids. Both the HI terms as constructed by {\bf GMS} and {\color{blue}Kim \& Karrila} retard the flow of the colloid particles in comparison with DDFTs without any interparticle HI shown in {\color{magenta}magenta}, which is what we expect from the standard descriptions of the effects of lubrication forces. This is in contrast to overdamped DDFT equations including long-range forces, essentially including HI terms corresponding to two-body $\bm{R}^{-1}$ which enhance collective motion \citet{goddard2016dynamical}.

Figures \ref{fig:gms_ddft}, \ref{fig:kk_ddft} show a substantial difference in the evolution of the density $\varrho$ (and flux) of the suspension, in particular  
{\color{blue}Kim \& Karrila} appears to underestimate the effect of the lubrication force on the overall dynamics of the density, where as {\bf GMS} shows the onset of extrusion in the density contours not visible using existing theory.

\section{Discussion and Open Problems}\label{sec:discussion}

The formula obtained in spherical bipolar coordinates is uniformly accurate for all separations, up to the particle contact point where the governing equations break down. In Figure \ref{fig:comparison_with_jeffrey_onishi_300_terms_beta_1}, the {\color{red} red} and {\bf{black}} curves differ substantially at surface separations equal to roughly 1 sphere radius; we therefore claim the spherical bipolar formalism would be particularly useful for simulations of colloidal flow with HI in the moderately-dense volume fraction regime.  
Additionally we expect our method to perform better for different particle radii as evidenced by Figure \ref{fig:comparison_with_jeffrey_onishi_15_terms_beta_4}, so the contributions of the present work go well into polydisperse particle systems.

We therefore expect that the derived formulae can be implemented in all numerical methods that incorporate the existing lubrication models and improve the simulation accuracy. We discuss, as examples, the potential application to and impact on a few different types of numerical methods. 

For methods solving particle dynamics using Newtonian equations, e.g., the discrete element method (DEM), the new formulae can be used to directly compute the hydrodynamic forces. Instead of using the existing formulae ($F_{z,l}$) with an arbitrary outer cut off~\cite{ness2016shear}, implementing either the exact $F^1_z$ or the asymptotic $F_{z}^e$ formulae could better capture the hydrodynamic interaction between $10^{-1}r_1$ and $10^0r_1$, as seen 
in Figure \ref{fig:force_biexact_asymp_kk}. This is expected to improve suspension viscosity predictions, compared to using $F_z$, which underestimates the viscosity especially at moderate concentrations \cite{ness2016shear}. Note however that by using $F_{z}^\ast$ for DEM, one requires accurate knowledge of the position of a hard cutoff of the asymptotic expressions for the force, if such a cutoff exists at all. 

Computational formalisms which use the closed asymptotic formula $F_z$ in-line can be trivially updated with the new asymptotic formulae $F^e_z$, meaning the applications of the presented results may also extend more generally to, e.g., lattice Boltzmann method \citet{Nguyen:02a} and Stokesian dynamics (SD) \citet{Brady:1988up}. SD takes into account singular lubrication interactions by making use of the explicit formulae $F_z$ between pairs of close particles without considering the lubrication many-body effects,  thus forgoing the large number of degrees of freedom required to resolve the lubrication flow of the interstitial fluid between particles. The missing many-body effects are considered in a more recent work \cite{lefebvre2015accurate} by decomposing the velocity field into a singular flow containing the short-range lubrication interactions and a remainder field which is regular and dealt with using a chosen fluid solver. Such methods may seek to use the present stream function $\psi$ for the decomposition. Meanwhile new approaches \cite{townsend2018anomalous} have been proposed to overcome unphysical results in pairwise lubrication models due the lost screening effects provided by neglected long-range HI. The present work can determine the deficit in lubrication beyond the critical interaction radius used in these methods.

Lastly, for continuum approaches such as dynamical density functional theory \cite{goddard2012general}, the inclusion of long range HI has been shown to produce qualitatively different colloidal fluid flows compared to systems without HI. So far the physical phenomena included in the governing fluid equations has extended to: inertial colloids with long range HI (including models of $\bm{R}^{-1}$) \cite{goddard2012unification} and without HI \cite{archer2009dynamical}, systems of multiple-species \cite{goddard2013multi} and particles with angular dependence \cite{duran2016dynamical}. Thus we expect natural numerical implementations of the present formulae to include lubrication interactions in the DDFT modelling formalism. In particular, for DDFT, since the terms corresponding to HI take the form of convolution integrals it is desirable to have explicit continuous integrands (and decay estimates) for the hydrodynamic interaction valid at all separations in order to ensure the convergence of these terms, which is what the current formalism provides.

Finally we remark that  the rate of convergence of the force asymptoting to unity at infinity will depend on $r_2/r_1$, as seen in Figure \ref{fig:force_r1r2},  and therefore we anticipate the novel study of bulk flow properties using the $F_z^1$, $F_z^2$ in the modelling of suspensions involving multiple species.

In this paper we have presented a new formula for the hydrodynamic force exerted on two converging spheres in viscous fluid in a functional form, as well as asymptotic formulae as the spheres are close, showing good agreement with the exact value even at centre to centre distances of $O(d^0)$. By construction, the derivation of this functional form provides the way for consideration of alternative boundary conditions. For the asymptotic results, the small argument limit newly derived shows better agreement with the exact solution compared to that from existing lubrication theory. The sphere plane limit may also be recovered more accurately. Additionally we have provided an analysis of the spectral properties of $\bm{R}$, demonstrating \emph{numerically} that the scalar resistance functions as determined by spherical bipolar coordinates preserves positivity in some, $N=2, 3$ and larger \emph{regular} sphere systems. Positivity is destroyed when using the perturbative functions of {\color{blue} Kim \& Karrila} without an arbitrary cut-off, meanwhile cut-offs may drastically underestimate the lubrication effect, as demonstrated by a numerical application in DDFT. It would be an interesting topic of future work to investigate the generality of this positive definiteness.

Furthermore we have shown that the scalar resistance functions obtained by {\color{red} Jeffrey \& Onishi}, while preserving positivity in the examined systems, are inaccurate compared to {\bf GMS} in inner regimes of flow (close particle surfaces) principally because they are based on multipole expansions which, intrinsic to the method, requires arbitrarily many terms as $h\to 0$, which for each $h>0$, become more computationally expensive to obtain. This property is an important consideration for dense particle systems. It would be an interesting topic of future work to investigate the generality of the positive definiteness obtained in this paper.

There are many promising extensions which may naturally be made to the theory presented here such as: alternative boundary conditions to model slippery particles and the shearing motion of two spheres converging perpendicular to their line of centres akin to \citet{goldman1966slow}. The former is generally important in liquid spreading problems \cite{cox1986dynamics}, in particular,  molecular dynamics simulations of Newtonian liquids have shown that there exists a nonlinear relationship between the amount of slip and the local shear rate of fluid at a solid surface\cite{thompson1997general}. 

\appendix
\section{Useful Formulae}\label{appA}
In Section \ref{subsec:linsys} we use the matrix $M=$ whose entries are defined by 
\begin{align}
\begin{split}
m_{11}=\cosh(n+\tfrac{3}{2})\eta_1, \quad m_{21}=\cosh(n+\tfrac{3}{2})\eta_2,\\
m_{12}=\sinh(n+\tfrac{3}{2})\eta_1, \quad m_{22}=\sinh(n+\tfrac{3}{2})\eta_2,\\
m_{13}=\cosh(n-\tfrac{1}{2})\eta_1, \quad m_{23}=\cosh(n-\tfrac{1}{2})\eta_2,\\
m_{14}=\sinh(n-\tfrac{1}{2})\eta_1, \quad m_{24}=\sinh(n-\tfrac{1}{2})\eta_2,
\end{split}
\end{align}
and
\begin{align}
\begin{split}
m_{31}=(n+\tfrac{3}{2})\sinh(n+\tfrac{3}{2})\eta_1, \quad m_{41}=(n+\tfrac{3}{2})\sinh(n+\tfrac{3}{2})\eta_2,\\
m_{32}=(n+\tfrac{3}{2})\cosh(n+\tfrac{3}{2})\eta_1, \quad m_{42}=(n+\tfrac{3}{2})\cosh(n+\tfrac{3}{2})\eta_2,\\
m_{33}=(n-\tfrac{1}{2})\sinh(n-\tfrac{1}{2})\eta_1, \quad m_{43}=(n-\tfrac{1}{2})\sinh(n-\tfrac{1}{2})\eta_2,\\
m_{34}=(n-\tfrac{1}{2})\cosh(n-\tfrac{1}{2})\eta_1, \quad m_{44}=(n-\tfrac{1}{2})\cosh(n-\tfrac{1}{2})\eta_2.
\end{split}
\end{align}

For the computation of \eqref{eq:stress_surface_integral} in Section \ref{sec:force} we find it useful to define the quantity 
\begin{align}
w_n(\eta):=\int_{-1}^{1}\tfrac{\mathrm{d}\mathfrak{x} \,Q_n(\mathfrak{x} )}{(\cosh\eta-\mathfrak{x} )^{1/2}} = 2\sqrt{2}[\tfrac{e^{\mp (n+3/2)\eta}}{2n+3}-\tfrac{e^{\mp(n-1/2)\eta}}{2n-1}]
\end{align}
where the signs are chosen according to each sphere.  
The first few derivatives of $w_n(\eta)$ are
\begin{align}
\begin{split}
w_n'=&-\tfrac{\sinh\eta}{2}\int_{-1}^{1}\tfrac{\mathrm{d}\mathfrak{x} \,Q_n(\mathfrak{x} )}{(\cosh\eta-\mathfrak{x} )^{3/2}},\\
w_n''=&\tfrac{3\,\sinh^2\eta}{4}\int_{-1}^{1}\tfrac{\mathrm{d}\mathfrak{x} \,Q_n(\mathfrak{x} )}{(\cosh\eta-\mathfrak{x} )^{5/2}}-\tfrac{\cosh\eta}{2}\int_{-1}^{1}\tfrac{\mathrm{d}\mathfrak{x} \,Q_n(\mathfrak{x} )}{(\cosh\eta-\mathfrak{x} )^{3/2}},\\
w_n^{(3)}=&-\tfrac{\sinh\eta}{2}\int_{-1}^{1}\tfrac{\mathrm{d}\mathfrak{x} \,Q_n(\mathfrak{x} )}{(\cosh\eta-\mathfrak{x} )^{3/2}}+\tfrac{9\, \sinh\eta\,\cosh\eta}{4}\int_{-1}^{1}\tfrac{\mathrm{d}\mathfrak{x} \,Q_n(\mathfrak{x} )}{(\cosh\eta-\mathfrak{x} )^{5/2}}\\
&-\tfrac{15\,\sinh^3\eta}{8}\int_{-1}^{1}\tfrac{\mathrm{d}\mathfrak{x} \,Q_n(\mathfrak{x} )}{(\cosh\eta-\mathfrak{x} )^{7/2}}.
\end{split}
\end{align}
The purpose of these expressions is to allow us to give explicit forms for certain integrals. In particular
let $p=2n+1$ for $n\in\mathbb{N}$ then for $\mathcal{I}_{p/2}:=\int_{-1}^{1}\tfrac{\mathrm{d}\mathfrak{x} \, Q_n(\mathfrak{x} )}{(\cosh\eta-\mathfrak{x} )^{p/2}}$ one has
the first few formulae
\begin{align}
\begin{split}
\mathcal{I}_{3/2}&=-2\csch\eta\, w_n'(\eta), \\
\mathcal{I}_{5/2}&=\tfrac{4\csch^2\eta}{3}\left[w_n''(\eta)+\tfrac{\cosh\eta}{2}\mathcal{I}_{3/ 2}^n\right],\\
\mathcal{I}_{7/2}&=-\tfrac{8\,\csch^3\eta}{15}\left[w_n^{(3)}(\eta)-\tfrac{9\,\cosh\eta\, \sinh \eta}{4}\mathcal{I}_{5/ 2}^n+\tfrac{\sinh\eta}{2}\mathcal{I}_{3/ 2}^n\right].
\end{split}
\end{align}
With the $\mathcal{I}_{p/2}$, the integral \eqref{eq:stress_surface_integral} may be computed explicitly.

\section{Small \& Large Argument Limits}\label{sec:asymptotics}
We divide this section into two cases: nondimensional separation going to zero and to infinity. First we identify a small parameter.
\subsection{Small Parameter}
Taking care that $\eta_2<0$, we have by the geometric properties of the bipolar coordinate system
\begin{align}\label{eq:trans_eqns_for_etas}
r_1\sinh\eta_1+r_2\sinh\eta_2=0, \qquad
d= r_1\cosh\eta_1+r_2\cosh\eta_2,
 \end{align} 
where $d$ is the centre to centre distance of the spheres. The equations \eqref{eq:trans_eqns_for_etas} constitute a coupled pair of transcendental equations in $\eta_1,\eta_2$. The determinant of the Jacobian associated to the system \eqref{eq:trans_eqns_for_etas} is always positive because $\sinh(\eta_1-\eta_2)>0$ and, given $d, r_1$, and $r_2$, it may be solved using a Newton iteration scheme.  In the case $r_1=r_2$ we may find $\eta_1$ (and $\eta_2$) explicitly. As $d$ approaches $r_1+r_2$ one obtains
\begin{align}
r_1\eta_1+r_2\eta_2\sim 0,\qquad
d\sim r_1(1+\tfrac{\eta_1^2}{2})+r_2(1+\tfrac{\eta_2^2}{2}).
\end{align}
Noting that $r_1+h+r_2=d$, the system may be solved with $\epsilon =  \tfrac{\eta_1^2}{2}\tfrac{\beta+1}{\beta}$ where $\epsilon = h/r_1$ and $\beta= r_2/r_1$. Thus we see, with an abuse of notation, by setting $a=r_1$ and $b=r_2$ that the gap distance may be written in terms of the average of the radii: $a\epsilon = \eta_1^2(a+b)/2$. This illuminates the relationship between the present small parameter $\eta_1$ and the lubrication theory small parameter  $\epsilon$ \cite{jeffrey1982low}. 

\subsection{Small Argument Behaviour}\label{sec:small_arg_behaviour}
We would like to examine the singular behaviour as $d$ approaches $r_1+r_2$ for unequal spheres. Firstly it will be seen that the limit $|\eta_j|\searrow 0$ for both $j=1,2$ may not be commuted with \eqref{eq:inf_series_force_expressions_1}, \eqref{eq:inf_series_force_expressions_2} because a divergent series is obtained despite (for physical reasons) the limit being well posed. This limit of the infinite series is hereby treated as a matched perturbation problem of Van Dyke type (see \citet{hinch1991perturbation}), whereby two series overlap in a shared regime of validity. We consider sphere 1 (a similar method can be applied to sphere 2), let $N$ be a large positive integer and nondimensionalise $F_{z}^1$. We now write $\mathcal{F}^{1}_z= \mathcal{F}_s+\mathcal{F}_r$
with
\begin{align}\label{eq:F_decomp_small_arg_lim}
\begin{split}
\mathcal{F}_s:=&-\tfrac{\sinh\eta_1}{3\sqrt{2}}\sum_{n=1}^{N}(2n+1)(a_n+b_n+c_n+d_n),\\
\mathcal{F}_r:=&-\tfrac{\sinh\eta_1}{3\sqrt{2}}\sum_{n=N+1}^{\infty}(2n+1)(a_n+b_n+c_n+d_n).
\end{split}
\end{align}
With this decomposition the difficulties arising in the limit $\eta_1\to 0$ may be avoided with proper care of the asymptotic parameter, summation index $n$ and the introduction of an intermediate variable in the shared regime of validity between $\mathcal{F}_s$ and $\mathcal{F}_r$. For the remaining calculations we set $\alpha=\eta_1$, $\eta_2 = -\beta^{-1}\alpha$ and proceed rigorously to the small limit by a parallel analysis to the asymptotic results of \citet{cox1967slow}.


Starting with $\mathcal{F}_s$ we write all the hyperbolic functions as power series in $\alpha$ obtaining
\begin{align}\label{eq:F_s_expansion_in_alpha}
\mathcal{F}_s = \alpha^{-2}\mathfrak{f}_1+\mathfrak{f}_2 + \alpha \mathfrak{f}_3+ O(\alpha^2)
\end{align}
where
\begin{align*}
\mathfrak{f}_1 &= \tfrac{128\,\beta^3}{(1+\beta)^3}\sum_{n=1}^{N}\tfrac{ n(n+1)}{(2n-1)^2(2n+1)(2n+3)^2}, \\
\mathfrak{f}_2 &= \tfrac{32\,\beta}{15(1+\beta)^3}\sum_{n=1}^{N}\tfrac{n(n+1)(15+12n+12n^2)}{(2n-1)^2(2n+1)(2n+3)^2}
\\
&\quad +\tfrac{32\,\beta^2}{15(1+\beta)^3}\sum_{n=1}^{N}\tfrac{n(n+1)(-15+84n+84n^2)}{(2n-1)^2(2n+1)(2n+3)^2}\\
&\quad +\tfrac{32\,\beta^3}{15(1+\beta)^3}\sum_{n=1}^{N}\tfrac{n(n+1)(25+12n+12n^2)}{(2n-1)^2(2n+1)(2n+3)^2},\\
\mathfrak{f}_3 &= -\tfrac{8(\beta^3+3\beta^2)}{3(1+\beta)^3}\sum_{n=1}^{N}\tfrac{ n(n+1)}{(2n-1)(2n+3)}.
\end{align*}

One may sum $\mathfrak{f}_1$ by expressing its summand in partial fractions and telescoping the resulting expression
\begin{align*}
&\sum_{n=1}^{N}\tfrac{ n(n+1)}{(2n-1)^2(2n+1)(2n+3)^2} =
\sum_{n=1}^{N}\tfrac{1}{64}\left[\tfrac{1}{(2n-1)(2n+1)}-\tfrac{1}{(2n+1)(2n+3)}\right]\\
& \qquad \qquad \qquad \qquad \qquad \qquad +\tfrac{3}{128}\left[\tfrac{1}{(2n-1)^2}-\tfrac{1}{(2n+3)^3}\right] \\
&=\tfrac{1}{64}\left[\tfrac{1}{3}-\tfrac{1}{(2N+1)(2N+3)}\right] 
+\tfrac{3}{128}\left[\tfrac{10}{9}-\tfrac{1}{(2N+1)^2}-\tfrac{1}{(2N+3)^2}\right].
\end{align*}
Therefore we have $$\mathfrak{f}_1 = [4-\tfrac{2}{(2N+1)(2N+3)}-3[\tfrac{1}{(2N+1)^2}+\tfrac{1}{(2N+3)^2}]]\tfrac{\beta^3}{(1+\beta)^3}$$ and
hence $\mathfrak{f}_1 = \tfrac{4\beta^3}{(1+\beta)^3}+\tfrac{2\beta^3N^{-2}}{(1+\beta)^3}$ as $N\to \infty$. 

Now notice that $\mathfrak{f}_2$ may be rewritten into the form
\begin{align}\label{eq:small_lim_f2_reexpressed}
\mathfrak{f}_2 &= \tfrac{3\cdot 32\,\beta}{15(1+\beta)^3}\sum_{n=1}^{N}\tfrac{n(n+1)((2n-1)(2n+3)+8)}{(2n-1)^2(2n+1)(2n+3)^2}
\\
&\quad +\tfrac{21\cdot 32\,\beta^2}{15(1+\beta)^3}\sum_{n=1}^{N}\tfrac{n(n+1)((2n-1)(2n+3)+8 -\tfrac{120}{21})}{(2n-1)^2(2n+1)(2n+3)^2}
\nonumber\\
&\quad +\tfrac{3\cdot 32\,\beta^3}{15(1+\beta)^3}\sum_{n=1}^{N}\tfrac{n(n+1)((2n-1)(2n+3)+8 + \tfrac{10}{3})}{(2n-1)^2(2n+1)(2n+3)^2}.
\nonumber
\end{align}
By use of the identity
\begin{multline}\label{eq:small_lim_partial_frac_decomp}
\tfrac{n(n+1)\left[(2n-1)(2n+3)+8\right]  }{(2n-1)^2(2n+1)(2n+3)^2}\\
=\tfrac{3}{32(2n-1)}+\tfrac{1}{16(2n+1)}
+\tfrac{3}{32(2n+3)}+\tfrac{8\,n(n+1)}{(2n-1)^2(2n+1)(2n+3)^2}
\end{multline}
we may sum \eqref{eq:small_lim_f2_reexpressed} explicitly. Notice that the last term on the right hand side of \eqref{eq:small_lim_partial_frac_decomp} is repeated from contributions to $\mathfrak{f}_1$. Observe too the identities
\begin{align*}
\sum_{n=1}^{N}\tfrac{1}{2n+1} &= \sum_{n=1}^{N}\tfrac{1}{2n-1}-1+\tfrac{1}{2N+1},\\
\sum_{n=1}^{N}\tfrac{1}{2n+3} &= \sum_{n=1}^{N}\tfrac{1}{2n-1}-\tfrac{4}{3}+\tfrac{1}{2N+1}+\tfrac{1}{2N+3}.
\end{align*}
Thus all contributions to $\mathfrak{f}_2$ may be written in terms of $\sum_{n=1}^{N}(2n-1)^{-1}$ and $\mathfrak{f}_1$, the former of which may be dealt with by asymptotics of partial summation expressions of the natural logarithm. Summing the identity \eqref{eq:small_lim_partial_frac_decomp} from $n=1$ to $n=N$ one obtains
\begin{multline}
\sum_{n=1}^{N}\tfrac{n(n+1)\left[(2n-1)(2n+3)+8\right]  }{(2n-1)^2(2n+1)(2n+3)^2}
=\tfrac{1}{4}\sum_{n=1}^{N}\tfrac{1}{2n-1}-\tfrac{3}{16}\\+\tfrac{5}{32(2N+1)}+\tfrac{3}{32(2N+3)}
-\tfrac{(1+\beta)^3}{16\beta^3}\mathfrak{f}_1.
\end{multline}
So that $\mathfrak{f}_2$ may be summed with use of $\mathfrak{f}_1$
\begin{multline*}
\mathfrak{f}_2 = \tfrac{32}{15(1+\beta)^3}[3\beta+21\beta^2+3\beta^3]
\left[ \tfrac{1}{4}\sum_{n=1}^{N}\tfrac{1}{2n-1}+\tfrac{1}{16} \right]
\\
+\tfrac{32}{15(1+\beta)^3}[-120\beta^2+10\beta^3] +O(N^{-1}).
\end{multline*}

Now from asymptotic expansions for large argument of the polygamma function,
$\sum_{n=1}^{N}\tfrac{1}{2n-1}\sim \tfrac{1}{2}(\gamma+\log N) + \log 2+\tfrac{1}{48\, N^2} + O(N^{-4})$
as $N\to \infty$, where $\gamma$ is the Euler-Masheroni constant. Thus we have
\begin{multline*}
\mathfrak{f}_2 = \tfrac{32}{15(1+\beta)^3}[3\beta+21\beta^2+3\beta^3][\tfrac{1}{8}(\gamma+\log N) + \tfrac{1}{4}\log 2+\tfrac{1}{16}]
\\+\tfrac{(-120\beta^2+10\beta^3)}{15(1+\beta)^3} + O(N^{-1})
\end{multline*}
as $N\to\infty$. 

Now consider $\mathfrak{f}_3$, by the identity 
\begin{align*}
\tfrac{n(n+1)}{(2n-1)(2n+3)} = \tfrac{1}{4}+\tfrac{3}{16}\left[\tfrac{1}{2n-1}-\tfrac{1}{2n+3}\right]
\end{align*}
and summing between $n=1$ and $N$ and telescoping we obtain
\begin{align}
\mathfrak{f}_3 = -\tfrac{8(1+3\beta)}{3(1+\beta)^3}\left[\tfrac{N}{4}+\tfrac{1}{4}+O(N^{-1})\right]
\end{align}
as $N\to\infty$.

Now for $\mathcal{F}_s$ all that remains is to order the error estimates. Returning to the decomposition \eqref{eq:F_decomp_small_arg_lim} we observe that $N$ is large and chosen such that in the shared regime of validity $N = O(\alpha^{-1})$ for the singular part, and $N = O(\alpha^{0})$ for the regular part. Since the former estimate holds for all $n\leq N$ we must have $N\to\infty$ as $\alpha\to 0$. Also by taking $\alpha$ to zero the tail $\mathcal{F}_r$ vanishes and $\mathcal{F}_s$ is an ever better infinite series approximation of $\mathcal{F}_1^z$. Note that the integer $N$ is arbitrary and must not appear in the final form, but it is permissible that $\mathcal{F}_s$ and $\mathcal{F}_r$ may depend on $N$ on their own. Typical of matched asymptotic problems the index $N$ is implicitly a function of $\alpha$, the natural choice being $N = \delta\alpha^{-(1-\theta)}$ for some $0<\theta<1$ with both $\delta$, $\theta$ independent of $\alpha$. With this, $N$ lies in the overlapping region and increases as $\alpha$ decreases. What is more, we have $N^{-1} = O(\alpha^{1-\theta}) = o(1)$ since $0<\theta<1$. One also has $\alpha N = O(\alpha^{\theta}) = o(1)$. Finally note that $O(\alpha)$ is higher than $o(1)$ with respect to $\alpha$ and may be neglected. Thus 
\begin{multline}\label{eq:lem_f_s_form}
\mathcal{F}_s\sim \tfrac{4\beta^3\alpha^{-2}}{(1+\beta)^3}+\tfrac{4\beta(1+7\beta+\beta^2)}{5(+\beta)^3}\,\left[\log N+\gamma+2\log 2+\tfrac{1}{2}\right]\\+\tfrac{(-120\beta^2+10\beta^3)}{15(1+\beta)^3} +\tfrac{2\alpha^{-2}N^{-2}\beta^3}{(1+\beta)^3} + o(1).
\end{multline}


For $\mathcal{F}_r$ the key idea here is to transform to a Riemann sum and hence to approximate it by an integral.  
Here the summation index is getting larger while $\alpha$ is going to zero so it is natural to introduce the intermediate variable $x = n\alpha$ where $\alpha\to 0$ with $x$ fixed making $n\to\infty$.  With this $\mathcal{F}_r$ takes the form
\begin{align*}
\mathcal{F}_r=&-\tfrac{\sinh\alpha}{3\sqrt{2}}\sum_{\substack{x = n\alpha \\ n = N+1}}^{\infty}(2n+1)(a_n+b_n+c_n+d_n).
\end{align*}
Expanding the summand for $x$ fixed and $\alpha$ small one obtains
\begin{align}
\mathcal{F}_r=&\tfrac{2}{3}(1+O(\alpha))\sum_{\substack{x = n\alpha \\ n = N+1}}^{\infty} \alpha\, \tfrac{f(x)}{g(x)}
\end{align}
where 
\begin{align*}
f(x)&:=-\beta ^2+\beta ^2(2 x^2+2 x+1) e^{\tfrac{2 (\beta +2) x}{\beta }}-e^{\tfrac{2x}{\beta }} (\beta ^2+2 x^2-2 \beta  x) \\
   &+e^{\tfrac{2 (\beta +1) x}{\beta }} (\beta ^2+4 (\beta +1) x^3+2 (\beta +1)^2 x^2+2 \beta  (\beta +1) x) \\
 g(x)&:=\beta ^2-2 e^{\tfrac{2 (\beta +1) x}{\beta }} (\beta ^2+2 (\beta +1)^2 x^2)+\beta ^2 e^{\tfrac{4 (\beta +1) x}{\beta }}.	
\end{align*}
Note that the summand is implicitly indexed by $n$ through the variable $x$. Note also that $\alpha = (n+1)\alpha-n\, \alpha =x_{n+1}-x_n =: \delta x$. Thus 
\begin{align}\label{eq:sum_delta_x_lem_small_arg}
\mathcal{F}_r=&\tfrac{2}{3}(1+O(\alpha))\sum_{x=X}^{\infty} \tfrac{f(x)}{g(x)} \delta x
\end{align}
where $X$ is the intermediate variable defined such that $N$ is the positive integer first less than $X/\alpha$. Thus $\alpha\to 0$ implies $X\to 0$.

Referring to Euler-Maclaurin \citet{kac2001quantum} one has
\begin{multline}
\sum_{x=X}^{\infty} \tfrac{f(x)}{g(x)}\delta x = \int_X^{\infty}\tfrac{f(x)}{g(x)}\,\mathrm{d}x + \tfrac{\alpha}{2}\left[\tfrac{f(\infty)}{g(\infty)}+\tfrac{f(X)}{g(X)}\right]\\ +\alpha\sum_{k=1}^{\infty}\tfrac{B_{2k}}{2k!}\left[\left(\tfrac{f}{g}\right)^{(2k-1)}(\infty)-\left(\tfrac{f}{g}\right)^{(2k-1)}(X)\right] 
\end{multline}
where $B_m$ is the $m\textsuperscript{th}$ Bernoulli number. It is now of importance to know the behaviour of the the function $l(x):=f/g\,(x)$ at $x=0$ and $x=\infty$. It is not hard to see that $l(x)\to 0$ as $x\to \infty$ due to the presence of the fourth exponential power in $g(x)$. Now as $x\to 0$ one has
\begin{align}
l(X) =\tfrac{6\beta^3}{(1+\beta)^3X^3}+\tfrac{6\beta(1+7\beta+\beta^2)}{5(1+\beta)^3X} -\tfrac{3\beta^2+\beta^3}{(1+\beta)^3} +O(X).
\end{align}
Therefore limiting the summation \eqref{eq:sum_delta_x_lem_small_arg} to the integral via Euler-Maclaurin one has
\begin{multline}\label{eq:sum_l_to_int_small_arg_lem}
\sum_{x=X}^{\infty} \tfrac{f(x)}{g(x)}\delta x\sim \int_X^{\infty}l(x)\,\mathrm{d}x +\alpha\tfrac{6\beta^3}{(1+\beta)^3X^3}\\
+\alpha\tfrac{6\beta(1+7\beta+\beta^2)}{5(1+\beta)^3X} - \alpha\tfrac{3\beta^2+\beta^3}{(1+\beta)^3} +O(\alpha\,X)
\end{multline}
where we have deemed the boundary term at infinity and terms of high order derivatives of $l(x)$ at infinity negligible, the latter of which may be justified by the persistence of the term $\exp(4kx)$ in the denominator at the $k\textsuperscript{th}$ derivative of $l(x)$. Additional terms in the regular expansion $\mathcal{F}_r$ may be obtained by considering the terms $l^{(k)}(X)$. 

Since $X\to 0$ as $\alpha\to 0$ it is natural to decompose the integrand in \eqref{eq:sum_l_to_int_small_arg_lem} into its small arguments, and the presence of $\log 2$ in \eqref{eq:lem_f_s_form}, suggests cutting the domain of integration as follows 
\begin{multline}\label{eq:int_l_decomposed_small_arg_lim}
\int_X^{\infty}l(x)\,\mathrm{d}x=\int_1^{\infty}k(x)\,\mathrm{d}x+\int_{X}^1j(x)\,\mathrm{d}x
\\
+\int_{2X}^1\tfrac{6\beta(1+7\beta+\beta^2)}{5(1+\beta)^3t}\,\mathrm{d}t+\int_{X}^{\infty}\tfrac{6\beta^3}{(1+\beta)^3x^3}\,\mathrm{d}x
\end{multline}
where $t=2x$, $j(x) := l(x)-\tfrac{6\beta^3}{(1+\beta)^3x^3}-\tfrac{6\beta(1+7\beta+\beta^2)}{5(1+\beta)^3x}$ and $k(x):=l(x)-\tfrac{6\beta^3}{(1+\beta)^3x^3}$. The third and fourth integrals in \eqref{eq:int_l_decomposed_small_arg_lim} are evaluated as
\begin{align*}
\int_{2X}^1\tfrac{6\beta(1+7\beta+\beta^2)}{5(1+\beta)^3t}\,\mathrm{d}x &=-\tfrac{6\beta(1+7\beta+\beta^2)}{5(1+\beta)^3}(\log2+\log X),\\
\int_{X}^{\infty}\tfrac{6\beta^3}{(1+\beta)^3x^3}\,\mathrm{d}x &=-\tfrac{3\beta^3}{(1+\beta)^3}X^{-2}.
\end{align*}
For the first two integrals, note that $\int_{X}^1j(x)\,\mathrm{d}x = \int_{0}^1j(x)\,\mathrm{d}x-\int_{0}^Xj(x)\,\mathrm{d}x$ and that $j(x)= O(x)$ as $x\to 0 $ so that $\int_{0}^Xj(x)\,\mathrm{d}x = O(X)$ as $\alpha\to 0$. Thus upon defining the constants (depending on $\beta$)
\begin{align}\label{eq:eqn_for_c1_c_2}
C_1 = \int_1^{\infty}k(x)\,\mathrm{d}x,\quad
C_2 =\int_{0}^1j(x)\,\mathrm{d}x
\end{align}
all the expanded leading terms of $\mathcal{F}_r$ have been integrated. 

It is elementary to show that both $C_1$ and $C_2$ are finite. For $C_1$, the contribution proportional to $x^{-3}$ converges on $[1,\infty]$ and $l(x)$ decays exponentially as $x\to\infty$. For $C_2$, we have the power series expansion as $x\to 0$
\begin{align}
j(x) = -\tfrac{3\beta^2(3+\beta)}{(1+\beta)^3}+\tfrac{4(8-19\beta+8\beta^2)x}{175\beta(1+\beta)}+O(x).
\end{align}
Therefore $j(x)$ is a continuous function at zero, moreover it is continuous on a closed interval and hence there must exist a finite bound $M>|j(x)|$ so that $C_1<M$. Therefore taking all the contributions together and with $X = N \alpha$ fixed,
\begin{multline}\label{eq:F_r_final_form}
\mathcal{F}_r\sim \tfrac{2}{3}(C_1+C_2)-\tfrac{4\beta(1+7\beta+\beta^2)}{5(1+\beta)^3}(\log X+\log 2)-\tfrac{2\beta^3}{(1+\beta)^3}X^{-2}\\
+O(\alpha X^{-3})+O(\alpha X^{-1}) + O(\alpha)
\end{multline}
as $\alpha\to 0$.
%
%
Now, adding together \eqref{eq:lem_f_s_form} and \eqref{eq:F_r_final_form} one sees that by writing $\log N = \log X -\log \alpha$ the $\log X$ terms cancel. Similarly with $X = N\alpha$ the $O(N^{-2}\alpha^{-2})$ terms cancel leaving the final expression for $\mathcal{F}$ as $\alpha\to 0$
\begin{align}\label{eq:alpha_asymptotics_Fz}
F_{z}^\ast = \tfrac{4\beta^3}{(1+\beta)^3}\alpha^{-2}-\tfrac{4\beta(1+7\beta+\beta^2)}{5(1+\beta)^3}\log \alpha +K_1 + o(1)
\end{align}
where $K_1 =\tfrac{4\beta(1+7\beta+\beta^2)}{5(1+\beta)^3}(\gamma+\log 2+\tfrac{1}{2})+\tfrac{(-120\beta^2+10\beta^3)}{15(1+\beta)^3}+\tfrac{2}{3}(C_1+C_2)$.
In Eulcidean units the force on sphere 1 reads
\begin{align}\label{eq:euclidean_asymptotics_Fz}
F_{z}^e = \tfrac{2\beta^2}{(1+\beta)^2}\epsilon^{-1}-\tfrac{2\beta(1+7\beta+\beta^2)}{5(1+\beta)^3}\log\epsilon +K_2 + o(1)
\end{align}
where where $K_2 =\tfrac{4\beta(1+7\beta+\beta^2)}{5(1+\beta)^3}(\gamma+\tfrac{1}{2}+\log 2+\tfrac{1}{2}\log \tfrac{2\beta}{1+\beta})+\tfrac{(-120\beta^2+10\beta^3)}{15(1+\beta)^3}+\tfrac{2}{3}(C_1+C_2)$.

\subsection{Large Argument Behaviour}
We note that for large separations it is sufficient to consider the symmetric case $\eta_1 = -\eta_2=:\alpha$, since by the inner analysis the force quickly decays for surface separations $\alpha$ not small. To this end we consider the asymptotic behaviour of the series
$$
\tfrac{\sinh \alpha}{3\sqrt{2}}\sum_{n=1}^{\infty}(2n+1)(b_n+c_n)
$$
as $\alpha \to \infty$ since $\alpha$ is a proxy for sphere distance. 
Expanding $|F^{j}_z|$ in an infinite series of exponential functions we have
\begin{align*}
\tfrac{|F^{j}_z(\alpha,-\alpha)|}{6 \pi \mu U r_j}&=\tfrac{\sinh\alpha}{3}\sum_{n=1}^{\infty}\tfrac{n(n+1)}{(2n+3)(2n-1)}\tfrac{s_n(\alpha)}{t_n(\alpha)}
\end{align*}
where $s_n(\alpha)=8 e^{\alpha}-2(2n-1)(2n+3)e^{2\alpha(n+1)}+(2n+1)(2n-1)e^{2\alpha n}
+(2n+3)(2n+1)e^{2\alpha(2+n)}$ and $
t_n(\alpha)={2(e^{\alpha}-e^{\alpha(4n+3)} )+(2n+1)(e^{2\alpha(n+1)}-e^{2\alpha n})}.
$ We observe that the limit of the summand as $\alpha\to \infty$ exists for each $n$ and the resulting series can be dominated by a second convergent series, thus the limit and the sum may be commuted, giving
\begin{align}\label{eq:inf_sum_under_lim}
\tfrac{\sinh\alpha}{3}\sum_{n=1}^{\infty}\tfrac{n(n+1)}{(2n+3)(2n-1)}\tfrac{\mathcal{C}_n(\alpha)}{\mathcal{D}_n(\alpha)}\sim \tfrac{1}{6}\sum_{n=1}^{\infty}\tfrac{n(n+1)(2n+1)}{2n-1}e^{-2\alpha(n-1)}.
\end{align}

Expanding the summation in \eqref{eq:inf_sum_under_lim} we have
\begin{align*}
\tfrac{e^{-2\alpha}|F^{j}_z(\alpha,-\alpha)|}{6 \pi \mu U r_j}\sim  \sum_{n=1}^{\infty}\tfrac{n^3e^{-2\alpha n}}{3(2n-1)}+\sum_{n=1}^{\infty}\tfrac{n^2e^{-2\alpha n}}{2(2n-1)}+\sum_{n=1}^{\infty}\tfrac{n e^{-2\alpha n}}{6(2n-1)}.
\end{align*}
We may bound this series in terms of known geometric and logarithmic summations as follows
\begin{multline*}
\sum_{n=1}^{\infty}\tfrac{ne^{-2\alpha n}}{3}+\tfrac{e^{-2\alpha n}}{2}+\tfrac{ e^{-2\alpha n}}{6n}\\\leq \tfrac{e^{-2\alpha}|F^{j}_z(\alpha,-\alpha)|}{6 \pi \mu U r_j} \leq \sum_{n=1}^{\infty}\tfrac{n^2e^{-2\alpha n}}{3}
+\tfrac{n e^{-2\alpha n}}{2}+\tfrac{ e^{-2\alpha n}}{6}.
\end{multline*}
Summing these lower and upper bounds we find
\begin{align*}
\tfrac{e^{4\alpha}}{3(e^{2\alpha}-1)^2}+\tfrac{e^{2\alpha}}{2(e^{2\alpha}-1)}-\tfrac{e^{2\alpha}}{6}\log(1-e^{-2\alpha}) \\
\leq \tfrac{|F^{j}_z(\alpha,-\alpha)|}{6 \pi \mu U r_j} \leq\tfrac{e^{4\alpha}1+e^{6\alpha}}{3(e^{2\alpha}-1)^3}
+\tfrac{e^{4\alpha}}{2(e^{2\alpha}-1)^2}+\tfrac{e^{2\alpha}}{6(e^{2\alpha}-1)}
\end{align*}
and upon taking the limit $\eta_1\to\infty$ the sandwich theorem gives 
$$
\lim_{\alpha\to\infty}\tfrac{|F^{j}_z(\alpha,-\alpha)|}{6 \pi \mu U r_j}=1.
$$


\section{Derivation of the Tangential Fields}\label{app:derivation_of_tang_fields}
\subsection{Tangential Field Equations}
We introduce four auxiliary functions: $W(r,z), X(r,z), Y(r,z), Z(r,z)$, and the governing equations \eqref{eq:p_r}, \eqref{eq:p_theta}, \eqref{eq:p_z} can be written in terms of known differential operators. Firstly consider the decomposition
\begin{align}
c\,p &=2 \mu U W(r,z)\cos\theta,\label{eq:p_decomp}\\
c\,u_r &= U\left[r W(r,z)+c\left(X(r,z)+Y(r,z)\right)  \right]\cos\theta,\label{eq:u_r_vel_decomp}\\
c\,u_\theta &= U\left[X(r,z)-Y(r,z)\right]\sin\theta,\label{eq:u_theta_vel_decomp}\\
c\,u_z &= U\left[z W(r,z)+2cZ(r,z)\right]\cos\theta.\label{eq:u_z_vel_decomp}
\end{align}

 the following equations 
\begin{align}
\partial_r W &= \left(\partial_r^2+r^{-1}\partial_r-2r^{-2}+\partial_z^2\right)\frac{c}{2}\left(X+Y\right)\nonumber\\    
&\quad+\left(\partial_r^2+r^{-1}\partial_r-2r^{-2}+\partial_z^2\right)\frac{rW}{2}-\frac{c\left(X-Y\right) }{r^2},\label{eq:eqn1}\\
0&=\left(\partial_r^2+r^{-1}\partial_r-2r^{-2}+\partial_z^2\right)\frac{c}{2}\left(X-Y\right)\nonumber\\
&\quad-\frac{c\left(X+Y\right) }{r^2},\label{eq:eqn2}\\
\partial_z W&= \left(\partial_r^2+r^{-1}\partial_r-r^{-2}+\partial_z^2\right)zW\nonumber\\    
&\quad +2c\left(\partial_r^2+r^{-1}\partial_r-r^{-2}+\partial_z^2\right)Z,\label{eq:eqn3}\\
0&=3 W + r\partial_r W +z\partial_z W +c\partial_r Y\nonumber\\
&\quad+c \partial_r X+2cr^{-1}X+2c\partial_z Z\label{eq:incomp_in_w_y_z}.
\end{align}

\begin{align}
L_1 W &=\frac{W}{r^2},\label{eq:eqn_for_W}\\
L_1 X &= \frac{4X}{r^2},\label{eq:eqn_for_X}\\
L_1 Y&=0,\label{eq:eqn_for_Y}\\
L_1 Z&=\frac{Z}{r^2},\label{eq:eqn_for_Z}\\
0&=3 W + r\partial_r W +z\partial_z W +c\partial_r Y\nonumber\\
&\quad +c \partial_r X+2cr^{-1}X+2c\partial_z Z\label{eq:incomp_rewrite}
\end{align}
where
\begin{align}
L_1=\partial_z^2+\partial_r^2+r^{-1}\partial_r.
\end{align}
Note that $L_1$ is a particular case of the differential operator $L_k$ given by $L_k=\partial_z^2+\partial_r^2+k\,r^{-1}\partial_r$ which is a closely studied operator in axially symmetric potential theory by those such as Weinstein\cite{weinstein1948discontinuous} and Payne \cite{payne1959representation} and in particular \cite{payne1960stokes} wherein explicit solutions for Stokes flow around classes of axially symmetric bodies are considered. Solutions $\omega_k$ to equations $L_k \omega_k(r,z) = 0$ are families of axially symmetric potential functions parametrised by $k$. In particular the homogeneous problem $L_1\omega_1=0$ in spherical bipolar coordinates has a solution expressible in a complete basis of Legendre polynomials.

By substituting \eqref{eq:p_decomp}--\eqref{eq:u_z_vel_decomp} into \eqref{eq:p_r}--\eqref{eq:p_z} we obtain \eqref{eq:eqn1}, \eqref{eq:eqn2}, \eqref{eq:eqn3} and \eqref{eq:incomp_in_w_y_z}. By adding \eqref{eq:eqn1} to equation \eqref{eq:eqn2}, subtracting \eqref{eq:eqn2} from equation \eqref{eq:eqn1} and along with \eqref{eq:eqn3} and \eqref{eq:incomp_in_w_y_z} we obtain equations for each of the scalar fields $W(r,z), X(r,z), Y(r,z), Z(r,z)$ in terms of the differential operator $L_1$, these are labelled \eqref{eq:eqn_for_W}, \eqref{eq:eqn_for_X}, \eqref{eq:eqn_for_Y}, \eqref{eq:eqn_for_Z} along with the incompressibility condition \eqref{eq:incomp_rewrite}. We obtain expressions for the auxiliary fields $W(r,z), X(r,z), Y(r,z), Z(r,z)$ in terms of special functions by transforming to spherical bipolar coordinates. The expression for $L_k$ for $k \in \mathbb{Z}$ is given by \eqref{eq:Lk_formula_in_SBC} and is a separable differential operator in the spherical bipolar coordinate system. The expressions for $W(r,z), X(r,z), Y(r,z), Z(r,z)$ may be obtained by a separation of variables procedure as described in \cite[Section 4-11]{happel2012low}. In particular, we derive the formulae \eqref{eq:eqn_for_W_SB}, \eqref{eq:eqn_for_X_SB}, \eqref{eq:eqn_for_Y_SB}, \eqref{eq:eqn_for_Z_SB} for a set summation coefficients $A_n$--$G_n$ which must be obtained by the boundary conditions on each sphere.

\subsection{Conversion to Spherical Bipolar Coordinates}
By making the transformation \eqref{eq:bipolar_trans} the generalised operator $L_k$ is given in bipolar spherical coordinates as
\begin{align}\label{eq:Lk_formula_in_SBC}
L_k = r^{-k}\mathfrak{h}^2\left\lbrace \partial_\xi\left(r^k\partial_\xi\right) +\partial_\eta\left(r^k\partial_\eta\right) \right\rbrace 
\end{align}
where $\mathfrak{h}=c/(\cosh\eta-\cos\xi)$ is the metrical coefficient. The derivation of this expression may be found in \citet{happel2012low}. We now use the operator definition \eqref{eq:Lk_formula_in_SBC} to solve \eqref{eq:eqn_for_W}-\eqref{eq:eqn_for_Z} in spherical bipolar coordinates.
\subsubsection{Equation for $Y$}\label{sec:eqn_for_Y}
We write $\bar{Y}(\xi,\eta) = Y(r(\xi,\eta),z(\xi,\eta))$ and find
\begin{align}
\bar{Y}(\xi,\eta) &=\sqrt{\cosh\eta-\cos\xi}\sum_{n=0}^{\infty}\left[D_n\cosh(n+\tfrac{1}{2})\eta\right. \\
&\left. \quad +E_n\sinh(n+\tfrac{1}{2})\eta\right]P_n(\cos\xi).  
\end{align}\label{eq:eqn_for_Y_SB}
\subsubsection{Equation for $W$ and $Z$}
We write $\hat{W}(\xi,\eta) = W(r(\xi,\eta),z(\xi,\eta))$, $\hat{Z}(\xi,\eta) = Z(r(\xi,\eta),z(\xi,\eta))$ and find
\begin{align}
\bar{W}(\xi,\eta)& =\sin\xi\sqrt{\cosh\eta-\cos\xi}\sum_{n=1}^{\infty}\left[B_n\cosh(n+\tfrac{1}{2})\eta\right.\nonumber\\
&\left.+C_n\sinh(n+\tfrac{1}{2})\eta\right]P_n'(\cos\xi),\label{eq:eqn_for_W_SB}\\
\bar{Z}(\xi,\eta)& =\sin\xi\sqrt{\cosh\eta-\cos\xi}\sum_{n=1}^{\infty}\left[A_n\cosh(n+\tfrac{1}{2})\eta\right.\nonumber\\
&\left.+H_n\sinh(n+\tfrac{1}{2})\eta\right]P_n'(\cos\xi).\label{eq:eqn_for_Z_SB}
\end{align}
We remark that the sums are to be taken starting $n=1,2,..$ because solutions to the associated Legendre equation are nonzero and nonsingular when $0\leq m = 1 \leq n$.
\subsubsection{Equation for $X$}
We write $\hat{X}(\xi,\eta) = X(r(\xi,\eta),z(\xi,\eta))$ and find
\begin{align}
\bar{X}(\xi,\eta)& =\sin^2\xi\sqrt{\cosh\eta-\cos\xi}\sum_{n=2}^{\infty}\left[F_n\cosh(n+\tfrac{1}{2})\eta\right.\nonumber\\
&\left.+G_n\sinh(n+\tfrac{1}{2})\eta\right]P_n''(\cos\xi)\label{eq:eqn_for_X_SB}
\end{align}
noting the sums are to be taken starting $n=2,3,..$ because solutions to the associated Legendre equation are nonzero and nonsingular when $0\leq m = 2 \leq n$. All that remains is to apply the boundary conditions \eqref{bc:shearing_spheres} transformed into spherical-bipolar coordinates to the appropriately combined general solutions $\hat{W}$, $\hat{X}$, $\hat{Y}$, $\hat{Z}$.

\subsection{Unequal Spheres}\label{app:unequal_sphere_derivation}
We now obtain the unknown constants for the case of equal spheres. In spherical bipolar coordinates this is equivalent to imposing
\begin{align*}
 z^{(1)}=\frac{c\,\sinh\eta_1}{\cosh\eta_1-\cos\xi}, &\qquad z^{(2)}=\frac{c\,\sinh\eta_2}{\cosh\eta_2-\cos\xi}\\
 r^{(1)}=\frac{c\,\sin\xi}{\cosh\eta_1-\cos\xi}, &\qquad r^{(2)}=\frac{c\,\sin\xi}{\cosh\eta_2-\cos\xi}.
\end{align*} 
We introduce the notation
\begin{align*}
c^k_\alpha = \cosh k \alpha, \qquad s^k_\alpha = \sinh k \alpha.
\end{align*}
By subtracting \eqref{bc:bcs_shearing_transformed_B_z} from \eqref{bc:bcs_shearing_transformed_A_z} we find
\begin{align}\label{eq:unequal_sub_zW_Z}
&\tfrac{s^1_{\eta_1}}{(c^1_{\eta_1}-\mathfrak{x} )^{1/2}}\sum_{n=1}^{\infty}[B_nc^{n+1/2}_{\eta_1}+C_ns^{n+1/2}_{\eta_1}]P_n'(\mathfrak{x} )\nonumber\\
&\quad+2(c^1_{\eta_1}-\mathfrak{x} )^{1/2}\sum_{n=1}^\infty A_nc^{n+1/2}_{\eta_1}P_n'(\mathfrak{x} )\nonumber\\
&-\tfrac{s^1_{\eta_2}}{(c^1_{\eta_2}-\mathfrak{x} )^{1/2}}\sum_{n=1}^{\infty}[B_nc^{n+1/2}_{\eta_2}+C_ns^{n+1/2}_{\eta_2}]P_n'(\mathfrak{x} )\nonumber\\
&\quad-2(c^1_{\eta_2}-\mathfrak{x} )^{1/2}\sum_{n=1}^\infty A_nc^{n+1/2}_{\eta_2}P_n'(\mathfrak{x} )=0.
\end{align}
We introduce the generating function for the Legendre polynomials 
\begin{align}\label{eq:gen_fun_legendre}
(\cosh\eta-\mathfrak{x} )^{-1/2}=\sum_{n=0}^\infty s_n(\eta)P_n(\mathfrak{x} )
\end{align}
where $s_n(\eta) =\sqrt{2}e^{\pm(n+\tfrac{1}{2})\eta}$ where the sign is chosen so that the exponential decays on each sphere. Using \eqref{eq:gen_fun_legendre} we integrate equation \eqref{eq:unequal_add_W_X_Y} over $\mathfrak{x} \in[-1,1]$. Firstly we note the identities
\begin{align}\label{eq:Pnprime_against_gen}
&\int_{-1}^1\mathrm{d}\mathfrak{x} \, \tfrac{P'_n(\mathfrak{x} )}{(c_{\eta_1}^1-\mathfrak{x} )^{1/2}} = \int_{-1}^1\mathrm{d}\mathfrak{x} \,\sum_{j=0}^\infty s_j(\eta_1)P_{j}(\mathfrak{x} )P'_{n}(\mathfrak{x} )\nonumber\\
&=2\sum_{j=0}^\infty s_j(\eta_1)\int_{-1}^1\mathrm{d}\mathfrak{x} \,P_{j}(\mathfrak{x} )\left[\tfrac{P_{n-1}(\mathfrak{x} )}{\|P_{n-1}\|^2}+\tfrac{P_{n-3}(\mathfrak{x} )}{\|P_{n-3}\|^2}+\cdots \right]\nonumber\\
&=2\sum_{j=0}^{n-1} s_j(\eta_1)\int_{-1}^1\mathrm{d}\mathfrak{x} \,P_{j}(\mathfrak{x} )\left[\tfrac{P_{n-1}(\mathfrak{x} )}{\|P_{n-1}\|^2}+\tfrac{P_{n-3}(\mathfrak{x} )}{\|P_{n-3}\|^2}+\cdots \right]\nonumber\\
&\quad+2\sum_{j= n}^\infty s_j(\eta_1)\int_{-1}^1\mathrm{d}\mathfrak{x} \,P_{j}(\mathfrak{x} )\left[\tfrac{P_{n-1}(\mathfrak{x} )}{\|P_{n-1}\|^2}+\tfrac{P_{n-3}(\mathfrak{x} )}{\|P_{n-3}\|^2}+\cdots \right]\nonumber\\
& = 2\left[s_{n-1}(\eta_1)+s_{n-3}(\eta_1)\cdots\right]  =:2t_n(\eta_1)
\end{align}
where the finite sum for $t_{n}$ is $\lceil \tfrac{n}{2}\rceil$ long. Additionally
\begin{align}\label{eq:Pnprime_against_oneover_gen}
&\int_{-1}^1\mathrm{d}\mathfrak{x} \, (c_{\eta_1}^1-\mathfrak{x} )^{1/2}P'_{n}(\mathfrak{x} )\nonumber\\
& = \int_{-1}^1\mathrm{d}\mathfrak{x} \, \frac{(c^1_{\eta_1}-\mathfrak{x} )P'_n(\mathfrak{x} )}{(c^1_{\eta_1}-\mathfrak{x} )^{1/2}}\nonumber\\
&=2c^1_{\eta_1}t_{n}(\eta_1) -  \int_{-1}^1\mathrm{d}\mathfrak{x} \, \frac{xP'_n(\mathfrak{x} )}{(c^1_{\eta_1}-\mathfrak{x} )^{1/2}}\nonumber\\
& = 2c^1_{\eta_1}t_{n}(\eta_1) - 2\sum_{j=0}^\infty s_j(\eta_1)\int_{-1}^1\mathrm{d}\mathfrak{x} \,xP_{j}(\mathfrak{x} )\left[\tfrac{P_{n-1}(\mathfrak{x} )}{\|P_{n-1}\|^2}+\cdots \right]\nonumber\\
& = 2c^1_{\eta_1}t_{n}(\eta_1)\nonumber\\
&\quad-2\sum_{j=0}^\infty s_j(\eta_1)\int_{-1}^1\mathrm{d}\mathfrak{x} \,\tfrac{(j+1)P_{j+1}(\mathfrak{x} )+jP_{j-1}(\mathfrak{x} )}{2j+1}\left[\tfrac{P_{n-1}(\mathfrak{x} )}{\|P_{n-1}\|^2}+\cdots \right]\nonumber\\
& =  2c^1_{\eta_1}t_{n}(\eta_1) - 2\sum_{j=0}^\infty s_j(\eta_1)\int_{-1}^1\mathrm{d}\mathfrak{x} \,\tfrac{(j+1)P_{j+1}(\mathfrak{x} )}{2j+1}\left[\tfrac{P_{n-1}(\mathfrak{x} )}{\|P_{n-1}\|^2}+\cdots \right] \nonumber\\
&\quad -2\sum_{j=0}^\infty s_j(\eta_1)\int_{-1}^1\mathrm{d}\mathfrak{x} \,\tfrac{jP_{j-1}(\mathfrak{x} )}{2j+1}\left[\tfrac{P_{n-1}(\mathfrak{x} )}{\|P_{n-1}\|^2}+\cdots \right] \nonumber\\
& = 2c^1_{\eta_1}t_{n}(\eta_1) -2 u_n(\eta_1) - 2v_n(\eta_1)
\end{align}
where 
\begin{align}
u_n(\eta_1) &= \tfrac{n-1}{2n-3}s_{n-2}(\eta_1) +  \tfrac{n-3}{2n-7}s_{n-4}(\eta_1)+\cdots,\\
v_n(\eta_1) &= \tfrac{n}{2n+1}s_{n}(\eta_1) +  \tfrac{n-2}{2n-3}s_{n-3}(\eta_1)+\cdots.
\end{align}
By using the formulas in \eqref{eq:Pnprime_against_gen} and \eqref{eq:Pnprime_against_oneover_gen}, equation becomes \eqref{eq:unequal_sub_zW_Z} becomes
\begin{align}\label{eq:unequal_sub_zW_Z_int}
0&=2s^1_{\eta_1}\sum_{n=1}^{\infty}[B_nc^{n+1/2}_{\eta_1}+C_ns^{n+1/2}_{\eta_1}]t_n(\eta_1)\nonumber\\
&+2\sum_{n=1}^\infty A_nc^{n+1/2}_{\eta_1}(c^1_{\eta_1}t_{n}(\eta_1) - u_n(\eta_1) - v_n(\eta_1))\nonumber\\
&-2s^1_{\eta_2}\sum_{n=1}^{\infty}[B_nc^{n+1/2}_{\eta_2}+C_ns^{n+1/2}_{\eta_2}]t_n(\eta_2)\nonumber\\
&-2\sum_{n=1}^\infty A_nc^{n+1/2}_{\eta_2}(c^1_{\eta_2}t_{n}(\eta_2) - u_n(\eta_2) - v_n(\eta_2))
\end{align}
Note that in the equal sphere case, $\eta_2=-\eta_1$ \eqref{eq:unequal_sub_zW_Z_int} implies
\begin{align*}
B_n\equiv 0 \qquad \forall n\in \mathbb{N}.
\end{align*}
By adding \eqref{bc:bcs_shearing_transformed_A_r} to \eqref{bc:bcs_shearing_transformed_B_r} one obtains
\begin{align}\label{eq:unequal_add_W_X_Y}
&\tfrac{(1-\mathfrak{x} ^2)}{(c^1_{\eta_1}-\mathfrak{x} )^{1/2}}\sum_{n=1}^\infty[B_n c^{n+1/2}_{\eta_1}+C_n s^{n+1/2}_{\eta_1}]P_n'(\mathfrak{x} )\nonumber\\
&+\tfrac{(1-\mathfrak{x} ^2)}{(c^1_{\eta_2}-\mathfrak{x} )^{1/2}}\sum_{n=1}^\infty[B_n c^{n+1/2}_{\eta_2}+C_n s^{n+1/2}_{\eta_2}]P_n'(\mathfrak{x} )\nonumber\\
&+(c^1_{\eta_1}-\mathfrak{x} )^{1/2}(1-\mathfrak{x} ^2)\sum_{n=2}^{\infty}[F_n c^{n+1/2}_{\eta_1}+G_n s^{n+1/2}_{\eta_1}]P_n''(\mathfrak{x} )\nonumber\\
&+(c^1_{\eta_2}-\mathfrak{x} )^{1/2}(1-\mathfrak{x} ^2)\sum_{n=2}^{\infty}[F_n c^{n+1/2}_{\eta_2}+G_n s^{n+1/2}_{\eta_2}]P_n''(\mathfrak{x} )\nonumber\\
&+(c^1_{\eta_1}-\mathfrak{x} )^{1/2}\sum_{n=0}^{\infty}[D_n c^{n+1/2}_{\eta_1}+E_n s^{n+1/2}_{\eta_1}]P_n(\mathfrak{x} )\nonumber\\
&+(c^1_{\eta_2}-\mathfrak{x} )^{1/2}\sum_{n=0}^{\infty}[D_n c^{n+1/2}_{\eta_2}+E_n s^{n+1/2}_{\eta_2}]P_n(\mathfrak{x} )=0.
\end{align}
To obtain an integral equation for \eqref{eq:unequal_add_W_X_Y} we consider additional identities. For example
\begin{align}
(1-\mathfrak{x} ^2)P'_n(\mathfrak{x} ) & = \tfrac{n(n+1)}{2n+1}(P_{n-1}-P_{n+1}),\\
(1-\mathfrak{x} ^2)P''_n(\mathfrak{x} )& = -n(n+1)P_{n}(\mathfrak{x} )\nonumber\\
& \quad +\tfrac{2}{2n+1}\left[(n+1)P'_{n-1}(\mathfrak{x} )+n P'_{n+1}(\mathfrak{x} )\right].  
\end{align}
We find that
\begin{align}
&\int_{-1}^1\mathrm{d}\mathfrak{x} \, (c^1_{\eta_1}-\mathfrak{x} )^{1/2}P_n(\mathfrak{x} ) 
= 2\tfrac{c^1_{\eta_1}s_n(\eta_1)}{2n+1}\nonumber\\
&\quad -2\tfrac{n+1}{2n+1}\tfrac{s_{n+1}(\eta_1)}{2n+3}-2\tfrac{n}{2n+1}\tfrac{s_{n-1}(\eta_1)}{2n-1},\\
&\int_{-1}^1\mathrm{d}\mathfrak{x} \, (c^1_{\eta_1}-\mathfrak{x} )^{1/2}(1-\mathfrak{x} ^2)P''_n(\mathfrak{x} ) \nonumber\\
& = -2n(n+1)\left[ \tfrac{c^1_{\eta_1}s_n(\eta_1)}{2n+1} -\tfrac{n+1}{2n+1}\tfrac{s_{n+1}(\eta_1)}{2n+3}-\tfrac{n}{2n+1}\tfrac{s_{n-1}(\eta_1)}{2n-1}\right]\nonumber\\
&\quad+\tfrac{4}{2n+1}\left[(n+1)(c^1_{\eta_1}t_{n-1}(\eta_1)-u_{n-1}(\eta_1)-v_{n-1}(\eta_1))\right .\nonumber\\
&\qquad\qquad\left. +n (c^1_{\eta_1}t_{n+1}(\eta_1)-u_{n+1}(\eta_1)-v_{n+1}(\eta_1))\right],\\
&\int_{-1}^1\mathrm{d}\mathfrak{x} \,\tfrac{(1-\mathfrak{x} ^2)P'_{n}(\mathfrak{x} )}{(c_{\eta_1}^1-\mathfrak{x} )^{1/2}} = 2\tfrac{n(n+1)}{2n+1}\left[\tfrac{s_{n-1}(\eta_1)}{2n-1}-\tfrac{s_{n+1}(\eta_1)}{2n+3}\right].
\end{align}
Note that in the equal sphere case, $\eta_2=-\eta_1$ \eqref{eq:unequal_add_W_X_Y} reduces to 
\begin{align*}
X_1+X_2+Y_1+Y_2=0
\end{align*}
and along with \eqref{bc:bcs_shearing_transformed_A_theta}, \eqref{bc:bcs_shearing_transformed_B_theta} this would imply $X_1+X_2=Y_1+Y_2\equiv 0$, that is $D_n=F_n\equiv 0$ for every $n\in\mathbb{N}\cup \{0\}$. By adding \eqref{bc:bcs_shearing_transformed_A_theta} to \eqref{bc:bcs_shearing_transformed_B_theta} one has
\begin{align}
&(1-\mathfrak{x} ^2)(c^1_{\eta_1}-\mathfrak{x} )^{1/2}\sum_{n=2}^{\infty}[F_n c^{n+1/2}_{\eta_1}+G_n s^{n+1/2}_{\eta_1}]P_n''(\mathfrak{x} )\nonumber\\
&+(1-\mathfrak{x} ^2)(c^1_{\eta_2}-\mathfrak{x} )^{1/2}\sum_{n=2}^{\infty}[F_n c^{n+1/2}_{\eta_2}+G_n s^{n+1/2}_{\eta_2}]P_n''(\mathfrak{x} )\nonumber\\
&-(c^1_{\eta_1}-\mathfrak{x} )^{1/2}\sum_{n=0}^{\infty}[D_n c^{n+1/2}_{\eta_1}+E_n s^{n+1/2}_{\eta_1}]P_n(\mathfrak{x} )\nonumber\\
&-(c^1_{\eta_2}-\mathfrak{x} )^{1/2}\sum_{n=0}^{\infty}[D_n c^{n+1/2}_{\eta_2}+E_n s^{n+1/2}_{\eta_2}]P_n(\mathfrak{x} )=0.
\end{align}
By adding \eqref{bc:bcs_shearing_transformed_B_z} to \eqref{bc:bcs_shearing_transformed_A_z} we find
\begin{align}\label{eq:unequal_add_zW_Z}
&\tfrac{s^1_{\eta_1}}{(c^1_{\eta_1}-\mathfrak{x} )^{1/2}}\sum_{n=1}^{\infty}[B_nc^{n+1/2}_{\eta_1}+C_ns^{n+1/2}_{\eta_1}]P_n'(\mathfrak{x} )\nonumber\\
&+2(c^1_{\eta_1}-\mathfrak{x} )^{1/2}\sum_{n=1}^\infty A_nc^{n+1/2}_{\eta_1}P_n'(\mathfrak{x} )\nonumber\\
&+\tfrac{s^1_{\eta_2}}{(c^1_{\eta_2}-\mathfrak{x} )^{1/2}}\sum_{n=1}^{\infty}[B_nc^{n+1/2}_{\eta_2}+C_ns^{n+1/2}_{\eta_2}]P_n'(\mathfrak{x} )\nonumber\\
&+2(c^1_{\eta_2}-\mathfrak{x} )^{1/2}\sum_{n=1}^\infty A_nc^{n+1/2}_{\eta_2}P_n'(\mathfrak{x} )=0.
\end{align}
Note that in the equal sphere case $\eta_2=-\eta_1$, by use of the relations
\begin{align*}
\mathfrak{x}  P_n'(\mathfrak{x} ) = \tfrac{n+1}{2n+1}P_{n-1}'(\mathfrak{x} )+\tfrac{n}{2n+1}P_{n+1}'(\mathfrak{x} )
\end{align*}
and
\begin{align*}
c^{n+3/2}_{\eta_1}=c^{n+1/2}_{\eta_1}c^{1}_{\eta_1}+s^{n+1/2}_{\eta_1}s^{1}_{\eta_1},\\
c^{n-1/2}_{\eta_2}=c^{n+1/2}_{\eta_1}c^{1}_{\eta_1}-s^{n+1/2}_{\eta_1}s^{1}_{\eta_1}
\end{align*}
relation \eqref{eq:unequal_add_zW_Z} reduces to 
\begin{align}
C_n = \left[2\tfrac{n-1}{2n-1}(\gamma_n-1)\right]A_{n-1}  -2\gamma_n A_n+ \left[2\tfrac{n+2}{2n+3}(\gamma_n+1)\right]A_{n+1}. 
\end{align}
Subtracting \eqref{bc:bcs_shearing_transformed_B_r} from \eqref{bc:bcs_shearing_transformed_A_r}, subtracting \eqref{bc:bcs_shearing_transformed_B_theta} from \eqref{bc:bcs_shearing_transformed_A_theta} and adding \eqref{bc:bcs_shearing_transformed_A_z} to \eqref{bc:bcs_shearing_transformed_B_z} yields
\begin{align}
\tfrac{r^{(1)}}{c}W_1-\tfrac{r^{(2)}}{c}W_2+X_1-X_2+Y_1-Y_2 &= 2,\label{eq:unequal_sub_rW_X_Y}\\
X_1-X_2-Y_1+Y_2 &= -2,\label{eq:unequal_sub_X_Y}\\
z_1W_1+z_2W_2+2c(Z_1+Z_2)&=0\label{eq:unequal_add_zW_Z_again}.
\end{align}
Now adding together \eqref{eq:unequal_sub_rW_X_Y} and \eqref{eq:unequal_sub_X_Y} one obtains
\begin{align}
&\tfrac{1}{(c^1_{\eta_1}-\mathfrak{x} )^{1/2}}\sum_{n=1}^\infty[B_n c^{n+1/2}_{\eta_1}+C_n s^{n+1/2}_{\eta_1}]P_n'(\mathfrak{x} )\nonumber\\
&-\tfrac{1}{(c^1_{\eta_2}-\mathfrak{x} )^{1/2}}\sum_{n=1}^\infty[B_n c^{n+1/2}_{\eta_2}+C_n s^{n+1/2}_{\eta_2}]P_n'(\mathfrak{x} )\\
&+2(c^1_{\eta_1}-\mathfrak{x} )^{1/2}\sum_{n=2}^\infty[F_n c^{n+1/2}_{\eta_1}+G_n s^{n+1/2}_{\eta_1}]P_n''(\mathfrak{x} )\nonumber\\
&-2(c^1_{\eta_2}-\mathfrak{x} )^{1/2}\sum_{n=2}^\infty[F_n c^{n+1/2}_{\eta_2}+G_n s^{n+1/2}_{\eta_2}]P_n''(\mathfrak{x} )=0.
\end{align}
Now subtracting \eqref{eq:unequal_sub_X_Y} from \eqref{eq:unequal_sub_rW_X_Y} one obtains
\begin{align}
&\tfrac{(1-\mathfrak{x} ^2)}{(c^1_{\eta_1}-\mathfrak{x} )^{1/2}}\sum_{n=1}^\infty[B_n c^{n+1/2}_{\eta_1}+C_n s^{n+1/2}_{\eta_1}]P_n'(\mathfrak{x} )\nonumber\\
&-\tfrac{(1-\mathfrak{x} ^2)}{(c^1_{\eta_2}-\mathfrak{x} )^{1/2}}\sum_{n=1}^\infty[B_n c^{n+1/2}_{\eta_2}+C_n s^{n+1/2}_{\eta_2}]P_n'(\mathfrak{x} )\\
&+2(c^1_{\eta_1}-\mathfrak{x} )^{1/2}\sum_{n=0}^\infty[D_n c^{n+1/2}_{\eta_1}+E_n s^{n+1/2}_{\eta_1}]P_n(\mathfrak{x} )\nonumber\\
&-2(c^1_{\eta_2}-\mathfrak{x} )^{1/2}\sum_{n=0}^\infty[D_n c^{n+1/2}_{\eta_2}+E_n s^{n+1/2}_{\eta_2}]P_n(\mathfrak{x} )=2.
\end{align}

\newpage


\section{List Of Notation}\label{sec:list_of_notation}

\begin{table}[!h]
\caption*{{\bf{Notation From Jeffrey \& Onishi \cite{jeffrey1984calculation}}}}
\begin{tabular}{lll}
$r$ & \quad & $d$ \\
$a_1$, $a_2$ & \quad & $r_1$, $r_2$ \\
$\lambda$ & \quad & $r_2/r_1$ \\
$\xi$ & \quad & $\frac{d-r_1-r_2}{\tfrac{1}{2}(r_1+r_2)}$ \\
$s$ & \quad & $\frac{d-r_1-r_2}{\tfrac{1}{2}(r_1+r_2)}+2$ 
\end{tabular}
\\ \vspace{1cm}
\caption*{{\bf{Abbreviations \& Mathematical Symbols}}}
\begin{tabular}{lll}
HI(s) & \quad & Hydrodynamic Interaction(s) \\
SD & \quad & Stokesian Dynamics \\
{\bf GMS} & \quad & Goddard, Mills, Sun (label for present work)\\
$\delta_{\vec{n}}$ & \quad & Directional derivative in $\vec{n}$\\
$\top$ & \quad & Transpose \\
$\nabla \cdot $ & \quad & Divergence \\
$\nabla^2 $ & \quad & Laplacian 
\end{tabular}
\end{table}

\begin{table}[!h]
\caption*{{\bf{Lower Case Greek}}}
\begin{tabular}{lll}
$\alpha$, $\eta_1$, $\eta_2$ & \quad & Spherical bipolar radial ordinate\\
$\beta$ & \quad & Sphere radii ratio $r_2/r_1$\\
$\gamma$ & \quad & Friction coefficient, Stokes constant per unit mass\\
$\gamma_n$ & \quad & $\coth\alpha\coth(n+\tfrac{1}{2})\alpha$\\
$\epsilon$ & \quad & Small nondimensional gap distance\\
$\varepsilon$ & \quad & Small perturbation from $X_\phi$\\
$(\eta,\xi,\theta)$ & \quad & Spherical bipolar coordinates\\
$\lambda$ & \quad & Separation constant in \eqref{eq:sep_of_vars_of_psi}\\
$\lambda_i$ & \quad & Eigenvalues of $\bm{R}$\\
$\mu$ & \quad & Dynamic viscosity\\
$\mu_c$ & \quad & Chemical potential\\
$\rho$ & \quad & (Constant) fluid density\\
$\varrho$ & \quad & Hard sphere fluid density\\
$\sigma$ & \quad & Sphere diameter\\
$\sigma_H$ & \quad & Hydrodynamic diameter\\
$\phi$ & \quad & Volume fraction\\
$\phi_g$ & \quad & Gauss packing fraction $\tfrac{\pi}{3\sqrt{2}}$\\
$\chi$ & \quad & Compact stream function\\
$\psi$,$\psi_1$,$\psi_2$ & \quad & Stream functions
\end{tabular}
\\ \vspace{1cm}
\caption*{{\bf{Upper Case Greek}}}
\begin{tabular}{lll}
$\Delta(n)$ & \quad & Solution coefficient in \eqref{eq:def_of_Dellta_of_n}\\
$\bm{\Gamma}$ & \quad & Dimensional $3N\times 3N$ friction tensor \\
$\tilde{\bm{\Gamma}}_{ij}$ & \quad & Nondimensional $3N\times 3N$ friction tensor \\
$\Omega$ & \quad & Fluid domain
\end{tabular}
\end{table}

\begin{table}[!h]
\caption*{{\bf{Lower Case Roman}}}
\begin{tabular}{lll}
$a_n$, $b_n$, $c_n$, $d_n$ & \quad & Squeeze summation coefficients in \eqref{eq:inf_series_force_expressions_1}--\eqref{eq:inf_series_force_expressions_2}\\
$a(\cdot)$, $b(\cdot)$ & \quad & Squeezing and shearing scalar resistance functions resp.\\
$c$ & \quad & Spherical bipolar focal length\\
$d$ & \quad & Centre to centre distance\\
$d_1$, $d_2$ & \quad & Centre distance to $O$, sphere 1 and 2 resp.\\
$d_{\min}$ & \quad & Minimal centre to centre distance\\
$f_m$, $g_m$ & \quad & Summation terms in \cite{jeffrey1982low}\\
$g(\vec{r},\vec{r}',[\varrho])$ & \quad & Two body correlation function\\
$h$ & \quad & Dimensional surface distance\\
$\mathfrak{h}$ & \quad & Metrical coefficient for spherical bipolar coordinates\\
$\vec{n}$ & \quad & Unit normal vector into fluid domain\\
$\vec{n}_{ij}$, ($\hat{\vec{n}}_{ij}$) & \quad & Intersphere (normalised) distance vector\\
$p$ & \quad & Fluid pressure\\
$r_1$, $r_2$ & \quad & Radii of sphere 1 and 2 resp.\\
$r$& \quad & $|\vec{r}|$\\
$\vec{r}_j$, $\vec{v}_j$ & \quad & Individual sphere momentum and position vectors resp. \\
$\vec{r}^N$, $\vec{v}^N$ & \quad & 3$N$ sphere momentum and position vectors resp. \\
$r_{ij}$ & \quad & $|\vec{r}_j-\vec{r}_i|$\\
$s$ & \quad & Nondimensional sphere distance in \cite{jeffrey1982low}\\
$\vec{u}$ & \quad & Velocity of Stokes fluid\\
$\vec{v}$ & \quad & Mean hard sphere fluid velocity\\
$\mathfrak{x}$ & \quad & $\cos \xi$\\
$(x,y,z)$ & \quad & Cartesian coordinates\\
$(z,r,\theta)$ & \quad & Cylindrical polar coordinates
\end{tabular}
\\ \vspace{1cm}
\caption*{{\bf{Upper Case Roman}}}
\begin{tabular}{lll}
$\bm{1}$ & \quad & $3\times 3$ identity matrix \\
$A_n$--$H_n$ & \quad & Shear summation coefficients in \eqref{eq:inf_series_force_expressions_tangential_1}--\eqref{eq:inf_series_force_expressions_tangential_2} \\
$A_{ij}^X$, $X_{ij}^A$ ,$Y_{ij}^A$ & \quad & Jeffrey \& Onishi scalar functions \\
$\bm{F}_x$ & \quad & Shearing force\\
$\bm{F}_z$ & \quad & Squeezing force\\
$\bm{F}^{\text{diss}}$ & \quad & Dissipative force \\
$\bm{F}$ & \quad & Non-dissipative force \\
$F_z$ & \quad & Non-dissipative force \\
$\mathcal{F}[\cdot]$ & \quad & Helmholtz free energy functional \\
$L$ & \quad & Characteristic length scale \\
$\bm{M}$ & \quad & Mass matrix\\
$N$ & \quad & Number of spheres \\
$P_n$ & \quad & Legendre polynomial of degree $n$\\
$Q_n$ & \quad & $P_{n+1}-P_{n-1}$\\
$\bm{R}$ & \quad & Resistance matrix \\
$Re$ & \quad & Reynolds number \\
$S_N$ & \quad & $N$--sphere configuration state space \\
$U$ & \quad & Sphere speed \\
$W$, $X$, $Y$, $Z$ & \quad & Auxiliary functions in \eqref{eq:p_decomp}--\eqref{eq:u_z_vel_decomp} \\
$\bm{Z}_1$, $\bm{Z}_2$ & \quad & Diagonal and off-diagonal matrices 
\end{tabular}
\end{table}

\newpage

\begin{acknowledgments}
BDG would like to acknowledge support from EPSRC EP/L025159/1. RDMW. is grateful to EPSRC for PhD funding. JS would like to acknowledge support from EPSRC EP/N025318/1, The Royal Academy of Engineering/The Leverhulme Trust Senior Research Fellowship LTSRF1617/13/2 and The National Natural Science Foundation of China grant 41728006.
\end{acknowledgments}

\section{Data Availability} The data that support the findings of this study are available from the corresponding author upon reasonable request,
and will be placed in an open access repository before final acceptance.

\bibliography{AIPBIB}

\begin{thebibliography}{53}%
\makeatletter
\providecommand \@ifxundefined [1]{%
 \@ifx{#1\undefined}
}%
\providecommand \@ifnum [1]{%
 \ifnum #1\expandafter \@firstoftwo
 \else \expandafter \@secondoftwo
 \fi
}%
\providecommand \@ifx [1]{%
 \ifx #1\expandafter \@firstoftwo
 \else \expandafter \@secondoftwo
 \fi
}%
\providecommand \natexlab [1]{#1}%
\providecommand \enquote  [1]{``#1''}%
\providecommand \bibnamefont  [1]{#1}%
\providecommand \bibfnamefont [1]{#1}%
\providecommand \citenamefont [1]{#1}%
\providecommand \href@noop [0]{\@secondoftwo}%
\providecommand \href [0]{\begingroup \@sanitize@url \@href}%
\providecommand \@href[1]{\@@startlink{#1}\@@href}%
\providecommand \@@href[1]{\endgroup#1\@@endlink}%
\providecommand \@sanitize@url [0]{\catcode `\\12\catcode `\$12\catcode
  `\&12\catcode `\#12\catcode `\^12\catcode `\_12\catcode `\%12\relax}%
\providecommand \@@startlink[1]{}%
\providecommand \@@endlink[0]{}%
\providecommand \url  [0]{\begingroup\@sanitize@url \@url }%
\providecommand \@url [1]{\endgroup\@href {#1}{\urlprefix }}%
\providecommand \urlprefix  [0]{URL }%
\providecommand \Eprint [0]{\href }%
\providecommand \doibase [0]{http://dx.doi.org/}%
\providecommand \selectlanguage [0]{\@gobble}%
\providecommand \bibinfo  [0]{\@secondoftwo}%
\providecommand \bibfield  [0]{\@secondoftwo}%
\providecommand \translation [1]{[#1]}%
\providecommand \BibitemOpen [0]{}%
\providecommand \bibitemStop [0]{}%
\providecommand \bibitemNoStop [0]{.\EOS\space}%
\providecommand \EOS [0]{\spacefactor3000\relax}%
\providecommand \BibitemShut  [1]{\csname bibitem#1\endcsname}%
\let\auto@bib@innerbib\@empty
\bibitem [{\citenamefont {Fall}\ \emph {et~al.}(2008)\citenamefont {Fall},
  \citenamefont {Huang}, \citenamefont {Bertrand}, \citenamefont {Ovarlez},\
  and\ \citenamefont {Bonn}}]{fall2008shear}%
  \BibitemOpen
  \bibfield  {author} {\bibinfo {author} {\bibfnamefont {A.}~\bibnamefont
  {Fall}}, \bibinfo {author} {\bibfnamefont {N.}~\bibnamefont {Huang}},
  \bibinfo {author} {\bibfnamefont {F.}~\bibnamefont {Bertrand}}, \bibinfo
  {author} {\bibfnamefont {G.}~\bibnamefont {Ovarlez}}, \ and\ \bibinfo
  {author} {\bibfnamefont {D.}~\bibnamefont {Bonn}},\ }\bibfield  {title}
  {\enquote {\bibinfo {title} {Shear thickening of cornstarch suspensions as a
  reentrant jamming transition},}\ }\href@noop {} {\bibfield  {journal}
  {\bibinfo  {journal} {Phys. Rev. Lett.}\ }\textbf {\bibinfo {volume} {100}},\
  \bibinfo {pages} {018301} (\bibinfo {year} {2008})}\BibitemShut {NoStop}%
\bibitem [{\citenamefont {Lin}\ \emph {et~al.}(2015)\citenamefont {Lin},
  \citenamefont {Guy}, \citenamefont {Hermes}, \citenamefont {Ness},
  \citenamefont {Sun}, \citenamefont {Poon},\ and\ \citenamefont
  {Cohen}}]{Lin:2015gf}%
  \BibitemOpen
  \bibfield  {author} {\bibinfo {author} {\bibfnamefont {N.~Y.~C.}\
  \bibnamefont {Lin}}, \bibinfo {author} {\bibfnamefont {B.~M.}\ \bibnamefont
  {Guy}}, \bibinfo {author} {\bibfnamefont {M.}~\bibnamefont {Hermes}},
  \bibinfo {author} {\bibfnamefont {C.}~\bibnamefont {Ness}}, \bibinfo {author}
  {\bibfnamefont {J.}~\bibnamefont {Sun}}, \bibinfo {author} {\bibfnamefont
  {W.~C.~K.}\ \bibnamefont {Poon}}, \ and\ \bibinfo {author} {\bibfnamefont
  {I.}~\bibnamefont {Cohen}},\ }\bibfield  {title} {\enquote {\bibinfo {title}
  {{Hydrodynamic and Contact Contributions to Continuous Shear Thickening in
  Colloidal Suspensions}},}\ }\href@noop {} {\bibfield  {journal} {\bibinfo
  {journal} {Phys. Rev. Lett.}\ }\textbf {\bibinfo {volume} {115}},\ \bibinfo
  {pages} {228304} (\bibinfo {year} {2015})}\BibitemShut {NoStop}%
\bibitem [{\citenamefont {Errill}(1969)}]{errill1969rheology}%
  \BibitemOpen
  \bibfield  {author} {\bibinfo {author} {\bibfnamefont {E.}~\bibnamefont
  {Errill}},\ }\bibfield  {title} {\enquote {\bibinfo {title} {Rheology of
  blood},}\ }\href@noop {} {\bibfield  {journal} {\bibinfo  {journal} {Physiol.
  Rev.}\ }\textbf {\bibinfo {volume} {49}},\ \bibinfo {pages} {863--888}
  (\bibinfo {year} {1969})}\BibitemShut {NoStop}%
\bibitem [{\citenamefont {Bagnold}(1962)}]{bagnold1962auto}%
  \BibitemOpen
  \bibfield  {author} {\bibinfo {author} {\bibfnamefont {R.~A.}\ \bibnamefont
  {Bagnold}},\ }\bibfield  {title} {\enquote {\bibinfo {title} {Auto-suspension
  of transported sediment; turbidity currents},}\ }\href@noop {} {\bibfield
  {journal} {\bibinfo  {journal} {Proc. R. Soc. Lond. A}\ }\textbf {\bibinfo
  {volume} {265}},\ \bibinfo {pages} {315--319} (\bibinfo {year}
  {1962})}\BibitemShut {NoStop}%
\bibitem [{\citenamefont {Goddard}\ \emph
  {et~al.}(2012{\natexlab{a}})\citenamefont {Goddard}, \citenamefont {Nold},
  \citenamefont {Savva}, \citenamefont {Pavliotis},\ and\ \citenamefont
  {Kalliadasis}}]{goddard2012general}%
  \BibitemOpen
  \bibfield  {author} {\bibinfo {author} {\bibfnamefont {B.~D.}\ \bibnamefont
  {Goddard}}, \bibinfo {author} {\bibfnamefont {A.}~\bibnamefont {Nold}},
  \bibinfo {author} {\bibfnamefont {N.}~\bibnamefont {Savva}}, \bibinfo
  {author} {\bibfnamefont {G.~A.}\ \bibnamefont {Pavliotis}}, \ and\ \bibinfo
  {author} {\bibfnamefont {S.}~\bibnamefont {Kalliadasis}},\ }\bibfield
  {title} {\enquote {\bibinfo {title} {General dynamical density functional
  theory for classical fluids},}\ }\href@noop {} {\bibfield  {journal}
  {\bibinfo  {journal} {Phys. Rev. Lett.}\ }\textbf {\bibinfo {volume} {109}},\
  \bibinfo {pages} {120603} (\bibinfo {year} {2012}{\natexlab{a}})}\BibitemShut
  {NoStop}%
\bibitem [{\citenamefont {Ball}\ and\ \citenamefont
  {Melrose}(1997)}]{ball1997simulation}%
  \BibitemOpen
  \bibfield  {author} {\bibinfo {author} {\bibfnamefont {R.~C.}\ \bibnamefont
  {Ball}}\ and\ \bibinfo {author} {\bibfnamefont {J.~R.}\ \bibnamefont
  {Melrose}},\ }\bibfield  {title} {\enquote {\bibinfo {title} {A simulation
  technique for many spheres in quasi-static motion under frame-invariant pair
  drag and {B}rownian forces},}\ }\href@noop {} {\bibfield  {journal} {\bibinfo
   {journal} {Physica A.}\ }\textbf {\bibinfo {volume} {247}},\ \bibinfo
  {pages} {444--472} (\bibinfo {year} {1997})}\BibitemShut {NoStop}%
\bibitem [{\citenamefont {Plischke}\ and\ \citenamefont
  {Bergersen}(1994)}]{plischke1994equilibrium}%
  \BibitemOpen
  \bibfield  {author} {\bibinfo {author} {\bibfnamefont {M.}~\bibnamefont
  {Plischke}}\ and\ \bibinfo {author} {\bibfnamefont {B.}~\bibnamefont
  {Bergersen}},\ }\href@noop {} {\emph {\bibinfo {title} {Equilibrium
  statistical physics}}}\ (\bibinfo  {publisher} {World Scientific Publishing
  Company},\ \bibinfo {year} {1994})\BibitemShut {NoStop}%
\bibitem [{\citenamefont {Rotne}\ and\ \citenamefont
  {Prager}(1969)}]{rotne1969variational}%
  \BibitemOpen
  \bibfield  {author} {\bibinfo {author} {\bibfnamefont {J.}~\bibnamefont
  {Rotne}}\ and\ \bibinfo {author} {\bibfnamefont {S.}~\bibnamefont {Prager}},\
  }\bibfield  {title} {\enquote {\bibinfo {title} {Variational treatment of
  hydrodynamic interaction in polymers},}\ }\href@noop {} {\bibfield  {journal}
  {\bibinfo  {journal} {J. Chem. Phys.}\ }\textbf {\bibinfo {volume} {50}},\
  \bibinfo {pages} {4831--4837} (\bibinfo {year} {1969})}\BibitemShut {NoStop}%
\bibitem [{\citenamefont {Bossis}\ and\ \citenamefont
  {Brady}(1984)}]{bossis1984dynamic}%
  \BibitemOpen
  \bibfield  {author} {\bibinfo {author} {\bibfnamefont {G.}~\bibnamefont
  {Bossis}}\ and\ \bibinfo {author} {\bibfnamefont {J.~F.}\ \bibnamefont
  {Brady}},\ }\bibfield  {title} {\enquote {\bibinfo {title} {Dynamic
  simulation of sheared suspensions. {I}. {G}eneral method},}\ }\href@noop {}
  {\bibfield  {journal} {\bibinfo  {journal} {J. Chem. Phys.}\ }\textbf
  {\bibinfo {volume} {80}},\ \bibinfo {pages} {5141--5154} (\bibinfo {year}
  {1984})}\BibitemShut {NoStop}%
\bibitem [{\citenamefont {Evans}\ and\ \citenamefont
  {Morriss}(1986)}]{evans1986shear}%
  \BibitemOpen
  \bibfield  {author} {\bibinfo {author} {\bibfnamefont {D.~J.}\ \bibnamefont
  {Evans}}\ and\ \bibinfo {author} {\bibfnamefont {G.~P.}\ \bibnamefont
  {Morriss}},\ }\bibfield  {title} {\enquote {\bibinfo {title} {Shear
  thickening and turbulence in simple fluids},}\ }\href@noop {} {\bibfield
  {journal} {\bibinfo  {journal} {Phys. Rev. Lett.}\ }\textbf {\bibinfo
  {volume} {56}},\ \bibinfo {pages} {2172} (\bibinfo {year}
  {1986})}\BibitemShut {NoStop}%
\bibitem [{\citenamefont {Goddard}\ \emph
  {et~al.}(2012{\natexlab{b}})\citenamefont {Goddard}, \citenamefont {Nold},
  \citenamefont {Savva}, \citenamefont {Yatsyshin},\ and\ \citenamefont
  {Kalliadasis}}]{goddard2012unification}%
  \BibitemOpen
  \bibfield  {author} {\bibinfo {author} {\bibfnamefont {B.}~\bibnamefont
  {Goddard}}, \bibinfo {author} {\bibfnamefont {A.}~\bibnamefont {Nold}},
  \bibinfo {author} {\bibfnamefont {N.}~\bibnamefont {Savva}}, \bibinfo
  {author} {\bibfnamefont {P.}~\bibnamefont {Yatsyshin}}, \ and\ \bibinfo
  {author} {\bibfnamefont {S.}~\bibnamefont {Kalliadasis}},\ }\bibfield
  {title} {\enquote {\bibinfo {title} {Unification of dynamic density
  functional theory for colloidal fluids to include inertia and hydrodynamic
  interactions: derivation and numerical experiments},}\ }\href@noop {}
  {\bibfield  {journal} {\bibinfo  {journal} {J. Phys.: Condens. Matter}\
  }\textbf {\bibinfo {volume} {25}},\ \bibinfo {pages} {035101} (\bibinfo
  {year} {2012}{\natexlab{b}})}\BibitemShut {NoStop}%
\bibitem [{\citenamefont {Kim}\ and\ \citenamefont
  {Karrila}(2013)}]{kim2013microhydrodynamics}%
  \BibitemOpen
  \bibfield  {author} {\bibinfo {author} {\bibfnamefont {S.}~\bibnamefont
  {Kim}}\ and\ \bibinfo {author} {\bibfnamefont {S.~J.}\ \bibnamefont
  {Karrila}},\ }\href@noop {} {\emph {\bibinfo {title} {Microhydrodynamics:
  principles and selected applications}}}\ (\bibinfo  {publisher} {Courier
  Corporation},\ \bibinfo {year} {2013})\BibitemShut {NoStop}%
\bibitem [{\citenamefont {Stimson}\ and\ \citenamefont
  {Jeffery}(1926)}]{stimson1926motion}%
  \BibitemOpen
  \bibfield  {author} {\bibinfo {author} {\bibfnamefont {M.}~\bibnamefont
  {Stimson}}\ and\ \bibinfo {author} {\bibfnamefont {G.}~\bibnamefont
  {Jeffery}},\ }\bibfield  {title} {\enquote {\bibinfo {title} {The motion of
  two spheres in a viscous fluid},}\ }\href@noop {} {\bibfield  {journal}
  {\bibinfo  {journal} {P. R. Soc. Lond. A-Conta.}\ }\textbf {\bibinfo {volume}
  {111}},\ \bibinfo {pages} {110--116} (\bibinfo {year} {1926})}\BibitemShut
  {NoStop}%
\bibitem [{\citenamefont {Goldman}, \citenamefont {Cox},\ and\ \citenamefont
  {Brenner}(1966)}]{goldman1966slow}%
  \BibitemOpen
  \bibfield  {author} {\bibinfo {author} {\bibfnamefont {A.}~\bibnamefont
  {Goldman}}, \bibinfo {author} {\bibfnamefont {R.}~\bibnamefont {Cox}}, \ and\
  \bibinfo {author} {\bibfnamefont {H.}~\bibnamefont {Brenner}},\ }\bibfield
  {title} {\enquote {\bibinfo {title} {The slow motion of two identical
  arbitrarily oriented spheres through a viscous fluid},}\ }\href@noop {}
  {\bibfield  {journal} {\bibinfo  {journal} {Chem. Eng. Sci.}\ }\textbf
  {\bibinfo {volume} {21}},\ \bibinfo {pages} {1151--1170} (\bibinfo {year}
  {1966})}\BibitemShut {NoStop}%
\bibitem [{\citenamefont {Happel}\ and\ \citenamefont
  {Brenner}(2012)}]{happel2012low}%
  \BibitemOpen
  \bibfield  {author} {\bibinfo {author} {\bibfnamefont {J.}~\bibnamefont
  {Happel}}\ and\ \bibinfo {author} {\bibfnamefont {H.}~\bibnamefont
  {Brenner}},\ }\href@noop {} {\emph {\bibinfo {title} {Low {R}eynolds number
  hydrodynamics: with special applications to particulate media}}},\
  Vol.~\bibinfo {volume} {1}\ (\bibinfo  {publisher} {Springer Science \&
  Business Media},\ \bibinfo {year} {2012})\BibitemShut {NoStop}%
\bibitem [{\citenamefont {Townsend}(2018)}]{townsend2018generating}%
  \BibitemOpen
  \bibfield  {author} {\bibinfo {author} {\bibfnamefont {A.~K.}\ \bibnamefont
  {Townsend}},\ }\bibfield  {title} {\enquote {\bibinfo {title} {Generating,
  from scratch, the near-field asymptotic forms of scalar resistance functions
  for two unequal rigid spheres in low-{R}eynolds-number flow},}\ }\href@noop
  {} {\bibfield  {journal} {\bibinfo  {journal} {arXiv preprint
  arXiv:1802.08226}\ } (\bibinfo {year} {2018})}\BibitemShut {NoStop}%
\bibitem [{\citenamefont {Jeffrey}\ and\ \citenamefont
  {Onishi}(1984)}]{jeffrey1984calculation}%
  \BibitemOpen
  \bibfield  {author} {\bibinfo {author} {\bibfnamefont {D.}~\bibnamefont
  {Jeffrey}}\ and\ \bibinfo {author} {\bibfnamefont {Y.}~\bibnamefont
  {Onishi}},\ }\bibfield  {title} {\enquote {\bibinfo {title} {Calculation of
  the resistance and mobility functions for two unequal rigid spheres in
  low-{R}eynolds-number flow},}\ }\href@noop {} {\bibfield  {journal} {\bibinfo
   {journal} {J. Fluid. Mech.}\ }\textbf {\bibinfo {volume} {139}},\ \bibinfo
  {pages} {261--290} (\bibinfo {year} {1984})}\BibitemShut {NoStop}%
\bibitem [{\citenamefont {Fax{\'e}n}(1927)}]{faxen1927}%
  \BibitemOpen
  \bibfield  {author} {\bibinfo {author} {\bibfnamefont {H.}~\bibnamefont
  {Fax{\'e}n}},\ }\bibfield  {title} {\enquote {\bibinfo {title} {Die
  geschwindigkeit zweier kugeln, die unter einwirkung der schwere in einer
  z{\"a}hen fl{\"u}ssigkeit fallen},}\ }\href@noop {} {\bibfield  {journal}
  {\bibinfo  {journal} {Z. Angew. Math. Mech.}\ }\textbf {\bibinfo {volume}
  {7}},\ \bibinfo {pages} {79--81} (\bibinfo {year} {1927})}\BibitemShut
  {NoStop}%
\bibitem [{\citenamefont {Bart}(1959)}]{bart1959interaction}%
  \BibitemOpen
  \bibfield  {author} {\bibinfo {author} {\bibfnamefont {E.~N.}\ \bibnamefont
  {Bart}},\ }\emph {\bibinfo {title} {Interaction of two spheres falling slowly
  in a viscous medium}},\ \href@noop {} {Ph.D. thesis} (\bibinfo {year}
  {1959})\BibitemShut {NoStop}%
\bibitem [{\citenamefont {Maude}(1961)}]{maude1961end}%
  \BibitemOpen
  \bibfield  {author} {\bibinfo {author} {\bibfnamefont {A.~D.}\ \bibnamefont
  {Maude}},\ }\bibfield  {title} {\enquote {\bibinfo {title} {End effects in a
  falling-sphere viscometer},}\ }\href@noop {} {\bibfield  {journal} {\bibinfo
  {journal} {British Journal of Applied Physics}\ }\textbf {\bibinfo {volume}
  {12}},\ \bibinfo {pages} {293} (\bibinfo {year} {1961})}\BibitemShut
  {NoStop}%
\bibitem [{\citenamefont {O'Neill}(1964)}]{o1964slow}%
  \BibitemOpen
  \bibfield  {author} {\bibinfo {author} {\bibfnamefont {M.~E.}\ \bibnamefont
  {O'Neill}},\ }\bibfield  {title} {\enquote {\bibinfo {title} {A slow motion
  of viscous liquid caused by a slowly moving solid sphere},}\ }\href@noop {}
  {\bibfield  {journal} {\bibinfo  {journal} {Mathematika}\ }\textbf {\bibinfo
  {volume} {11}},\ \bibinfo {pages} {67--74} (\bibinfo {year}
  {1964})}\BibitemShut {NoStop}%
\bibitem [{\citenamefont {O'Neill}\ and\ \citenamefont
  {Majumdar}(1970)}]{o1970asymmetrical}%
  \BibitemOpen
  \bibfield  {author} {\bibinfo {author} {\bibfnamefont {M.~E.}\ \bibnamefont
  {O'Neill}}\ and\ \bibinfo {author} {\bibfnamefont {R.}~\bibnamefont
  {Majumdar}},\ }\bibfield  {title} {\enquote {\bibinfo {title} {Asymmetrical
  slow viscous fluid motions caused by the translation or rotation of two
  spheres. part {I}: The determination of exact solutions for any values of the
  ratio of radii and separation parameters},}\ }\href@noop {} {\bibfield
  {journal} {\bibinfo  {journal} {Z. Angew. Math. Physik.}\ }\textbf {\bibinfo
  {volume} {21}},\ \bibinfo {pages} {164--179} (\bibinfo {year}
  {1970})}\BibitemShut {NoStop}%
\bibitem [{\citenamefont {Cox}\ and\ \citenamefont
  {Brenner}(1967)}]{cox1967slow}%
  \BibitemOpen
  \bibfield  {author} {\bibinfo {author} {\bibfnamefont {R.~G.}\ \bibnamefont
  {Cox}}\ and\ \bibinfo {author} {\bibfnamefont {H.}~\bibnamefont {Brenner}},\
  }\bibfield  {title} {\enquote {\bibinfo {title} {The slow motion of a sphere
  through a viscous fluid towards a plane surface ii small gap widths,
  including inertial effects},}\ }\href@noop {} {\bibfield  {journal} {\bibinfo
   {journal} {Chem. Eng. Sci.}\ }\textbf {\bibinfo {volume} {22}},\ \bibinfo
  {pages} {1753--1777} (\bibinfo {year} {1967})}\BibitemShut {NoStop}%
\bibitem [{\citenamefont {Hansford}(1970)}]{hansford1970converging}%
  \BibitemOpen
  \bibfield  {author} {\bibinfo {author} {\bibfnamefont {R.~E.}\ \bibnamefont
  {Hansford}},\ }\bibfield  {title} {\enquote {\bibinfo {title} {On converging
  solid spheres in a highly viscous fluid},}\ }\href@noop {} {\bibfield
  {journal} {\bibinfo  {journal} {Mathematika}\ }\textbf {\bibinfo {volume}
  {17}},\ \bibinfo {pages} {250--254} (\bibinfo {year} {1970})}\BibitemShut
  {NoStop}%
\bibitem [{\citenamefont {Brenner}(1961)}]{brenner1961slow}%
  \BibitemOpen
  \bibfield  {author} {\bibinfo {author} {\bibfnamefont {H.}~\bibnamefont
  {Brenner}},\ }\bibfield  {title} {\enquote {\bibinfo {title} {The slow motion
  of a sphere through a viscous fluid towards a plane surface},}\ }\href@noop
  {} {\bibfield  {journal} {\bibinfo  {journal} {Chem. Eng. Sci.}\ }\textbf
  {\bibinfo {volume} {16}},\ \bibinfo {pages} {242--251} (\bibinfo {year}
  {1961})}\BibitemShut {NoStop}%
\bibitem [{\citenamefont {Papavassiliou}\ and\ \citenamefont
  {Alexander}(2017)}]{papavassiliou2017exact}%
  \BibitemOpen
  \bibfield  {author} {\bibinfo {author} {\bibfnamefont {D.}~\bibnamefont
  {Papavassiliou}}\ and\ \bibinfo {author} {\bibfnamefont {G.~P.}\ \bibnamefont
  {Alexander}},\ }\bibfield  {title} {\enquote {\bibinfo {title} {Exact
  solutions for hydrodynamic interactions of two squirming spheres},}\
  }\href@noop {} {\bibfield  {journal} {\bibinfo  {journal} {J. Fluid. Mech.}\
  }\textbf {\bibinfo {volume} {813}},\ \bibinfo {pages} {618--646} (\bibinfo
  {year} {2017})}\BibitemShut {NoStop}%
\bibitem [{\citenamefont {Wacholder}\ and\ \citenamefont
  {Weihs}(1972)}]{wacholder1972slow}%
  \BibitemOpen
  \bibfield  {author} {\bibinfo {author} {\bibfnamefont {E.}~\bibnamefont
  {Wacholder}}\ and\ \bibinfo {author} {\bibfnamefont {D.}~\bibnamefont
  {Weihs}},\ }\bibfield  {title} {\enquote {\bibinfo {title} {Slow motion of a
  fluid sphere in the vicinity of another sphere or a plane boundary},}\
  }\href@noop {} {\bibfield  {journal} {\bibinfo  {journal} {Chem. Eng. Sci.}\
  }\textbf {\bibinfo {volume} {27}},\ \bibinfo {pages} {1817--1828} (\bibinfo
  {year} {1972})}\BibitemShut {NoStop}%
\bibitem [{\citenamefont {Haber}, \citenamefont {Hetsroni},\ and\ \citenamefont
  {Solan}(1973)}]{haber1973low}%
  \BibitemOpen
  \bibfield  {author} {\bibinfo {author} {\bibfnamefont {S.}~\bibnamefont
  {Haber}}, \bibinfo {author} {\bibfnamefont {G.}~\bibnamefont {Hetsroni}}, \
  and\ \bibinfo {author} {\bibfnamefont {A.}~\bibnamefont {Solan}},\ }\bibfield
   {title} {\enquote {\bibinfo {title} {On the low {R}eynolds number motion of
  two droplets},}\ }\href@noop {} {\bibfield  {journal} {\bibinfo  {journal}
  {Int. J. Multiphas. Flow.}\ }\textbf {\bibinfo {volume} {1}},\ \bibinfo
  {pages} {57--71} (\bibinfo {year} {1973})}\BibitemShut {NoStop}%
\bibitem [{\citenamefont {Jeffrey}(1982)}]{jeffrey1982low}%
  \BibitemOpen
  \bibfield  {author} {\bibinfo {author} {\bibfnamefont {D.~J.}\ \bibnamefont
  {Jeffrey}},\ }\bibfield  {title} {\enquote {\bibinfo {title}
  {Low-{R}eynolds-number flow between converging spheres},}\ }\href@noop {}
  {\bibfield  {journal} {\bibinfo  {journal} {Mathematika}\ }\textbf {\bibinfo
  {volume} {29}},\ \bibinfo {pages} {58--66} (\bibinfo {year}
  {1982})}\BibitemShut {NoStop}%
\bibitem [{\citenamefont {Townsend}\ and\ \citenamefont
  {Wilson}(2018)}]{townsend2018anomalous}%
  \BibitemOpen
  \bibfield  {author} {\bibinfo {author} {\bibfnamefont {A.}~\bibnamefont
  {Townsend}}\ and\ \bibinfo {author} {\bibfnamefont {H.}~\bibnamefont
  {Wilson}},\ }\bibfield  {title} {\enquote {\bibinfo {title} {Anomalous effect
  of turning off long-range mobility interactions in stokesian dynamics},}\
  }\href@noop {} {\bibfield  {journal} {\bibinfo  {journal} {Phys. Fluids.}\
  }\textbf {\bibinfo {volume} {30}},\ \bibinfo {pages} {077103} (\bibinfo
  {year} {2018})}\BibitemShut {NoStop}%
\bibitem [{\citenamefont {Payne}\ and\ \citenamefont
  {Pell}(1960)}]{payne1960stokes}%
  \BibitemOpen
  \bibfield  {author} {\bibinfo {author} {\bibfnamefont {L.~E.}\ \bibnamefont
  {Payne}}\ and\ \bibinfo {author} {\bibfnamefont {W.}~\bibnamefont {Pell}},\
  }\bibfield  {title} {\enquote {\bibinfo {title} {The stokes flow problem for
  a class of axially symmetric bodies},}\ }\href@noop {} {\bibfield  {journal}
  {\bibinfo  {journal} {J. Fluid. Mech.}\ }\textbf {\bibinfo {volume} {7}},\
  \bibinfo {pages} {529--549} (\bibinfo {year} {1960})}\BibitemShut {NoStop}%
\bibitem [{\citenamefont {Jeffrey}({\natexlab{a}})}]{jeffrey_online_code}%
  \BibitemOpen
  \bibfield  {author} {\bibinfo {author} {\bibfnamefont {D.~J.}\ \bibnamefont
  {Jeffrey}},\ }\href@noop {} {\enquote {\bibinfo {title} {{Programs for Stokes
  Resistance Functions}},}\ }\bibinfo {howpublished}
  {\url{https://www.uwo.ca/apmaths/faculty/jeffrey/research/Resistance.html}}
  ({\natexlab{a}}),\ \bibinfo {note} {[Accessed 01-March-2019]}\BibitemShut
  {NoStop}%
\bibitem [{\citenamefont {Jeffrey}(1992)}]{jeffrey1992calculation}%
  \BibitemOpen
  \bibfield  {author} {\bibinfo {author} {\bibfnamefont {D.}~\bibnamefont
  {Jeffrey}},\ }\bibfield  {title} {\enquote {\bibinfo {title} {The calculation
  of the low {R}eynolds number resistance functions for two unequal spheres},}\
  }\href@noop {} {\bibfield  {journal} {\bibinfo  {journal} {Phys. Fluids.
  A-Fluid.}\ }\textbf {\bibinfo {volume} {4}},\ \bibinfo {pages} {16--29}
  (\bibinfo {year} {1992})}\BibitemShut {NoStop}%
\bibitem [{\citenamefont {Jeffrey}({\natexlab{b}})}]{jeffrey_online_code_300}%
  \BibitemOpen
  \bibfield  {author} {\bibinfo {author} {\bibfnamefont {D.~J.}\ \bibnamefont
  {Jeffrey}},\ }\href@noop {} {\enquote {\bibinfo {title} {{Programs for Stokes
  Resistance Functions, first 300 terms of $X^A_{11}$ eq (3.13)}},}\ }\bibinfo
  {howpublished}
  {\url{https://www.uwo.ca/apmaths/faculty/jeffrey/research/resistancefunctions/xa/rxa300.dat.txt}}
  ({\natexlab{b}}),\ \bibinfo {note} {[Accessed 01-March-2019]}\BibitemShut
  {NoStop}%
\bibitem [{\citenamefont {Trefethen}\ and\ \citenamefont
  {Bau~III}(1997)}]{trefethen1997numerical}%
  \BibitemOpen
  \bibfield  {author} {\bibinfo {author} {\bibfnamefont {L.~N.}\ \bibnamefont
  {Trefethen}}\ and\ \bibinfo {author} {\bibfnamefont {D.}~\bibnamefont
  {Bau~III}},\ }\href@noop {} {\emph {\bibinfo {title} {Numerical linear
  algebra}}},\ Vol.~\bibinfo {volume} {50}\ (\bibinfo  {publisher} {Siam},\
  \bibinfo {year} {1997})\BibitemShut {NoStop}%
\bibitem [{\citenamefont {Rosenfeld}(1989)}]{rosenfeld1989free}%
  \BibitemOpen
  \bibfield  {author} {\bibinfo {author} {\bibfnamefont {Y.}~\bibnamefont
  {Rosenfeld}},\ }\bibfield  {title} {\enquote {\bibinfo {title} {Free-energy
  model for the inhomogeneous hard-sphere fluid mixture and density-functional
  theory of freezing},}\ }\href@noop {} {\bibfield  {journal} {\bibinfo
  {journal} {Phys. Rev. Lett.}\ }\textbf {\bibinfo {volume} {63}},\ \bibinfo
  {pages} {980} (\bibinfo {year} {1989})}\BibitemShut {NoStop}%
\bibitem [{\citenamefont {Roth}(2010)}]{roth2010fundamental}%
  \BibitemOpen
  \bibfield  {author} {\bibinfo {author} {\bibfnamefont {R.}~\bibnamefont
  {Roth}},\ }\bibfield  {title} {\enquote {\bibinfo {title} {Fundamental
  measure theory for hard-sphere mixtures: a review},}\ }\href@noop {}
  {\bibfield  {journal} {\bibinfo  {journal} {J. Phys.: Condens. Matter}\
  }\textbf {\bibinfo {volume} {22}},\ \bibinfo {pages} {063102} (\bibinfo
  {year} {2010})}\BibitemShut {NoStop}%
\bibitem [{\citenamefont {Goddard}, \citenamefont {Nold},\ and\ \citenamefont
  {Kalliadasis}(2013)}]{goddard2013multi}%
  \BibitemOpen
  \bibfield  {author} {\bibinfo {author} {\bibfnamefont {B.}~\bibnamefont
  {Goddard}}, \bibinfo {author} {\bibfnamefont {A.}~\bibnamefont {Nold}}, \
  and\ \bibinfo {author} {\bibfnamefont {S.}~\bibnamefont {Kalliadasis}},\
  }\bibfield  {title} {\enquote {\bibinfo {title} {Multi-species dynamical
  density functional theory},}\ }\href@noop {} {\bibfield  {journal} {\bibinfo
  {journal} {J. Chem. Phys.}\ }\textbf {\bibinfo {volume} {138}},\ \bibinfo
  {pages} {144904} (\bibinfo {year} {2013})}\BibitemShut {NoStop}%
\bibitem [{\citenamefont {Goddard}, \citenamefont {Nold},\ and\ \citenamefont
  {Kalliadasis}(2017)}]{DDFTCode}%
  \BibitemOpen
  \bibfield  {author} {\bibinfo {author} {\bibfnamefont {B.~D.}\ \bibnamefont
  {Goddard}}, \bibinfo {author} {\bibfnamefont {A.}~\bibnamefont {Nold}}, \
  and\ \bibinfo {author} {\bibfnamefont {S.}~\bibnamefont {Kalliadasis}},\
  }\href@noop {} {\enquote {\bibinfo {title} {{2DChebClass [Software]}},}\
  }\bibinfo {howpublished} {http://dx.doi.org/10.7488/ds/1991} (\bibinfo {year}
  {2017})\BibitemShut {NoStop}%
\bibitem [{\citenamefont {Nold}\ \emph {et~al.}(2017)\citenamefont {Nold},
  \citenamefont {Goddard}, \citenamefont {Yatsyshin}, \citenamefont {Savva},\
  and\ \citenamefont {Kalliadasis}}]{nold2017pseudospectral}%
  \BibitemOpen
  \bibfield  {author} {\bibinfo {author} {\bibfnamefont {A.}~\bibnamefont
  {Nold}}, \bibinfo {author} {\bibfnamefont {B.~D.}\ \bibnamefont {Goddard}},
  \bibinfo {author} {\bibfnamefont {P.}~\bibnamefont {Yatsyshin}}, \bibinfo
  {author} {\bibfnamefont {N.}~\bibnamefont {Savva}}, \ and\ \bibinfo {author}
  {\bibfnamefont {S.}~\bibnamefont {Kalliadasis}},\ }\bibfield  {title}
  {\enquote {\bibinfo {title} {Pseudospectral methods for density functional
  theory in bounded and unbounded domains},}\ }\href@noop {} {\bibfield
  {journal} {\bibinfo  {journal} {J. Comput. Phys.}\ }\textbf {\bibinfo
  {volume} {334}},\ \bibinfo {pages} {639--664} (\bibinfo {year}
  {2017})}\BibitemShut {NoStop}%
\bibitem [{\citenamefont {Goddard}, \citenamefont {Nold},\ and\ \citenamefont
  {Kalliadasis}(2016)}]{goddard2016dynamical}%
  \BibitemOpen
  \bibfield  {author} {\bibinfo {author} {\bibfnamefont {B.}~\bibnamefont
  {Goddard}}, \bibinfo {author} {\bibfnamefont {A.}~\bibnamefont {Nold}}, \
  and\ \bibinfo {author} {\bibfnamefont {S.}~\bibnamefont {Kalliadasis}},\
  }\bibfield  {title} {\enquote {\bibinfo {title} {Dynamical density functional
  theory with hydrodynamic interactions in confined geometries},}\ }\href@noop
  {} {\bibfield  {journal} {\bibinfo  {journal} {J. Chem. Phys.}\ }\textbf
  {\bibinfo {volume} {145}},\ \bibinfo {pages} {214106} (\bibinfo {year}
  {2016})}\BibitemShut {NoStop}%
\bibitem [{\citenamefont {Ness}\ and\ \citenamefont
  {Sun}(2016)}]{ness2016shear}%
  \BibitemOpen
  \bibfield  {author} {\bibinfo {author} {\bibfnamefont {C.}~\bibnamefont
  {Ness}}\ and\ \bibinfo {author} {\bibfnamefont {J.}~\bibnamefont {Sun}},\
  }\bibfield  {title} {\enquote {\bibinfo {title} {Shear thickening regimes of
  dense non-{B}rownian suspensions},}\ }\href@noop {} {\bibfield  {journal}
  {\bibinfo  {journal} {Soft Matter}\ }\textbf {\bibinfo {volume} {12}},\
  \bibinfo {pages} {914--924} (\bibinfo {year} {2016})}\BibitemShut {NoStop}%
\bibitem [{\citenamefont {Nguyen}\ and\ \citenamefont
  {Ladd}(2002)}]{Nguyen:02a}%
  \BibitemOpen
  \bibfield  {author} {\bibinfo {author} {\bibfnamefont {N.}~\bibnamefont
  {Nguyen}}\ and\ \bibinfo {author} {\bibfnamefont {A.}~\bibnamefont {Ladd}},\
  }\bibfield  {title} {\enquote {\bibinfo {title} {{Lubrication corrections for
  lattice-Boltzmann simulations of particle suspensions}},}\ }\href@noop {}
  {\bibfield  {journal} {\bibinfo  {journal} {Phys. Rev. E}\ }\textbf {\bibinfo
  {volume} {66}},\ \bibinfo {pages} {046708--046708} (\bibinfo {year}
  {2002})}\BibitemShut {NoStop}%
\bibitem [{\citenamefont {Brady}\ and\ \citenamefont
  {Bossis}(1988)}]{Brady:1988up}%
  \BibitemOpen
  \bibfield  {author} {\bibinfo {author} {\bibfnamefont {J.~F.}\ \bibnamefont
  {Brady}}\ and\ \bibinfo {author} {\bibfnamefont {G.}~\bibnamefont {Bossis}},\
  }\bibfield  {title} {\enquote {\bibinfo {title} {{Stokesian dynamics}},}\
  }\href@noop {} {\bibfield  {journal} {\bibinfo  {journal} {Annu. Rev. Fluid.
  Mech.}\ }\textbf {\bibinfo {volume} {20}},\ \bibinfo {pages} {111--157}
  (\bibinfo {year} {1988})}\BibitemShut {NoStop}%
\bibitem [{\citenamefont {Lefebvre-Lepot}, \citenamefont {Merlet},\ and\
  \citenamefont {Nguyen}(2015)}]{lefebvre2015accurate}%
  \BibitemOpen
  \bibfield  {author} {\bibinfo {author} {\bibfnamefont {A.}~\bibnamefont
  {Lefebvre-Lepot}}, \bibinfo {author} {\bibfnamefont {B.}~\bibnamefont
  {Merlet}}, \ and\ \bibinfo {author} {\bibfnamefont {T.}~\bibnamefont
  {Nguyen}},\ }\bibfield  {title} {\enquote {\bibinfo {title} {An accurate
  method to include lubrication forces in numerical simulations of dense
  {S}tokesian suspensions},}\ }\href@noop {} {\bibfield  {journal} {\bibinfo
  {journal} {J. Fluid. Mech.}\ }\textbf {\bibinfo {volume} {769}},\ \bibinfo
  {pages} {369--386} (\bibinfo {year} {2015})}\BibitemShut {NoStop}%
\bibitem [{\citenamefont {Archer}(2009)}]{archer2009dynamical}%
  \BibitemOpen
  \bibfield  {author} {\bibinfo {author} {\bibfnamefont {A.~J.}\ \bibnamefont
  {Archer}},\ }\bibfield  {title} {\enquote {\bibinfo {title} {Dynamical
  density functional theory for molecular and colloidal fluids: A microscopic
  approach to fluid mechanics},}\ }\href@noop {} {\bibfield  {journal}
  {\bibinfo  {journal} {J. Chem. Phys.}\ }\textbf {\bibinfo {volume} {130}},\
  \bibinfo {pages} {014509} (\bibinfo {year} {2009})}\BibitemShut {NoStop}%
\bibitem [{\citenamefont {Dur{\'a}n-Olivencia}, \citenamefont {Goddard},\ and\
  \citenamefont {Kalliadasis}(2016)}]{duran2016dynamical}%
  \BibitemOpen
  \bibfield  {author} {\bibinfo {author} {\bibfnamefont {M.}~\bibnamefont
  {Dur{\'a}n-Olivencia}}, \bibinfo {author} {\bibfnamefont {B.}~\bibnamefont
  {Goddard}}, \ and\ \bibinfo {author} {\bibfnamefont {S.}~\bibnamefont
  {Kalliadasis}},\ }\bibfield  {title} {\enquote {\bibinfo {title} {Dynamical
  density functional theory for orientable colloids including inertia and
  hydrodynamic interactions},}\ }\href@noop {} {\bibfield  {journal} {\bibinfo
  {journal} {J. Stat. Phys.}\ }\textbf {\bibinfo {volume} {164}},\ \bibinfo
  {pages} {785--809} (\bibinfo {year} {2016})}\BibitemShut {NoStop}%
\bibitem [{\citenamefont {Cox}(1986)}]{cox1986dynamics}%
  \BibitemOpen
  \bibfield  {author} {\bibinfo {author} {\bibfnamefont {R.}~\bibnamefont
  {Cox}},\ }\bibfield  {title} {\enquote {\bibinfo {title} {The dynamics of the
  spreading of liquids on a solid surface. part 1. viscous flow},}\ }\href@noop
  {} {\bibfield  {journal} {\bibinfo  {journal} {J. Fluid. Mech.}\ }\textbf
  {\bibinfo {volume} {168}},\ \bibinfo {pages} {169--194} (\bibinfo {year}
  {1986})}\BibitemShut {NoStop}%
\bibitem [{\citenamefont {Thompson}\ and\ \citenamefont
  {Troian}(1997)}]{thompson1997general}%
  \BibitemOpen
  \bibfield  {author} {\bibinfo {author} {\bibfnamefont {P.}~\bibnamefont
  {Thompson}}\ and\ \bibinfo {author} {\bibfnamefont {S.}~\bibnamefont
  {Troian}},\ }\bibfield  {title} {\enquote {\bibinfo {title} {A general
  boundary condition for liquid flow at solid surfaces},}\ }\href@noop {}
  {\bibfield  {journal} {\bibinfo  {journal} {Nature}\ }\textbf {\bibinfo
  {volume} {389}},\ \bibinfo {pages} {360} (\bibinfo {year}
  {1997})}\BibitemShut {NoStop}%
\bibitem [{\citenamefont {Hinch}(1991)}]{hinch1991perturbation}%
  \BibitemOpen
  \bibfield  {author} {\bibinfo {author} {\bibfnamefont {E.~J.}\ \bibnamefont
  {Hinch}},\ }\href@noop {} {\emph {\bibinfo {title} {Perturbation methods}}}\
  (\bibinfo  {publisher} {Cambridge university press},\ \bibinfo {year}
  {1991})\BibitemShut {NoStop}%
\bibitem [{\citenamefont {Kac}\ and\ \citenamefont
  {Cheung}(2001)}]{kac2001quantum}%
  \BibitemOpen
  \bibfield  {author} {\bibinfo {author} {\bibfnamefont {V.}~\bibnamefont
  {Kac}}\ and\ \bibinfo {author} {\bibfnamefont {P.}~\bibnamefont {Cheung}},\
  }\href@noop {} {\emph {\bibinfo {title} {Quantum calculus}}}\ (\bibinfo
  {publisher} {Springer Science \& Business Media},\ \bibinfo {year}
  {2001})\BibitemShut {NoStop}%
\bibitem [{\citenamefont {Weinstein}(1948)}]{weinstein1948discontinuous}%
  \BibitemOpen
  \bibfield  {author} {\bibinfo {author} {\bibfnamefont {A.}~\bibnamefont
  {Weinstein}},\ }\bibfield  {title} {\enquote {\bibinfo {title} {Discontinuous
  integrals and generalized potential theory},}\ }\href@noop {} {\bibfield
  {journal} {\bibinfo  {journal} {T. Am. Math. Soc.}\ }\textbf {\bibinfo
  {volume} {63}},\ \bibinfo {pages} {342--354} (\bibinfo {year}
  {1948})}\BibitemShut {NoStop}%
\bibitem [{\citenamefont {Payne}(1959)}]{payne1959representation}%
  \BibitemOpen
  \bibfield  {author} {\bibinfo {author} {\bibfnamefont {L.~E.}\ \bibnamefont
  {Payne}},\ }\bibfield  {title} {\enquote {\bibinfo {title} {Representation
  formulas for solutions of a class of partial differential equations},}\
  }\href@noop {} {\bibfield  {journal} {\bibinfo  {journal} {J. Math. Phys.
  Camb.}\ }\textbf {\bibinfo {volume} {38}},\ \bibinfo {pages} {145--149}
  (\bibinfo {year} {1959})}\BibitemShut {NoStop}%
\end{thebibliography}%


%

\end{document}